\documentclass[twocolumn]{aastex62}

\usepackage{longtable} 
\usepackage{graphicx} 
\usepackage{threeparttablex}
\usepackage{multirow}

\DeclareRobustCommand{\ion}[2]{%
\relax\ifmmode
\ifx\testbx\f@series
{\mathbf{#1\,\mathsc{#2}}}\else
{\mathrm{#1\,\mathsc{#2}}}\fi
\else\textup{#1\,{\mdseries\textsc{#2}}}%
\fi}

\newcommand{\hst}{\textit{HST}~}
\newcommand{\fuse}{\textit{FUSE}~}
\newcommand{\ebv}{{$E(B-V)~$}}

\newcommand{\fnz}{f(N_{\textrm{\ion{H}{i}}}, z)}
\newcommand{\lmfp}{\lambda_{\rm mfp}^{912}}

\newcommand{\Tl}{T_\lambda}

\newcommand{\zabs}{z_{\rm abs}}
\newcommand{\mbh}{M_{\rm BH}}
\newcommand{\lbol}{L_{\rm bol}}
\newcommand{\lumedd}{L_{\rm Edd}}
\newcommand{\nh}{N_{\rm HI}}
\newcommand{\taue}{\tau_{\lambda,\rm eff}}
\newcommand{\taull}{\tau^{\rm LL}_{\rm eff}}
\newcommand{\tauly}{\tau^{\rm Ly}_{\rm eff}}
\newcommand{\taulls}{\tau^{\rm LL}_{912}}

\newcommand{\ledd}{\lambda_{\rm Edd}}

\newcommand{\mi}{M_i(z=2)}

\newcommand{\mnhi}{N_{\rm HI}}

\newcommand{\mkms}{{\rm km~s^{-1}}}
\newcommand{\h}{h_{70}^{-1}}

\newcommand{\rev}[1]{{ #1}}
\newcommand{\revs}[1]{{ #1}}

\graphicspath{{./}{figures/}}

\received{February 26, 2018}
\accepted{May 4, 2018}
\submitjournal{ApJ}

\shorttitle{HST/WFC3 survey of quasar pairs}
\shortauthors{Lusso et al.}

\begin{document}

\title{THE SPECTRAL AND ENVIRONMENT PROPERTIES OF  \lowercase{$z\sim2.0-2.5$} QUASAR PAIRS}

\correspondingauthor{Elisabeta Lusso}
\email{elisabeta.lusso@durham.ac.uk}
\author{Elisabeta Lusso}
\affil{Centre for Extragalactic Astronomy, Department of Physics, Durham University, South Road, Durham, DH1 3LE, UK}
\author{Michele Fumagalli}
\affiliation{Centre for Extragalactic Astronomy, Department of Physics, Durham University, South Road, Durham, DH1 3LE, UK}
\affiliation{Institute for Computational Cosmology, Durham University, South Road, Durham DH1 3LE, UK}
\author{Marc Rafelski}
\affiliation{Space Telescope Science Institute, 3700 San Martin Drive, Baltimore, MD 21218, USA}
\affiliation{Department of Physics \& Astronomy, Johns Hopkins University, Baltimore, MD 21218, USA}
\author{Marcel Neeleman}
\affiliation{University of California Observatories, Lick Observatory, 1156 High Street, Santa Cruz, CA 95064, USA}
\affiliation{Max-Planck-Institut fuer Astronomie, Koenigstuhl 17, D-69117 Heidelberg, Germany}
\author{Jason X. Prochaska}
\affiliation{Department of Astronomy and Astrophysics, University of California, 1156 High Street, Santa Cruz, CA 95064, USA}
\affiliation{University of California Observatories, Lick Observatory, 1156 High Street, Santa Cruz, CA 95064, USA}
\author{Joseph F. Hennawi}
\affiliation{Department of Physics, Broida Hall, University of California, Santa Barbara, CA 93106, USA}
\affiliation{Max-Planck-Institut fuer Astronomie, Koenigstuhl 17, D-69117 Heidelberg, Germany}
\author{John M. O'Meara}
\affiliation{Department of Physics, Saint Michael's College, One Winooski Park, Colchester, VT 05439, USA}
\author{Tom Theuns}
\affiliation{Institute for Computational Cosmology, Durham University, South Road, Durham DH1 3LE, UK}



\begin{abstract}
We present the first results from our survey of intervening and proximate Lyman limit systems (LLSs) at $z$$\sim$2.0--2.5 using the Wide Field Camera 3 on-board the {\it Hubble Space Telescope}. The quasars in our sample are projected pairs with proper transverse separations $R_\perp$$\leq$150 kpc and line of sight velocity separations $\lesssim$11,000 km/s. We construct a stacked ultraviolet (rest-frame wavelengths 700--2000\AA) spectrum of pairs corrected for the intervening Lyman forest and Lyman continuum absorption. The observed spectral composite presents a moderate flux excess for the most prominent broad emission lines, a $\sim$30\% decrease in flux at $\lambda$=800--900\AA\ compared to a stack of brighter quasars not in pairs at similar redshifts, and lower values of the mean free path of the HI ionizing radiation for pairs ($\lambda_{\rm mfp}^{912}=140.7\pm20.2~h_{70}^{-1}$Mpc) compared to single quasars ($\lambda_{\rm mfp}^{912}=213.8\pm28~h_{70}^{-1}$Mpc) at the average redshift $z\simeq2.44$.
From the modelling of LLS absorption in these pairs, we find a higher ($\sim$20\%) incidence of proximate LLSs with $\log N_{\rm HI}\geq17.2$ at $\delta v$$<$5,000 km/s compared to single quasars ($\sim$6\%). These two rates are different at the 5$\sigma$ level. Moreover, we find that optically-thick absorbers are equally shared between foreground and background quasars. 
Based on these pieces of evidence, we conclude that there is a moderate excess of gas absorbing Lyman continuum photons in our closely-projected quasar pairs compared to single quasars. We argue that this gas arises mostly within large-scale structures or partially neutral regions inside the dark matter haloes where these close pairs reside.
\end{abstract}



\section{Introduction} 
\label{sec:intro}
Quasars represent the brightest phase of the active galactic nuclei (AGN) population, with optical-ultraviolet luminosities in the range $\sim10^{44}-10^{48}$ erg/s. To support these luminosities, a significant mass of gas must flow from kilo-parsec scales to the centre of the galaxy at sub-parsec scales. One possible mechanism to drive gas to the galaxy's centre is through gas-rich major mergers (e.g. \citealt{2005Natur.433..604D,2005ApJ...630..705H,2005MNRAS.361..776S,2006ApJS..163....1H,2008ApJS..175..356H}); but minor mergers (e.g. \citealt{2000ApJ...536L..73C}) as well as secular processes (e.g. \citealt{2011ApJ...726...57C}) are also probable mechanisms in funneling gas towards the central supermassive black hole (SMBH). 
Models of structure formation can reproduce the observed large-scale quasar properties (e.g. clustering, environment measurements) if the bright and short-lived quasar phase within the galaxy lifetime is triggered by mergers (see \citealt{2012NewAR..56...93A} and references therein). 

Pair\footnote{In the following, we will refer to a {\it pair} as a system of two quasars with small projected and, spectroscopically confirmed, redshift separations.} (or dual) quasars at (projected) separation at tens of Mpc to several hundreds of kpc have become particularly interesting in the last decade as these systems could reside in the same cosmological structure, thus tracing the large-scale quasar environment (e.g. \citealt{hennawi2006,hennawi2010,2014MNRAS.444.1835S,2017MNRAS.468...77E,2018MNRAS.474.4925S}). The detection of these systems in the optical and mid-infrared, mostly from the Sloan Digital Sky Survey (SDSS) and the Wide-field Infrared Survey Explorer (WISE), down to a few tens of kpc reinforce the idea of gas-rich mergers mutually triggering the active nuclear phase likely in both quasars (e.g. \citealt{2008ApJ...678..635M,2009ApJ...693.1554F,2014MNRAS.441.1297S,2017ApJ...848..126S,2017MNRAS.464.3882W}). 
Quasar pairs at similar redshifts, with projected separations less than a few hundreds of kpc, are thus ideal probes of the large-scale environment, since this is where mergers are more likely to occur, thereby providing possible tracers of massive proto-clusters (e.g. \citealt{2007ApJ...662L...1D,2011ApJ...736L...7L,2013MNRAS.431.1019F,2014Natur.511...57D,2015Sci...348..779H}).  

In this paper, we further investigate the large-scale quasar environment by analysing the spectral properties (e.g. ionizing continuum, emission line fluxes) and associated absorbers of quasar pairs with proper transverse separation $R_\perp\leq 150$ kpc and line of sight velocities $<$11,000 km/s in the redshift interval $z\simeq2.0-2.5$. Our sample consists of \rev{47 relatively close quasar pairs} at similar redshifts observed during our survey for Lyman limit systems (LLSs, i.e. optically-thick absorption line systems) using the Wide Field Camera 3 onboard the {\it Hubble Space Telescope} (\hst; Proposal ID: 14127).  
\revs{Our survey also includes 6 lensed quasars, 2 field single quasars (SDSS J133905.25$+$374755.3 and SDSS J154815.42$+$284452.6), and a projected pair (i.e. the system SDSS J172855.24$+$263449.1 and SDSS J172855.31$+$263458.1 with a line of sight velocity separations $>$11,000 km/s) which will be discussed in a separate paper, leading to a total of 104 single sources (111 observations).} 
By comparing the ionising spectral continuum of quasar pairs at $z>2$ as a function of luminosity to similar quantities of single quasars at comparable redshifts, we can provide constraints on the structure of the intergalactic medium (IGM) at 10--100 kpc scale (in the transverse direction) where these systems reside.

From modelling the associated absorbers (LLSs and damped Ly$\alpha$ systems, DLAs, with $\log \nh\geq20.3$), we investigate the interplay between quasars and their environment, as well as constrain the evolution of the ultraviolet background. Quasars indeed provide significant flux of ionising photons that regulate both the ionisation state and the temperature of the IGM at $z\sim3$ \citep[e.g.][]{haardt96,haardt12,meiksin03,faucher09}. Whilst not numerous enough at $z\ga6$ to have significantly contributed to the \ion{H}{i} reionisation \citep[e.g.][]{meiksin05,jiang08,shankar07,willott10,fontanot12,fontanot14}, they are the main sources responsible for the reionisation of \ion{He}{ii} at $z\sim3$ \citep{miralda00,faucher08,furlanetto09,mcquinn09,haardt12,compostella13}. The common denominator of all these studies is that they rely upon the parameterizations of the quasar continuum at rest-frame UV wavelengths.

The composite spectrum of quasars also provides a wealth of additional information. Observationally, composite spectra of AGN were previously constructed by taking advantage of major surveys, covering a relatively large range of redshifts: the Large Bright Quasar survey (LBQS, \citealt{1991ApJ...373..465F}), Faint Images of the Radio Sky at Twenty-cm (FIRST, \citealt{2001ApJ...546..775B}), {\it Sloan Digital Sky Survey} (SDSS, \citealt{vandenberk2001}), {\it Hubble Space Telescope} (\hst, \citealt{zheng97,2002ApJ...565..773T,2012ApJ...752..162S,2014ApJ...794...75S,2015MNRAS.449.4204L,2016ApJ...817...56T}), and the Far Ultraviolet Spectroscopic Explorer (\fuse, \citealt{2004ApJ...615..135S}). 
\rev{The composites in these studies indicate that the optical continuum can be described by a power law of the form $f_\nu \propto \nu^{\alpha_\nu}$, with a slope spanning a rather wide range of values (e.g. $-0.83\la\alpha_\nu\la-0.61$ in the rest-frame wavelength range 1200--2000~\AA; \citealt{2002ApJ...565..773T,2012ApJ...752..162S,2014ApJ...794...75S,2015MNRAS.449.4204L}).}

The quasar composites also show a softening in the far-ultraviolet (blueward of \ion{Ly}{$\alpha$}), which is interpreted as comptonization of the thermal disc emission in a soft X-ray corona above the disc \citep{1987ApJ...321..305C,1997ApJ...477...93L,zheng97}. 
\rev{However, \citet[S04 hereafter]{2004ApJ...615..135S}, who considered more than 100 AGN at $z<0.1$ observed with \fuse, found that the quasar composite does not display any break and/or softening of the continuum, but a significantly hard slope with $\alpha_\nu=-0.56$ at the rest-frame wavelength range 630$-$1100\AA. \citet{2014ApJ...794...75S} investigated possible reasons for this difference (see their Figs.~7 and 8), concluding that the \fuse spectral stack was affected by quasar broad emission lines in the wavelength range covered by \fuse. Additionally, the \fuse survey considered low-redshift quasars, and the rest-frame wavelengths longer than 1100\AA\ were not covered.}

The observed quasar spectra can also be used to trace the evolution of the ionization state of the IGM through the estimate of the effective opacity in the Lyman continuum ($\taue$), which is often represented by the mean free path, $\lmfp$. The $\lmfp$ parameter is defined as the physical distance a packet of ionizing photons can travel before encountering an $e^{-1}$ attenuation \citep[e.g.][]{worseck14}. As such, the $\lmfp$ should approach zero as the redshift increases towards the epoch of reionization. 
The redshift evolution of the $\lmfp$ is thus a key cosmological parameter that constrains the distribution of neutral hydrogen in the Universe, whilst the estimates of the attenuation $\taue$ ($\propto1/\lmfp$; \citealt{2009ApJ...705L.113P}) is a key parameter in constraining the extragalactic UV background (e.g. \citealt{2015ApJ...813L...8M}, and references therein).
Direct estimates of the mean free path have been obtained through the analysis of composite quasar spectra in the rest-frame at $z\ga4.4$ using high signal-to-noise, low-resolution spectra taken from the Gemini Multi Object Spectrometers \citep{worseck14}, and at $z=2-4$ with both space and ground-based facilities \citep{2009ApJ...705L.113P,2013ApJ...775...78F,2013ApJ...765..137O}. These studies find that the mean free path increases with decreasing redshift, from $\sim10$ $\h$ Mpc at $z\simeq5$ to more than 200 $\h$ Mpc at $z\simeq2.4$. Yet, only two direct $\lmfp$ estimates are available in the redshift range $z=2-3$ (i.e. \citealt{2013ApJ...775...78F,2013ApJ...765..137O}), due to the fact that, at $z<2.5$, one must consider spaced-based spectroscopy to cover the rest-frame wavelength bluewards of 912\AA.

The structure of this paper is as follows. We discuss the sample, the selection criteria, and the data reduction of the quasar pairs in our \hst survey in Section~\ref{The data-set}. In Section~\ref{Composite construction} we describe the technique to construct the stacked spectrum, and the IGM transmission curves adopted to correct the observed average spectrum are presented in Section~\ref{IGM transmission correction}, where we also describe our IGM corrected stack along with its uncertainties. The formalism considered in the estimate of the mean free path to ionising photons is presented in Section~\ref{Mean free path}. Section~\ref{Fitting for LLS absorbers} describes how we model absorbers in our sample, and the discussion on the implications of our analysis and conclusions are presented in Section~\ref{Discussion and Conclusion}.

We adopt a concordance flat $\Lambda$-cosmology with $H_0=70\, \rm{km \,s^{-1}\, Mpc^{-1}}$, $\Omega_\mathrm{m}=0.3$, and $\Omega_\Lambda=0.7$. 
Unless noted otherwise, we will distinguish between the ionizing and non-ionizing part of the spectrum as $\lambda<912$~\AA\ and 912--2000~\AA, respectively.

\section{The data set}
\label{The data-set}
The sample of quasars observed in our \hst programme is drawn from a compilation of quasar pairs with $g^* < 21$ mag (\citealt{2006ApJ...651...61H, hennawi2010,2018arXiv180408624F}), selected from the SDSS/BOSS footprints in the redshift range $z\simeq2.0-2.5$. Six quasar lenses with comparable magnitudes and redshifts are also observed during our campaign  but excluded for this analysis. \rev{Our \hst programme includes 47 quasar pairs and four additional single quasars observed within the survey, leading to a total of 104 quasars (111 observations\footnote{\rev{We obtained 94 spectra for the quasar pairs, 13 spectra for the lenses and 4 additional spectra for the single quasars, for a total of 111 observations.}})}. Roughly 50\% of the quasars within the \rev{survey} have a spectroscopic redshifts from SDSS, whilst the rest have redshift measurements from our follow-up optical campaign. \rev{Table~\ref{tab:sample} lists the 111 observations for the whole WFC3 sample of 104 quasars. In the present analysis, we will focus on the properties of the 47 quasar pairs only. We will present further results on the properties of the intervening absorbers, including their autocorrelation function, from the entire survey in a forthcoming publication.}  

\begin{figure}
 \includegraphics[width=\linewidth]{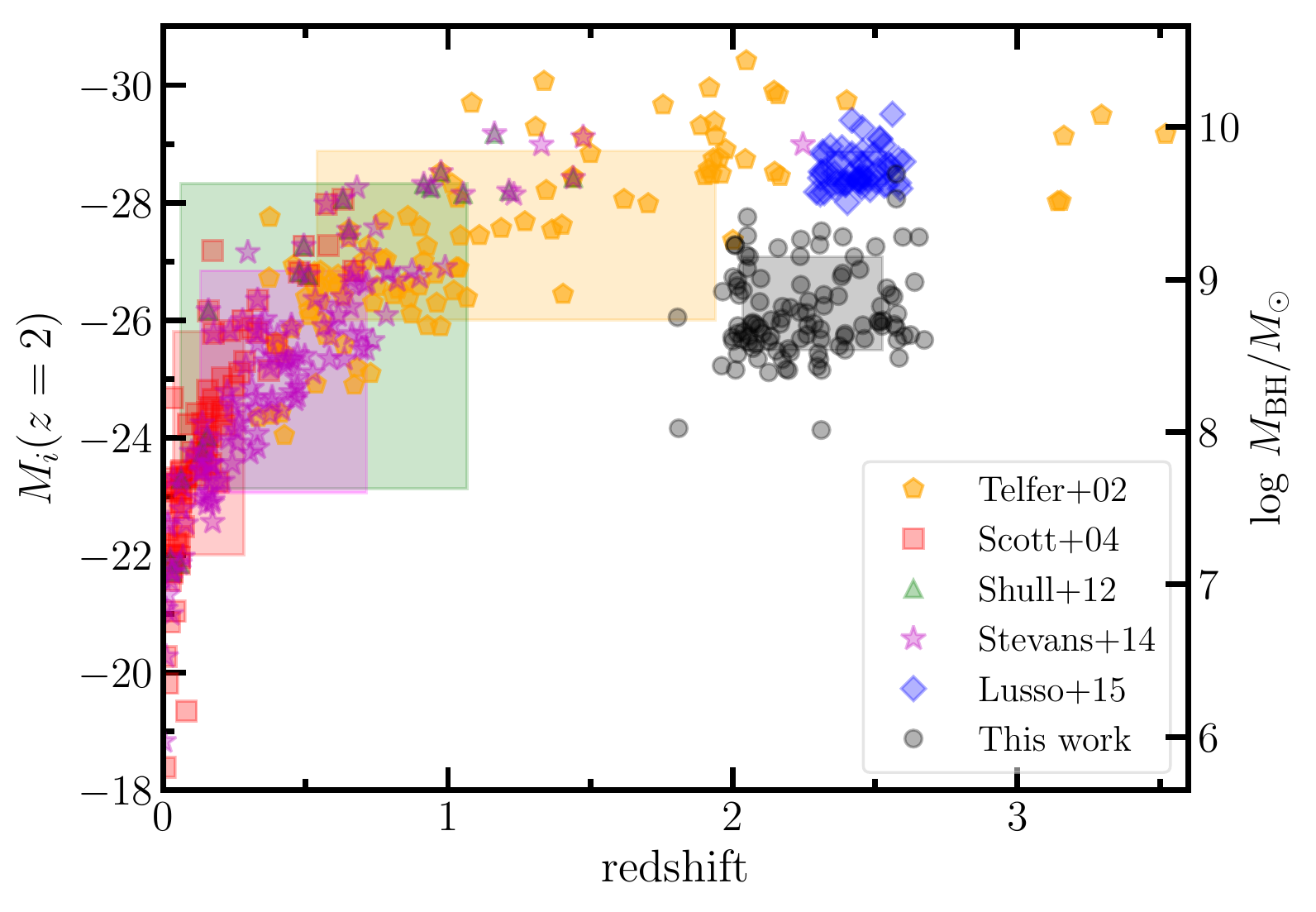}
 \caption{\rev{Absolute $i$-band magnitude (normalized at $z=2$, K-corrected following \citealt{2006AJ....131.2766R}) as a function of redshift. Symbols represent different literature samples: T02 (orange pentagon), S04 (red square), S12 (green triangle), S14 (magenta star), L15 WFC3 sample (blue diamond), and our new WFC3 quasar pair sample (black circle). Shaded areas indicate the redshift and magnitude ranges for the different samples, estimated from the $16^{\rm th}$ and $84^{\rm th}$ percentiles. Approximate values of the black hole masses (in units of $M_\odot$) are plotted on the $y-$axis on the right.}}
 \label{magi_z}
\end{figure}

To compare our new WFC3 quasar pair sample with previous works in the literature, we computed the absolute $i-$band magnitude, $\mi$, from the observed SDSS $i^\ast$, normalized at $z=2$, and $K-$corrected following \citealt{2006AJ....131.2766R}. Figure~\ref{magi_z} shows the distribution of $\mi$ as a function of redshift for several single quasar samples from the literature, from which the composite AGN spectra were constructed. We also plotted indicative values of the black hole masses (in units of $M_\odot$) on the $y-$axis on the right, where $\mbh$ is estimated via $\ledd = \lbol/\lumedd$ assuming an average $\ledd=0.35$.  The relation between $\lbol$ and $\mi$ has been estimated to be $\log \lbol = -10.03~\mi/26 + 36.33$ by fitting the quasars in the SDSS-DR7 quasar catalog \citep{2011ApJS..194...45S}.
The samples included in this comparison are the \citet[L15 hereafter]{2015MNRAS.449.4204L} WFC3 sample (blue diamond), \citet[S14 hereafter]{2014ApJ...794...75S} (magenta star), \citet[S12 hereafter]{2012ApJ...752..162S}  (green triangle), \citet[orange pentagon]{2002ApJ...565..773T}, and \citet[S04 hereafter]{2004ApJ...615..135S} (red square). Shaded areas indicate the redshift and magnitude ranges for the different samples, estimated from the $16^{\rm th}$ and $84^{\rm th}$ percentiles. 
Our new WFC3 quasar pair data-set is at a similar redshift range ($1.961\leq z\leq2.673$) as that considered by L15 ($z\simeq2.440$, \citealt{2011ApJS..195...16O,2013ApJ...765..137O}), with a mean (median) redshift of $\langle z\rangle\simeq2.256~(2.237)$, while probing a lower luminosity quasars.

\subsection{Data reduction}
\label{Data reduction}
Each quasar pair was observed with HST WFC3/UVIS for one orbit between September 2016 and March 12 2017 (Cycle 23, program ID 14127, PI: Fumagalli). 
Every visit consisted of one F300X direct images of 100 seconds and two G280 dispersed images of 1200 seconds each, to enable cleaning the images of cosmic rays. 
The only exceptions are J224136$+$230909 and J210329$+$064653, which have just one direct image of 240 seconds each. 
To correct for the charge transfer efficiency (CTE) degradation of the detector, the raw images were processed with \textsc{wfc3uv\_ctereverse}\footnote{http://www.stsci.edu/hst/wfc3/tools/cte\_tools} for a pixel-based CTE correction based on modelling of hot pixels \citep{2010PASP..122.1035A,2010MNRAS.401..371M}. 

We created custom dark reference files to correct for dark current structure and to mitigate hot pixels. Specifically, for optimal dark subtraction and hot pixel identification we created super dark files as detailed in \citet{2015AJ....150...31R} and \citet{2016ApJ...831...38V}. These super dark files are similar to those currently produced at the {\it Space Telescope Science Institute} (STScI, WFC3 ISR 2016--08), but they also include the use of concurrent darks as the observations and improved hot pixel rejection, which is important for our program due to the small number of exposures obtained. In particular, our methodology models the dark background with a $3^{\rm rd}$ order polynomial to remove the background gradient temporarily before identifying hot pixels, enabling the detection of a uniform number of hot pixels both far and close to the readout of each chip. The resultant science files are CTE corrected with reduced background gradients, blotchy patterns, and appropriate hot pixel flagging \citep{2015AJ....150...31R}. 

Cosmic ray rejection was performed by building an association table for each pair of exposures and using the \textsc{calwf3} built-in cosmic ray rejection called \textsc{wf3rej}. No cosmic ray rejection was done for the direct images of J224136$+$230909 and J210329$+$064653, and centers of the pairs for these two targets were identified manually to avoid issues related to cosmic rays. The resultant calibrated and cosmic ray cleaned images are utilised in the extraction described below. 

\subsection{Wavelength and Flux calibration}
\label{Flux calibration}
\begin{figure*}
 \plotone{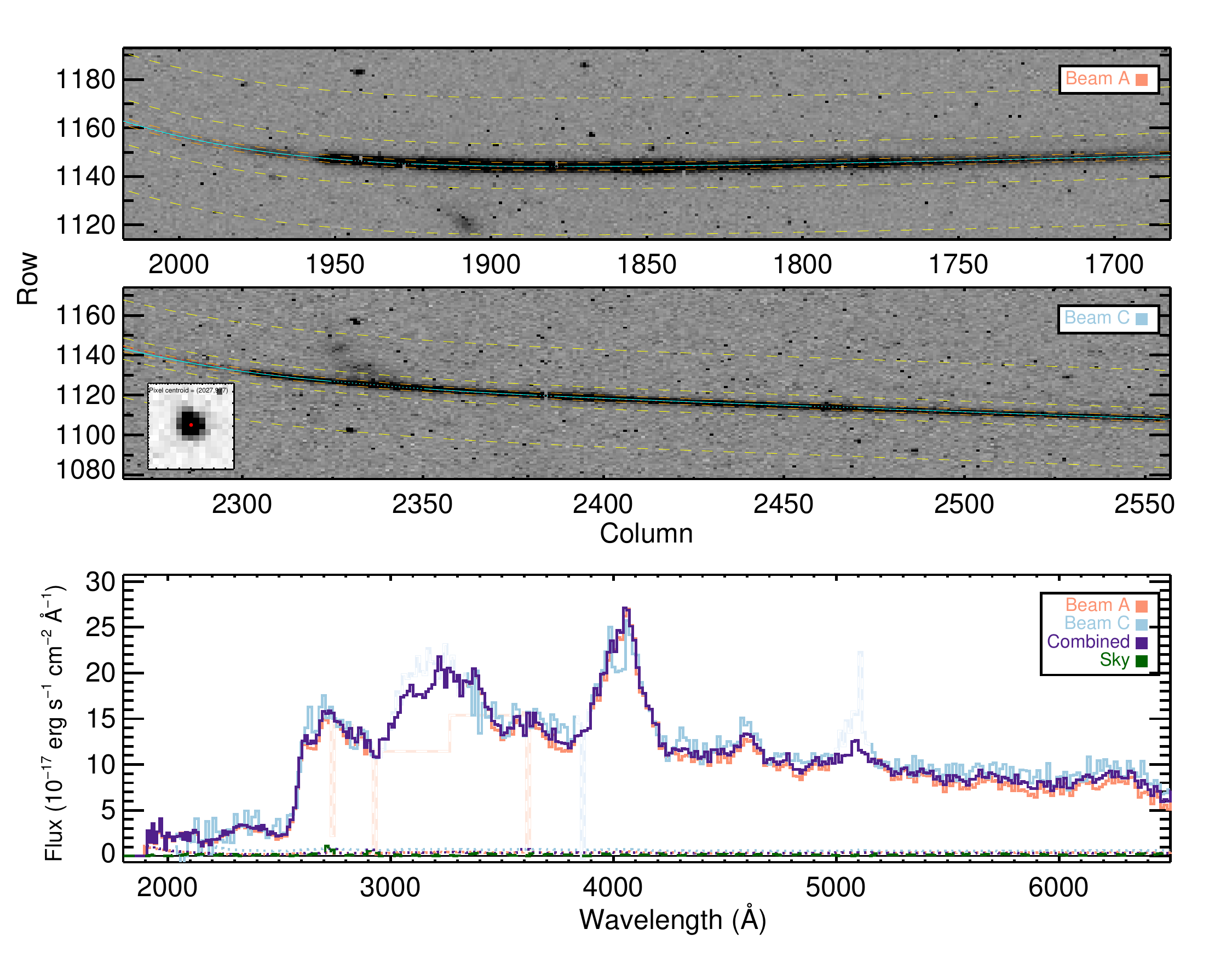}
 \caption{WFC3/G280 grism exposure and flux calibrated spectra of the quasar SDSSJ100233.9$+$353127.5. {Upper two panels}: the $1^{\rm st}$ order dispersed spectrum on the top panel (Beam A) and the --$1^{\rm st}$order spectrum on the bottom (Beam C). The zeroth-order image of the quasar is shown on the bottom-left side of the Beam C panel. The bottom panel shows the fluxed 1D spectra for beam A (red), beam C (cyan), and  for the combined beams (purple). \\(The complete set of figures (111 images) are available in the online journal.)}
 \label{spectrum_example}
\end{figure*}
To extract 1D spectra from the G280 dispersed images, we used an updated version of the pipeline discussed in \citet{2011ApJS..195...16O}. The pipeline makes use of the latest calibration files provided by the aXe team\footnote{http://www.stsci.edu/hst/wfc3/analysis/grism\_obs/\\calibrations/wfc3\_g280.html}. Major changes to the pipeline include the use of both $\pm 1$ orders (which we will refer to as beam A and beam C in accordance with the nomenclature used in the literature), improved wavelength and trace solutions away from the chip center, and an improved extraction algorithm. The pipeline is provided as part of the publicly available \textsc{XIDL} software package\footnote{https://github.com/profxj/xidl.git}.

In detail, the pipeline first considers the direct image to find the center of the emission around a user-supplied right ascension and declination using the IDL routine \textit{cntrd}. The position of the center of emission is then employed to calculate an initial trace and wavelength solution for the individual beams using the updated calibrations supplied by the aXe team. These solutions are 6$^{\rm th}$ and 4$^{\rm th}$ order polynomial functions that vary smoothly as a function of position on the chip. 

After finding these initial trace solutions, the pipeline calculates the offset between this trace and the Gaussian centroid of the data for each individual column. The median offset for each individual beam is calculated and the trace is offset by this amount. For beam C, below a wavelength of $\sim 3000$~\AA, we found significant deviations between the centroid of the data and the adjusted trace. For these data columns, we fit the residual offsets with a 3$^{\rm rd}$ order polynomial and apply this offset. We note that this step produces errors in the wavelength calibration on the order of $\sim$10~\%, and when co-adding the beams we use beam C mainly as a substitute when the primary beam A is affected by chance superposition with other sources or detector artifacts.

Sky subtraction is performed on a 20 pixel wide region of blank sky above and below the trace of each of the individual beams. In case of close quasar pairs or lensed quasars, only a single sky region (the region not containing the spectrum of the other sightline) is used. All features above 2.5$\sigma$ in the sky region are clipped and the 1D sky spectrum is smoothed by a zeroth-order SAVGOL filter. This sky model is then subtracted from all pixels in the corresponding beam. 

A variance image is created, assuming Gaussian statistics and a read noise of 3.3 electrons per exposure. The final 1D spectrum for each of the beams is extracted from this sky-subtracted image using optimal extraction, which assumes a Gaussian profile for the spectrum. The resultant 1D spectrum for each of the individual beams is then fluxed using the calibration files supplied by the aXe team. 

Next, the two fluxed 1D spectra of the individual beams are visually inspected, and regions of the spectra containing bad pixels or interloping zeroth order emission from unrelated galaxies are masked. Finally, the two beams are combined using the \textsc{XIDL} routine \textit{long\_combspec}. This routine interpolates the data onto a common wavelength grid, clips any outliers and performs an average of the two beams weighting by the signal-to-noise (S/N) ratio.
Figure~\ref{spectrum_example} presents the 2D spectral image (sky-subtracted) and the fluxed 1D spectrum for beam A and beam C for one source. The fluxed 1D spectrum for the combined beams is also shown. 

To further check our flux calibration, we have estimated the observed $g^\ast$ band magnitude from the WFC3 spectra and compared this value with the one obtained from either the SDSS or the BOSS survey for all quasars with an optical spectrum\footnote{For quasars with multiple spectra we have selected the one with the highest S/N per resolution element.} (53 quasars). The difference between these magnitudes ($\Delta g^\ast = g^\ast_{\rm WFC3} - g^\ast$) does not display strong systematics with a mean (median) $\Delta g^\ast$ of about 0.05 (0.03), and a dispersion around the mean of 0.16 dex. SDSS$/$BOSS observations have been carried out between 2002 and 2013, therefore part of this scatter may be due to intrinsic long-term UV variability \citep{2012ApJ...753..106M}. 

\begin{figure}
\includegraphics[width=8.5cm,clip]{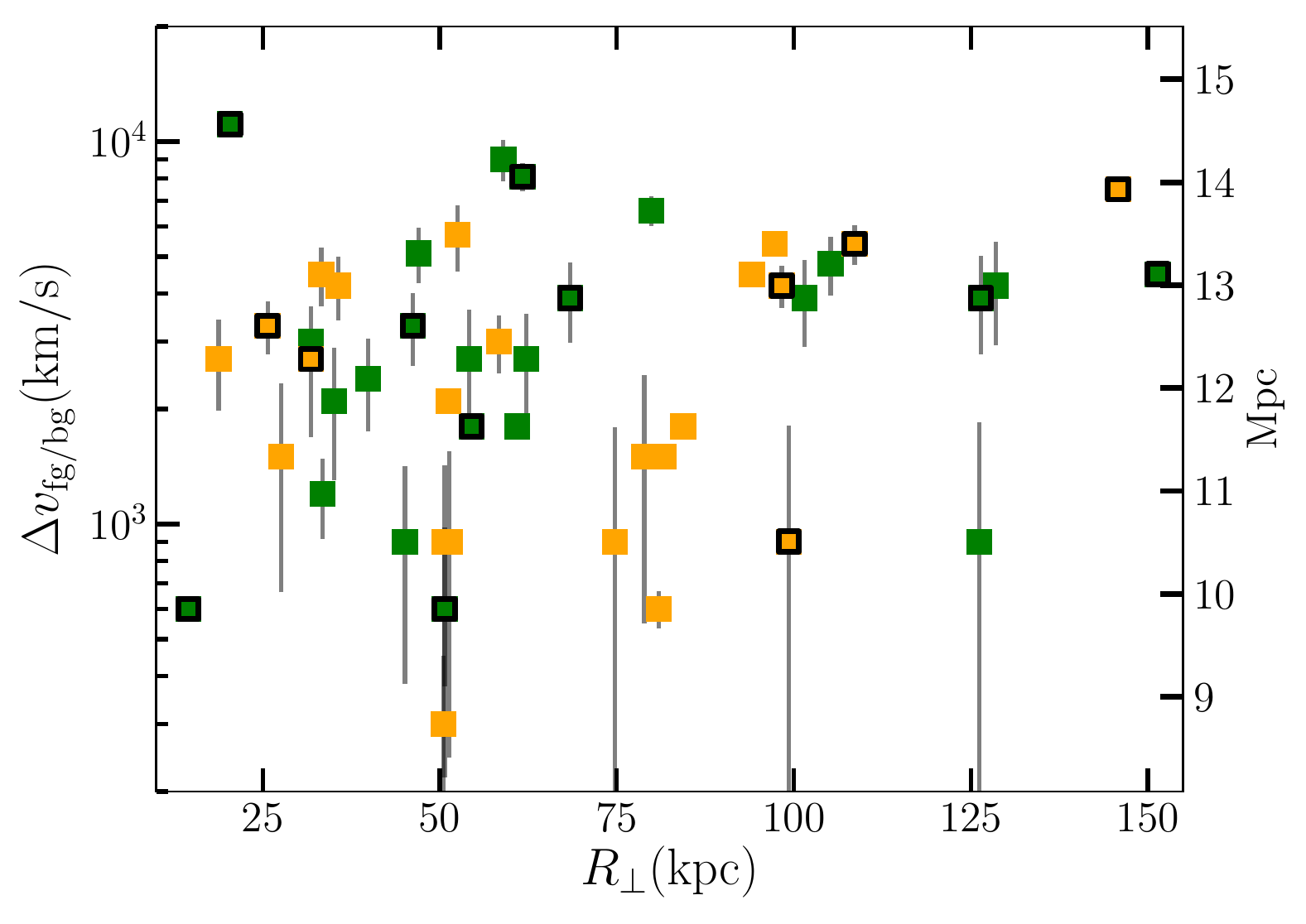}
\caption{Range of relative velocities between the foreground and background quasar pair  ($\Delta v_{\rm fg/bg}= c |z_{\rm fg}-z_{\rm bg}|$) and their proper transverse separations. The physical distances in Mpc are plotted on the y-axis on the right. The green and orange squares represent the low ($\langle z\rangle=2.09$) and high redshift ($\langle z\rangle=2.44$) quasar subsamples, respectively. The quasars with at least one absorber ($\log\nh>17.2$) within 5,000 km/s are marked with empty black squares (see discussion in \S\ref{Fitting for LLS absorbers} and \S\ref{Implication for the quasar environment}).}
 \label{deltavrp}
\end{figure}

\subsection{Redshift estimates}
\label{Redshift estimates}
Optical spectra were obtained for a fraction of our quasar pair from a variety of instruments and thus have different wavelength coverage and resolution. 
High S/N spectra were taken for 27 quasars using the Echellette Spectrograph and Imager (ESI) at the Keck II telescope, and the moderate-resolution Magellan Echellette (MagE) optical spectrograph. 
Twelve quasars have medium/low resolution optical spectroscopy from several telescopes, such as the 6.5m Multiple Mirror Telescope (MMT), Calar Alto Observatory (CAHA), Keck, the 2.1m telescope at the Kitt Peak National Observatory (KPNO), and Gemini (see Table~\ref{tab:sample}). All of the observed quasars have both the \ion{Ly}{$\alpha$} and \ion{C}{iv} lines covered. The ESI spectra are always considered for redshift determination when available, otherwise we use BOSS/SDSS redshifts (50\% of the sample) or the low S/N spectra.
Details on the observations and data reduction will be provided in a separate paper, as here we only use  these spectra for redshift determination of the quasars and the absorbers.
At the time of writing, eleven quasars do not have any other spectra than those taken with HST WFC3/UVIS. The redshifts of these objects were determined from the \ion{C}{iv} line observed by WFC3, thus these have the least precise measurements ($\sigma_z\simeq800-1000$ km s$^{-1}$).
For the two lensed quasars HE0230$-$2130 and Q1017$-$207, the redshift was taken from the literature \citep{1996A&A...305L...9C,1997A&A...327L...1S,2008A&A...481..615A}.

To compute the redshift, we have followed a similar procedure as the one described by \citet{2006ApJ...651...61H}. Lines were fitted as the sum of a Gaussian plus a linear local continuum using a custom made IDL code. 
Strong absorption and/or noisy features were masked. For the majority of the sources, the redshift was estimated from the \ion{C}{iv} line only\footnote{\rev{We note that redshifts estimated using high-ionization broad emission lines, such as \ion{C}{iv} and \ion{Ly}{$\alpha$}, could be shifted blueward with respect to the systemic (e.g. $>500-1,000$ km/s;  \citealt{1982ApJ...263...79G,1989ApJ...342..666E,1990ApJ...357..346C}).}}, but there were cases (19 quasars) in which that line was used in combination with other emission lines, like the \ion{Si}{iv} and the semi-forbidden \ion{C}{iii}] lines.
For six quasars in which the \ion{Mg}{ii} broad emission line was also covered, we have computed the redshift from that line only, as this is considered a better tracer of the systemic redshift \citep{2002AJ....124....1R}.

The distribution of radial velocity differences between the foreground and background quasars in the pair ($\Delta v_{\rm fg/bg}= c |z_{\rm fg}-z_{\rm bg}|$) as a function of their proper transverse separations is shown in Figure~\ref{deltavrp}. Given the sample selection (i.e. we observed the closest projected pairs with similar spectroscopic redshift), the bulk of the quasar pair sample is clustered at small velocity differences ($\Delta v_{\rm fg/bg}<4,000$km/s), which translates into physical distances of $<15$Mpc.
The redshifts and the emission lines considered for their estimates are listed in Table~\ref{tab:sample} along with the associated (statistical) uncertainties.

\begin{longrotatetable}
\begin{deluxetable}{ccccccccccccccl}
\tablecaption{Full sample of WFC3 quasars\label{tab:sample}}
\tablewidth{1000pt}
\tabletypesize{\scriptsize}
\tablehead{
\colhead{Name} & 
\colhead{R.A. (J2000)} & \colhead{Decl. (J2000)} & 
\colhead{$u^\ast$} & \colhead{$g^\ast$} & 
\colhead{$r^\ast$} & \colhead{$i^\ast$} & 
\colhead{$z^\ast$} & \colhead{$R_\perp$} & \colhead{$\Delta\theta$} &
\colhead{$z$} & \colhead{$\sigma_z$} & 
\colhead{em. line$^{\rm a}$} & \colhead{Instr.$^{\rm b}$} & \colhead{Note}\\ 
\colhead{} & 
\colhead{(degrees)} & \colhead{(degrees)} &
\colhead{} & \colhead{} & \colhead{} & \colhead{} & \colhead{} &
\colhead{kpc} & \colhead{$\arcsec$} & \colhead{} & \colhead{km s$^{-1}$} &
\colhead{} & \colhead{} & \colhead{}
} 
\startdata
  SDSS J002423.88$-$012827.6   &   6.099520   & $-$1.474347  &  19.11  &  18.91  &  18.76  &  18.67  &  18.51  & 46.2   & 5.34  & 2.047  &  275   &  MgII           & MagE   &                  \\
  SDSS J002424.21$-$012825.7   &   6.100912   & $-$1.473832  &  19.03  &  18.93  &  18.84  &  18.72  &  18.52  & 46.2   & 5.34  & 2.058  &  657   &  CIV-CIII]      & MagE   &                  \\
  SDSS J011707.51$+$315341.1   &   19.281323  & 31.894773    &  20.83  &  20.03  &  19.81  &  19.79  &  19.49  & 94.1   & 11.32 & 2.639  &  28    &                 & BOSS   &                  \\
  SDSS J011708.38$+$315338.6   &   19.284939  & 31.894074    &  21.66  &  20.78  &  20.60  &  20.68  &  20.39  & 94.1   & 11.32 & 2.624  &  40    &                 & BOSS   &                  \\
  SDSS J013458.85$+$243050.5   &   23.745247  & 24.514045    &  21.03  &  20.43  &  20.18  &  20.09  &  19.77  & 31.8   & 3.69  & 2.105  &  56    &                 & BOSS   &                  \\
  SDSS J013459.01$+$243047.5   &   23.745903  & 24.513213    &  20.28  &  19.85  &  19.60  &  19.53  &  19.28  & 31.8   & 3.69  & 2.095  &  53    &                 & BOSS   &                  \\
       HE0230$-$2130           &   38.138119  & $-$21.290540 &  ---    &  ---    &  ---    &  ---    &  ---    & ---    & ---   & 2.163  &  284   &                 & ref    &   lense$^\star$  \\
       HE0230$-$2130           &   38.138473  & $-$21.290031 &  ---    &  ---    &  ---    &  ---    &  ---    & ---    & ---   & 2.163  &  284   &                 & ref    &   lense$^\star$  \\
       HE0230$-$2130           &   38.138325  & $-$21.290467 &  ---    &  ---    &  ---    &  ---    &  ---    & ---    & ---   & 2.163  &  284   &                 & ref    &   lense          \\
  SDSS J034406.64$+$101509.8   &   56.027689  & 10.252729    &  20.85  &  20.52  &  20.12  &  19.84  &  19.51  & 105.2  & 12.11 & 2.018  &  656   &  CIV-CIII]      & KPNO   &                  \\
  SDSS J034407.03$+$101520.5   &   56.029300  & 10.255699    &  19.94  &  19.54  &  19.20  &  18.98  &  18.72  & 105.2  & 12.11 & 2.002  &  519   &  SiIV-CIV-CIII] & KPNO   &                  \\
  SDSS J073522.43$+$295710.2   &   113.843476 & 29.952837    &  21.05  &  20.53  &  20.38  &  20.24  &  20.21  & 47.0   & 5.44  & 2.082  &  798   &  CIV            & ESI    &                  \\
  SDSS J073522.55$+$295705.0   &   113.843980 & 29.951404    &  20.77  &  20.37  &  20.07  &  19.87  &  19.56  & 47.0   & 5.44  & 2.065  &  274   &  MgII           & ESI    &                  \\
  SDSS J074653.04$+$440351.4   &   116.720790 & 44.064326    &  18.87  &  18.81  &  18.75  &  18.45  &  18.15  & 8.7    & 1.08  & 2.008  &  27    &                 & SDSS   &   lense$^\star$  \\
  SDSS J074653.04$+$440351.4   &   116.721160 & 44.064181    &  18.87  &  18.81  &  18.75  &  18.45  &  18.15  & 8.7    & 1.08  & 2.008  &  27    &                 & SDSS   &   lense          \\
  SDSS J081329.49$+$101405.2   &   123.372873 & 10.234790    &  19.67  &  19.41  &  19.34  &  19.14  &  18.90  & 61.7   & 7.13  & 2.098  &  629   &  SiIV-CIV       & MagE   &                  \\
  SDSS J081329.70$+$101411.6   &   123.373778 & 10.236561    &  20.13  &  20.09  &  20.03  &  19.85  &  19.75  & 61.7   & 7.13  & 2.071  &  273   &  MgII           & MagE   &                  \\
  SDSS J084624.33$+$270958.3   &   131.601376 & 27.166211    &  21.08  &  20.82  &  20.67  &  20.54  &  20.26  & 39.9   & 4.65  & 2.203  &  655   &  CIV-CIII]      & ESI    &                  \\
  SDSS J084624.50$+$271002.4   &   131.602116 & 27.167341    &  20.94  &  20.37  &  20.48  &  20.47  &  20.18  & 39.9   & 4.65  & 2.195  &  27    &                 & BOSS   &                  \\
  SDSS J085230.22$+$350003.6   &   133.125924 & 35.001018    &  20.71  &  20.36  &  20.36  &  20.36  &  20.02  & 45.1   & 5.27  & 2.238  &  518   &  SiIV-CIV-CIII] & ESI    &                  \\
  SDSS J085230.53$+$350000.0   &   133.127220 & 35.000008    &  20.49  &  19.97  &  19.93  &  19.90  &  19.52  & 45.1   & 5.27  & 2.235  &  22    &                 & BOSS   &                  \\
  SDSS J092056.00$+$131102.6   &   140.233359 & 13.184074    &  20.00  &  19.22  &  19.23  &  19.19  &  18.94  & 52.5   & 6.21  & 2.427  &  796   &  CIV            & MagE   &                  \\
  SDSS J092056.23$+$131057.4   &   140.234319 & 13.182618    &  19.89  &  19.31  &  19.49  &  19.42  &  19.09  & 52.5   & 6.21  & 2.446  &  792   &  CIV            & MagE   &                  \\
  SDSS J093747.24$+$150928.0   &   144.446866 & 15.157782    &  20.78  &  20.04  &  19.98  &  19.95  &  19.59  & 98.3   & 11.75 & 2.555  &  523   &  SiIV-CIV-CIII] & MagE   &                  \\
  SDSS J093747.40$+$150939.5   &   144.447531 & 15.160981    &  20.42  &  19.92  &  19.80  &  19.83  &  19.59  & 98.3   & 11.75 & 2.541  &  35    &                 & BOSS   &                  \\
  SDSS J100233.90$+$353127.5   &   150.641268 & 35.524327    &  19.37  &  18.86  &  18.82  &  18.84  &  18.62  & 33.3   & 3.91  & 2.305  &  39    &                 & BOSS   &                  \\
  SDSS J100234.21$+$353128.6   &   150.642548 & 35.524634    &  20.84  &  20.05  &  20.02  &  19.90  &  19.76  & 33.3   & 3.91  & 2.320  &  795   &  CIV            & ESI    &                  \\
  SDSS J101652.88$+$222412.1   &   154.220347 & 22.403368    &  20.24  &  20.29  &  20.35  &  20.08  &  19.77  & 128.5  & 14.81 & 2.031  &  791   &  CIV            & CAHA   &                  \\
  SDSS J101653.94$+$222413.7   &   154.224775 & 22.403806    &  21.01  &  20.94  &  20.64  &  20.31  &  20.00  & 128.5  & 14.81 & 2.017  &  994   &  CIV            & WCF3   &                  \\
       Q1017$-$207             &   154.349680 & $-$20.783085 &  ---    &  ---    &  ---    &  ---    &  ---    &  ---   & ---   & 2.545  &  254   &                 & ref    &   lense$^\star$  \\
       Q1017$-$207             &   154.349930 & $-$20.783091 &  ---    &  ---    &  ---    &  ---    &  ---    &  ---   & ---   & 2.545  &  254   &                 & ref    &   lense          \\
  SDSS J103424.76$+$330624.1   &   158.603203 & 33.106711    &  20.97  &  20.73  &  20.67  &  20.57  &  20.10  & 101.5  & 11.71 & 2.010  &  996   &  CIV            & WFC3   &                  \\
  SDSS J103425.08$+$330635.1   &   158.604527 & 33.109769    &  19.44  &  19.35  &  19.33  &  19.32  &  19.17  & 101.5  & 11.71 & 2.023  &  79    &                 & SDSS   &                  \\
  SDSS J104533.31$+$404137.9   &   161.388830 & 40.693886    &  20.90  &  20.41  &  20.43  &  20.32  &  20.13  & 145.8  & 17.05 & 2.261  &  27    &                 & BOSS   &                  \\
  SDSS J104533.54$+$404121.1   &   161.389760 & 40.689207    &  21.03  &  20.26  &  20.24  &  20.16  &  19.92  & 145.8  & 17.05 & 2.286  &  40    &                 & BOSS   &                  \\
  SDSS J105644.88$-$005933.4   &   164.187008 & $-$0.992616  &  20.48  &  20.09  &  20.02  &  19.90  &  19.66  & 62.2   & 7.21  & 2.135  &  660   &  CIII]          & MagE   &                  \\
  SDSS J105645.24$-$005938.1   &   164.188530 & $-$0.993918  &  21.18  &  20.96  &  20.83  &  20.78  &  20.58  & 62.2   & 7.21  & 2.126  &  518   &  SiIV-CIV-CIII] & MagE   &                  \\
  SDSS J110430.00$+$290753.4   &   166.125030 & 29.131523    &  20.74  &  20.35  &  20.26  &  20.18  &  19.92  & 54.5   & 6.32  & 2.133  &  38    &                 & BOSS   &                  \\
  SDSS J110430.34$+$290749.0   &   166.126455 & 29.130283    &  21.14  &  20.49  &  20.29  &  20.25  &  20.16  & 54.5   & 6.32  & 2.127  &  0     &                 & BOSS   &                  \\
  SDSS J111641.79$+$650717.2   &   169.174126 & 65.121465    &  21.71  &  21.06  &  21.01  &  22.00  &  20.44  & 51.4   & 6.04  & 2.311  &  53    &                 & BOSS   &                  \\
  SDSS J111642.55$+$650720.9   &   169.177312 & 65.122475    &  21.27  &  20.54  &  20.62  &  20.28  &  19.90  & 51.4   & 6.04  & 2.308  &  652   &  CIV-CIII]      & LRIS   &                  \\
  SDSS J112455.24$+$571056.5   &   171.230202 & 57.182383    &  19.35  &  18.56  &  18.74  &  18.62  &  18.35  & 18.8   & 2.21  & 2.311  &  26    &                 & BOSS   &                  \\
  SDSS J112455.44$+$571058.1   &   171.230999 & 57.182811    &  20.54  &  19.92  &  19.83  &  19.71  &  19.33  & 18.8   & 2.21  & 2.320  &  722   &  SiIV-CIV       & MMT    &                  \\
  SDSS J113947.06$+$414351.1   &   174.946097 & 41.730875    &  20.89  &  20.22  &  20.03  &  19.78  &  19.37  & 20.4   & 2.39  & 2.202  &  58    &                 & BOSS   &                  \\
  SDSS J113947.25$+$414352.1   &   174.946912 & 41.731139    &  20.67  &  19.98  &  19.80  &  19.44  &  19.03  & 20.4   & 2.39  & 2.239  &  648   &  CIV            & ESI    &                  \\
  SDSS J114504.35$+$285713.0   &   176.268160 & 28.953626    &  21.14  &  20.68  &  20.73  &  20.63  &  20.68  & 35.1   & 4.09  & 2.174  &  42    &                 & BOSS   &                  \\
  SDSS J114504.66$+$285712.6   &   176.269454 & 28.953521    &  20.78  &  20.35  &  20.10  &  20.01  &  19.87  & 35.1   & 4.09  & 2.167  &  795   &  CIV            & KPNO   &                  \\
  SDSS J115031.14$+$045353.2   &   177.629760 & 4.898127     &  21.48  &  20.58  &  20.37  &  20.37  &  20.14  & 58.4   & 6.96  & 2.527  &  518   &  SiIV-CIV-CIII] & MagE   &                  \\
  SDSS J115031.54$+$045356.8   &   177.631427 & 4.899124     &  21.20  &  20.52  &  20.38  &  20.39  &  20.18  & 58.4   & 6.96  & 2.517  &  28    &                 & BOSS   &                  \\
  SDSS J121645.92$+$352941.5   &   184.191355 & 35.494881    &  20.50  &  20.40  &  20.26  &  19.96  &  19.78  & ---    & ---   & 2.017  &  52    &                 & SDSS   &   lense$^\star$  \\
  SDSS J121646.04$+$352941.5   &   184.191862 & 35.494861    &  19.52  &  19.38  &  19.19  &  19.06  &  18.85  & ---    & ---   & 2.017  &  52    &                 & SDSS   &   lense          \\
  SDSS J122545.23$+$564445.0   &   186.438466 & 56.745842    &  21.33  &  20.70  &  20.62  &  20.50  &  19.98  & 51.2   & 6.05  & 2.393  &  40    &                 & BOSS   &                  \\
  SDSS J122545.73$+$564440.5   &   186.440546 & 56.744607    &  20.13  &  19.37  &  19.47  &  19.42  &  19.16  & 51.2   & 6.05  & 2.386  &  1     &                 & BOSS   &                  \\
  SDSS J123635.14$+$522058.8   &   189.146430 & 52.349670    &  21.20  &  20.63  &  20.42  &  20.47  &  20.35  & 25.7   & 3.07  & 2.567  &  521   &  SiIV-CIV-CIII] & ESI    &                  \\
  SDSS J123635.42$+$522057.0   &   189.147585 & 52.349185    &  21.59  &  20.49  &  20.51  &  20.75  &  20.18  & 25.7   & 3.07  & 2.578  &  14    &                 & BOSS   &                  \\
  SDSS J125420.52$+$610435.7   &   193.585516 & 61.076602    &  19.67  &  19.57  &  19.41  &  19.26  &  19.09  & 151.4  & 17.60 & 2.036  &  39    &                 & BOSS   &                  \\
  SDSS J125421.98$+$610421.7   &   193.591607 & 61.072701    &  19.25  &  19.07  &  18.99  &  18.90  &  18.71  & 151.4  & 17.60 & 2.051  &  70    &                 & SDSS   &                  \\
  SDSS J133145.98$+$033546.2   &   202.941610 & 3.596176     &  20.85  &  20.31  &  20.29  &  20.30  &  20.26  & 27.6   & 3.30  & 2.579  &  37    &                 & BOSS   &                  \\
  SDSS J133146.19$+$033545.4   &   202.942482 & 3.595964     &  21.84  &  21.12  &  21.14  &  21.05  &  21.39  & 27.6   & 3.30  & 2.584  &  836   &  Ly$\alpha$        & MMT    &                  \\
  SDSS J133209.26$+$252301.3   &   203.038616 & 25.383711    &  20.31  &  20.13  &  20.02  &  19.90  &  19.65  & 68.4   & 7.92  & 2.080  &  652   &  CIV-CIII]      & ESI    &                  \\
  SDSS J133209.69$+$252306.8   &   203.040375 & 25.385231    &  20.07  &  20.09  &  20.12  &  19.97  &  19.62  & 68.4   & 7.92  & 2.093  &  649   &  CIV-CIII]      & ESI    &                  \\
  SDSS J133221.71$+$471721.3   &   203.090460 & 47.289273    &  21.52  &  20.76  &  20.87  &  20.98  &  20.61  & 80.9   & 9.50  & 2.312  &  34    &                 & BOSS   &                  \\
  SDSS J133222.03$+$471712.4   &   203.091828 & 47.286801    &  21.19  &  20.65  &  20.53  &  20.46  &  20.32  & 80.9   & 9.50  & 2.310  &  58    &                 & BOSS   &                  \\
  SDSS J133831.53$+$001056.2   &   204.631394 & 0.182282     &  21.35  &  20.59  &  20.73  &  20.68  &  20.46  & 99.3   & 11.64 & 2.300  &  908   &  CIV            & WFC3   &                  \\
  SDSS J133831.96$+$001105.9   &   204.633186 & 0.184976     &  21.23  &  20.92  &  20.96  &  20.90  &  20.78  & 99.3   & 11.64 & 2.297  &  33    &                 & BOSS   &                  \\
  SDSS J133904.10$+$374742.3   &   204.767099 & 37.795098    &  21.34  &  20.90  &  20.80  &  20.81  &  20.66  & 54.2   & 6.32  & 2.187  &  941   &                 & CAHA   &                  \\
  SDSS J133904.46$+$374737.7   &   204.768612 & 37.793813    &  21.66  &  21.02  &  20.92  &  20.84  &  20.53  & 54.2   & 6.32  & 2.196  &  61    &                 & BOSS   &                  \\
  SDSS J133905.25$+$374755.3   &   204.771883 & 37.798711    &  21.55  &  21.60  &  21.68  &  21.25  &  21.96  &  ---   & ---   & 1.810  &  1067  &  CIV            & WFC3   &   $\dagger$      \\
  SDSS J133907.13$+$131039.6   &   204.779743 & 13.177685    &  19.14  &  18.64  &  18.77  &  18.67  &  18.76  & 14.5   & 1.70  & 2.239  &  28    &                 & SDSS   &                  \\
  SDSS J133907.23$+$131038.7   &   204.780142 & 13.177416    &  19.58  &  19.16  &  19.00  &  18.95  &  18.51  & 14.5   & 1.70  & 2.237  &  21    &                 & BOSS   &                  \\
  SDSS J134543.64$+$262506.9   &   206.431832 & 26.418598    &  20.28  &  20.27  &  20.26  &  19.96  &  19.70  & 79.9   & 9.22  & 2.038  &  276   &  MgII           & ESI    &                  \\
  SDSS J134544.31$+$262505.3   &   206.434650 & 26.418155    &  19.96  &  19.97  &  19.51  &  19.16  &  18.94  & 79.9   & 9.22  & 2.016  &  517   &  SiIV-CIV-CIII] & ESI    &   BAL       \\
  SDSS J140052.07$+$123235.2   &   210.216976 & 12.543120    &  20.54  &  20.36  &  20.31  &  20.17  &  19.95  & 126.4  & 14.60 & 2.058  &  794   &  CIV            & WFC3   &                  \\
  SDSS J140052.55$+$123248.0   &   210.218986 & 12.546672    &  20.66  &  20.49  &  20.47  &  20.29  &  19.97  & 126.4  & 14.60 & 2.071  &  791   &  CIV            & WFC3   &                  \\
  SDSS J140953.74$+$392000.1   &   212.473921 & 39.333362    &  20.28  &  20.17  &  20.15  &  20.07  &  19.97  & 59.0   & 6.82  & 2.058  &  794   &  CIV            & WFC3   &                  \\
  SDSS J140953.87$+$391953.4   &   212.474488 & 39.331517    &  21.05  &  20.78  &  20.82  &  20.49  &  20.39  & 59.0   & 6.82  & 2.088  &  796   &  CIV            & WFC3   &                  \\
  SDSS J142148.79$+$163017.5   &   215.453308 & 16.504886    &  21.02  &  20.40  &  20.35  &  20.38  &  20.26  & 84.4   & 10.02 & 2.457  &  113   &                 & BOSS   &                  \\
  SDSS J142149.00$+$163027.1   &   215.454200 & 16.507535    &  21.62  &  20.76  &  20.56  &  20.62  &  20.42  & 84.4   & 10.02 & 2.463  &  29    &                 & BOSS   &                  \\
  SDSS J143104.64$+$270524.6   &   217.769363 & 27.090177    &  20.57  &  19.80  &  19.90  &  19.76  &  19.57  & 50.7   & 5.94  & 2.266  &  14    &                 & BOSS   &                  \\
  SDSS J143104.97$+$270528.6   &   217.770735 & 27.091286    &  20.91  &  20.26  &  20.28  &  20.17  &  19.97  & 50.7   & 5.94  & 2.263  &  524   &  SiIV-CIV-CIII] & ESI    &                  \\
  SDSS J144254.60$+$405535.0   &   220.727519 & 40.926407    &  20.38  &  19.12  &  18.68  &  18.34  &  17.99  & ---    &  ---  & 2.575  &  11    &                 & BOSS   &   lense$^\star$  \\
  SDSS J144254.78$+$405535.5   &   220.728257 & 40.926553    &  19.60  &  18.51  &  18.12  &  17.92  &  17.61  & ---    &  ---  & 2.575  &  11    &                 & BOSS   &   lense          \\
  SDSS J144320.92$+$200825.4   &   220.837190 & 20.140400    &  20.12  &  19.32  &  19.14  &  19.03  &  18.98  & 97.3   & 11.73 & 2.654  &  18    &                 & BOSS   &                  \\
  SDSS J144321.03$+$200813.8   &   220.837665 & 20.137169    &  21.66  &  20.80  &  20.64  &  20.79  &  20.37  & 97.3   & 11.73 & 2.672  &  20    &                 & BOSS   &                  \\
  SDSS J151538.47$+$151134.8   &   228.910324 & 15.193007    &  19.01  &  18.58  &  18.56  &  18.35  &  18.21  &  ---   & ---   & 2.051  &  11    &                 & BOSS   &    lense$^\star$ \\
  SDSS J151538.59$+$151135.9   &   228.910805 & 15.193315    &  18.28  &  18.26  &  18.22  &  18.03  &  17.79  &  ---   & ---   & 2.051  &  11    &                 & BOSS   &    lense         \\
  SDSS J154815.42$+$284452.6   &   237.064263 & 28.747970    &  21.07  &  20.49  &  20.62  &  20.79  &  20.82  &  ---   & ---   & 2.305  &  38    &                 & BOSS   &    $\dagger$     \\
  SDSS J154938.17$+$313646.8   &   237.409048 & 31.613024    &  21.10  &  20.12  &  20.18  &  20.12  &  19.83  & 108.6  & 12.96 & 2.520  &  15    &                 & BOSS   &                  \\
  SDSS J154938.49$+$313634.6   &   237.410398 & 31.609612    &  20.12  &  19.29  &  19.09  &  19.09  &  18.92  & 108.6  & 12.96 & 2.502  &  651   &  CIV-CIII]      & ESI    &                  \\
  SDSS J161301.69$+$080806.0   &   243.257052 & 8.135014     &  20.17  &  19.56  &  19.53  &  19.47  &  19.29  & 81.6   & 9.64  & 2.382  &  20    &                 & BOSS   &                  \\
  SDSS J161302.03$+$080814.2   &   243.258469 & 8.137295     &  19.51  &  18.91  &  18.84  &  18.80  &  18.61  & 81.6   & 9.64  & 2.387  &  17    &                 & BOSS   &                  \\
  SDSS J163700.87$+$263613.7   &   249.253855 & 26.602753    &  20.99  &  20.70  &  20.68  &  20.43  &  20.23  & 33.5   & 3.85  & 1.961  &  273   &                 & ESI    &                  \\
  SDSS J163700.92$+$263609.9   &   249.253654 & 26.603810    &  19.67  &  19.40  &  19.26  &  19.17  &  19.07  & 33.5   & 3.85  & 1.965  &  84    &                 & SDSS   &                  \\
  SDSS J171945.87$+$254951.2   &   259.941135 & 25.830905    &  20.26  &  19.87  &  19.75  &  19.73  &  19.56  & 126.2  & 14.68 & 2.175  &  944   &                 & WFC3   &                  \\
  SDSS J171946.66$+$254941.1   &   259.944429 & 25.828104    &  20.34  &  20.05  &  19.99  &  19.92  &  19.68  & 126.2  & 14.68 & 2.172  &  39    &                 & BOSS   &                  \\
  SDSS J172855.24$+$263449.1   &   262.230176 & 26.580311    &  20.11  &  19.82  &  19.59  &  19.35  &  19.24  & 77.5   & 9.07  & 1.806  &  1068  &                 & WFC3   &   $\dagger$    \\
  SDSS J172855.31$+$263458.1   &   262.230485 & 26.582816    &  20.65  &  20.14  &  20.01  &  19.94  &  19.76  & 77.5   & 9.07  & 2.260  &  31    &                 & SDSS   &    $\dagger$   \\
  SDSS J210329.25$+$064653.3   &   315.871874 & 6.781478     &  21.47  &  20.60  &  20.44  &  20.53  &  20.26  & 31.8   & 3.80  & 2.574  &  629   &  SiIV-CIII]     & MagE   &                  \\
  SDSS J210329.37$+$064649.9   &   315.872385 & 6.780550     &  21.23  &  20.22  &  20.07  &  20.00  &  19.66  & 31.8   & 3.80  & 2.565  &  790   &  SiIV           & MagE   &                  \\
  SDSS J221426.79$+$132652.3   &   333.611628 & 13.447874    &  20.87  &  20.58  &  20.36  &  20.07  &  19.78  & 50.7   & 5.84  & 2.000  &  270   &  MgII           & GMOS   &                  \\
  SDSS J221427.03$+$132657.0   &   333.612634 & 13.449171    &  20.57  &  20.34  &  20.25  &  20.00  &  19.55  & 50.7   & 5.84  & 1.998  &  270   &  MgII           & GMOS   &                  \\
  SDSS J224136.99$+$230909.8   &   340.404141 & 23.152724    &  20.58  &  19.87  &  19.88  &  19.79  &  19.59  & 74.7   & 8.81  & 2.371  &  53    &                 & BOSS   &                  \\
  SDSS J224137.03$+$230901.0   &   340.404323 & 23.150281    &  21.48  &  20.77  &  20.59  &  20.64  &  20.14  & 74.7   & 8.81  & 2.374  &  889   &                 & WFC3   &                  \\
  SDSS J224204.37$+$055828.6   &   340.518218 & 5.974627     &  21.58  &  21.05  &  20.68  &  20.47  &  20.09  & 35.7   & 4.27  & 2.525  &  791   &  CIV            & GMOS   &                  \\
  SDSS J224204.63$+$055830.4   &   340.519297 & 5.975129     &  21.41  &  20.61  &  20.40  &  20.37  &  20.34  & 35.7   & 4.27  & 2.511  &  28    &                 & BOSS   &                  \\
  SDSS J224325.04$-$061350.3   &   340.854357 & $-$6.230640  &  21.29  &  20.83  &  20.77  &  20.69  &  20.45  & 78.9   & 9.46  & 2.602  &  791   &  CIV            & GMOS   &                  \\
  SDSS J224325.67$-$061350.9   &   340.856994 & $-$6.230824  &  19.80  &  19.13  &  19.09  &  19.00  &  18.68  & 78.9   & 9.46  & 2.597  &  525   &  SiIV-CIV-CIII] & MagE   &                  \\
  SDSS J224856.83$+$030700.2   &   342.236798 & 3.116728     &  21.66  &  20.87  &  20.75  &  20.73  &  20.66  & 50.5   & 5.96  & 2.394  &  43    &                 & BOSS   &                  \\
  SDSS J224857.22$+$030659.5   &   342.238446 & 3.116531     &  21.40  &  20.61  &  20.49  &  20.46  &  20.30  & 50.5   & 5.96  & 2.395  &  147   &                 & BOSS   &                  \\
  SDSS J234819.19$+$005717.5   &   357.079959 & 0.954877     &  20.98  &  20.63  &  20.54  &  20.45  &  20.28  & 60.9   & 7.07  & 2.153  &  52    &                 & BOSS   &                  \\
  SDSS J234819.58$+$005721.4   &   357.081604 & 0.955965     &  19.19  &  18.84  &  18.75  &  18.71  &  18.50  & 60.9   & 7.07  & 2.159  &  44    &                 & SDSS   &                  \\
\enddata
\end{deluxetable}
\noindent\footnotesize{{\bf Note}: \revs{All magnitudes quoted here are SDSS magnitudes (no Galactic extinction applied)}. The column labelled $R_\perp$ is the transverse proper separation in kpc, while $\Delta\theta$ is the angular separation in arcseconds. Redshifts along with their uncertainties (in km s$^{-1}$) are listed in the column labeled $z$ and $\sigma_z$, respectively.}
\begin{tablenotes}
\item {$^{\rm a}$\footnotesize Emission lines considered for the redshift estimates.}
\item {$^{\rm b}$\footnotesize Survey/instrument/telescope considered for the quasar redshift. The relative reference is provided in the case a source has already a redshift available from the literature. *=\citet{2008A&A...481..615A}, **=\citet{1996A&A...305L...9C,1997A&A...327L...1S}.}
\item {$^\star$\footnotesize Lower signal-to-noise lense spectrum that has been excluded from the stacking analysis.}
\item {$\dagger$\footnotesize \revs{SDSS J133905.25$+$374755.3 and SDSS J154815.42$+$284452.6 are single field quasars observed during our survey. The system SDSS J172855.24$+$263449.1 and SDSS J172855.31$+$263458.1 is a projected pair with a line of sight velocity separations $>$11,000 km/s, thus not included in quasar pair analysis.}}
\end{tablenotes}
\end{longrotatetable}

\begin{figure*}
 \includegraphics[width=0.5\textwidth,clip]{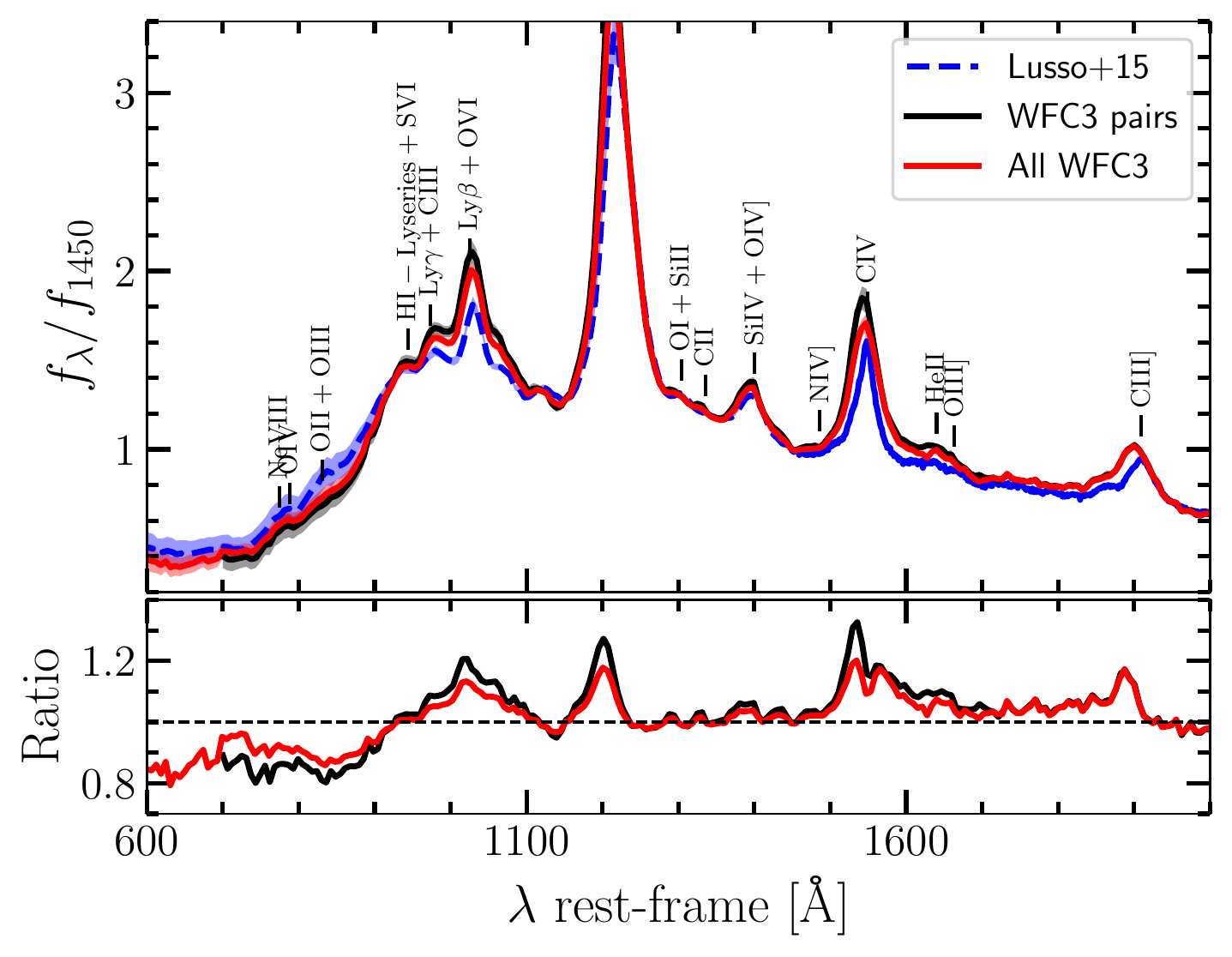}
 \includegraphics[width=0.515\textwidth,clip]{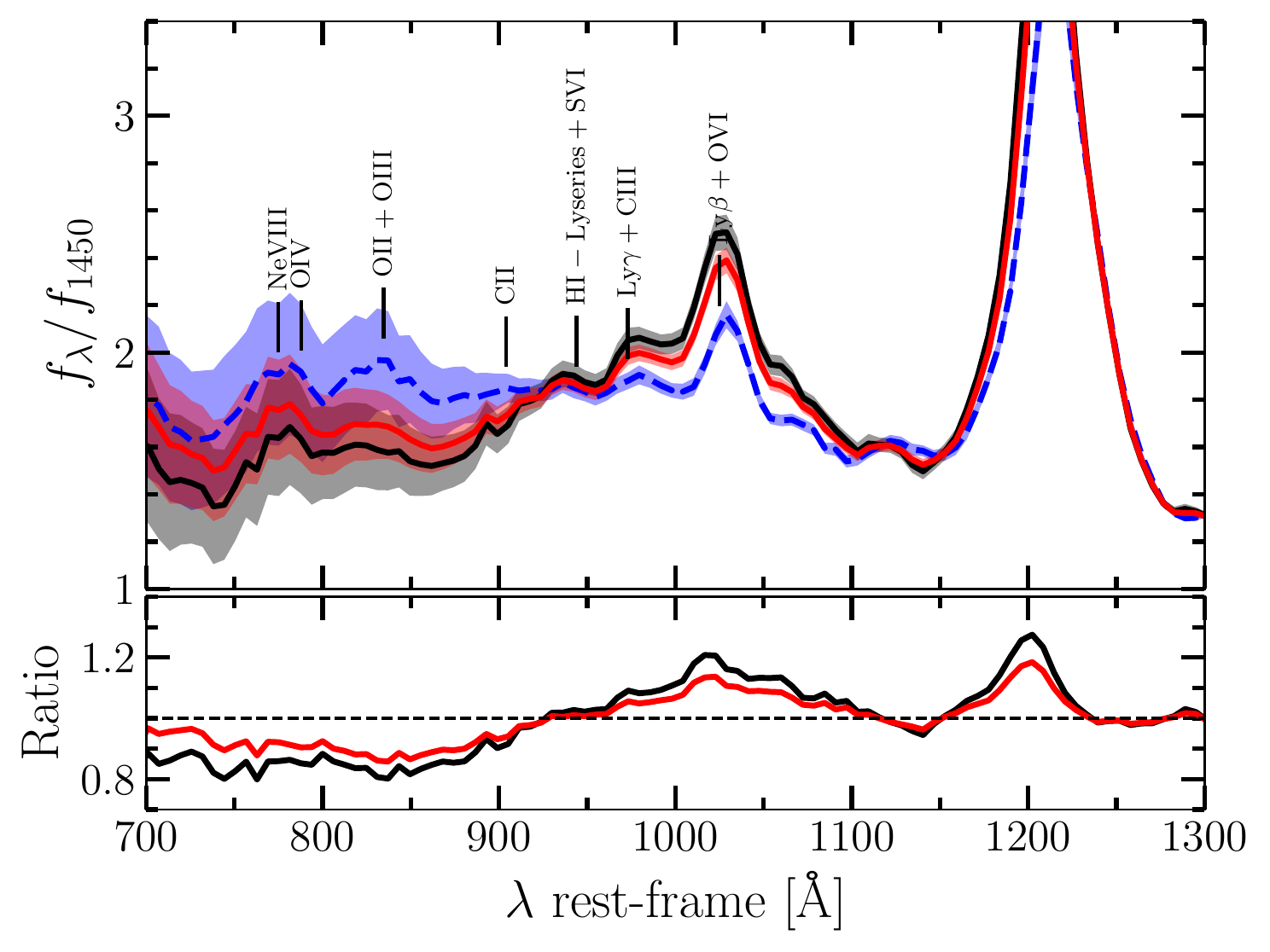}
\caption{{\it Left panel:} Mean observed quasar spectrum for the new WFC3 quasar pair sample \rev{(94 quasars, black solid line)} obtained from 10000 bootstraps compared to the one obtained from a stack of 53 WFC3 spectra at redshift $z\simeq2.44$ (L15, blue dashed line). The spectral composite considering all WFC3 quasars (157 objects) is plotted with the green solid line. All spectra are normalized to unit flux at 1450\AA. The new WFC3 to the L15 spectral stack ratio is shown in the bottom panel with the black solid line. We also show the ratio between the stack obtained with the whole WFC3 sample (157 objects) and the L15 stack (green solid line) for completeness. 
{\it Right panel}: Zoom-in of the mean IGM corrected quasar spectrum at short wavelengths with uncertainties from bootstrap (shaded area) for the WFC3 pair sample (black solid line), for the L15 (blue dashed line) sample, and considering all WFC3 quasars (157 objects).
}
 \label{obs_corr_comparison}
\end{figure*}

\section{Composite quasar spectrum}
\label{Composite construction}
The spectral stack for the WFC3 quasar pairs is constructed following a similar approach to the one in L15. 
In the WFC3 data, the observed wavelengths shorter than $\sim2100$ \AA\ and longer than 6500\AA\ are trimmed because the sensitivity of the \rev{G280 detector declines} rapidly at those wavelengths, leading to complicated systematic effects and artifacts. To construct the quasar spectral stack, we use the observed spectrum obtained combining the two beams as the reference. 
The procedure we follow is outlined below.
\begin{enumerate}

 \item We correct the quasar flux density\footnote{In the following we will use the word ``flux" to mean the flux density (i.e. flux per unit wavelength).} ($f_\lambda$) for Galactic reddening by adopting the \ebv estimates from \citet{2011ApJ...737..103S}, whose median reddening value is \ebv$\simeq 0.03$ mag, and the Galactic extinction curve from \citet{1999PASP..111...63F} with $R_V=3.1$. We do not correct the spectra for intrinsic dust absorption, as this is a relatively high redshift ($z>2$), optically-selected quasar sample, and thus intrinsic reddening is expected to be small.
 
 \item We generate a rest-frame wavelength array with fixed dispersion $\Delta\lambda$. The dispersion value was set to be large enough to include at least one entire pixel from the WFC3/UVIS-G280 spectra at rest wavelengths $\lambda < 1215$\AA~(i.e. $\Delta\lambda\simeq6.2$\AA).

 \item Each quasar spectrum was shifted to the rest-frame and rebinned  over the common rest-frame wavelength array\footnote{Wavelengths are divided by $(1+z)$ to shift the spectra into the source rest frame, while fluxes in $f_\lambda$ are multiplied by $(1+z)$.}. Given our adopted masking ($2100<\lambda_{\rm obs}<6500$\AA), the final rest-frame wavelength range where almost all objects are contributing in each flux bin is restricted to 700--2000\AA. 

 \item We normalized individual spectra by their flux at rest wavelength $\lambda =1450$\AA.

 \item All the flux values at each wavelength were then averaged (mean) to produce the stacked spectrum normalized to unity at $\lambda=1450$\AA.  
\end{enumerate}
Uncertainties on the observed stack are estimated through a bootstrap resampling technique. We created 10000 random samples of the quasar spectra with replacement, and we applied the same procedure as described above.

\subsection{Comparison to the L15 WFC3 composite for single quasars}
The quasar sample employed by L15 was drawn from a similar survey performed with \hst using the low-resolution  WFC3/UVIS-G280 grism. \hst observations and reduction procedures are described in detail in \citet{2011ApJS..195...16O} (see also \citealt{2013ApJ...765..137O}). The \citet{2011ApJS..195...16O} survey consists of 53 single quasars selected from SDSS-DR5 with $g^\ast < 18.5$ mag, with an average redshift of $\langle z \rangle\simeq2.44$. These data were taken specifically for the scientific goal of surveying the abundance of strong \ion{H}{i} Lyman limit absorption features at $z\ga 1.5$. The WFC3/UVIS-G280 spectra utilized in L15 have relatively high signal-to-noise ratio (S/N$\sim$20) per pixel down to $\lambda_{\rm obs} \sim 2000$\AA\footnote{The data generally have S/N exceeding 10 pixel$^{-1}$ at all wavelengths $\lambda_{\rm obs} >$ 2000\AA.}~(with a full width at half-maximum $FWHM\sim60$\AA~at $\lambda_{\rm obs}=2500$\AA). In comparison, our new survey has been designed to characterize the small-scale structure of optically-thick gas in the cirumgalactic medium (CGM) by observing 54 double sightlines at impact parameters in the range of 10--250 kpc \citep{2014ApJ...780...74F}.
Whilst both the L15 and the quasar pair sample are selected based on a similar SDSS optical colour selection, the pair sample is roughly two magnitudes fainter ($g^\ast=18.3-21.6$) and spans a wider redshift range ($1.9 < z < 2.7$, with $\langle z \rangle\simeq2.3$) than the L15 single quasar sample. Our quasar pair spectra also have by comparison lower signal-to-noise on average, with S/N$<$10 per pixel down to $\lambda \sim 2000$\AA\ than the L15 sample.

The spectral quasar composite (not corrected for IGM absorption) obtained with our new WFC3 quasar pair sample (94 objects) is presented in Figure~\ref{obs_corr_comparison} together with the one published by L15 for single quasars. 
The new WFC3 quasar pair composite is shown as the solid black line, while the resulting uncertainties on the stacked spectrum are plotted with a shaded area. 

The shapes of the two composites are similar overall, yet the new quasar pair stack shows {\it (i)} a moderate emission line flux excess and {\it (ii)} a $\sim$30\% decrease in flux at $\lambda=800-900$\AA\ with respect to the L15 one. 
The latter may be due to relatively small differences in the ionising continua of  quasar pairs with respect to single quasars, differences in the environment, or a combination of the two. We will discuss this in detail in Sections~\ref{Spectral fit} and \ref{Mean free path}.

Regarding the first point, the observed flux excess can be ascribed to the classical {\it Baldwin effect}, i.e., the anti-correlation between the rest-frame equivalent width (EW) of a broad-emission line and the continuum quasar luminosity \citep{baldwin1977}.
This anti-correlation becomes steeper toward higher quasar luminosities and for broad-emission lines with higher ionization potentials \citep{2002ApJ...581..912D}. 
Such behaviour is also consistent with the shape of the ionizing continuum becoming softer for more luminous quasars \citep[e.g.,][]{1982ApJ...254...22M,1992MNRAS.254...15N,1993ApJ...415..517Z,1996ApJ...467...61G,zheng97,2015MNRAS.449.4204L}.


Finally, we also constructed for completeness the composite considering all the WFC3 quasars: L15 $+$ the WFC3 quasars presented here considering lenses and the two field quasars, \revs{as well as the projected pair}, for a total of 157 sources.  For the lenses, we only considered the spectrum with the higher $S/N$. The resulting stack obtained from 20,000 bootstraps is plotted in Figure~\ref{obs_corr_comparison} with the red solid line. 
This composite, as expected, has a similar shape to the one resulting from the new WFC3 quasar pair sample, with only a few minor differences in the emission line flux. By including  the additional WFC3 single quasars from L15, the $S/N$ at 700\AA\ of the resulting stack is improved by roughly a factor of 2.

\subsection{Stack in redshift intervals}
\begin{figure}
\includegraphics[width=8.5cm,clip]{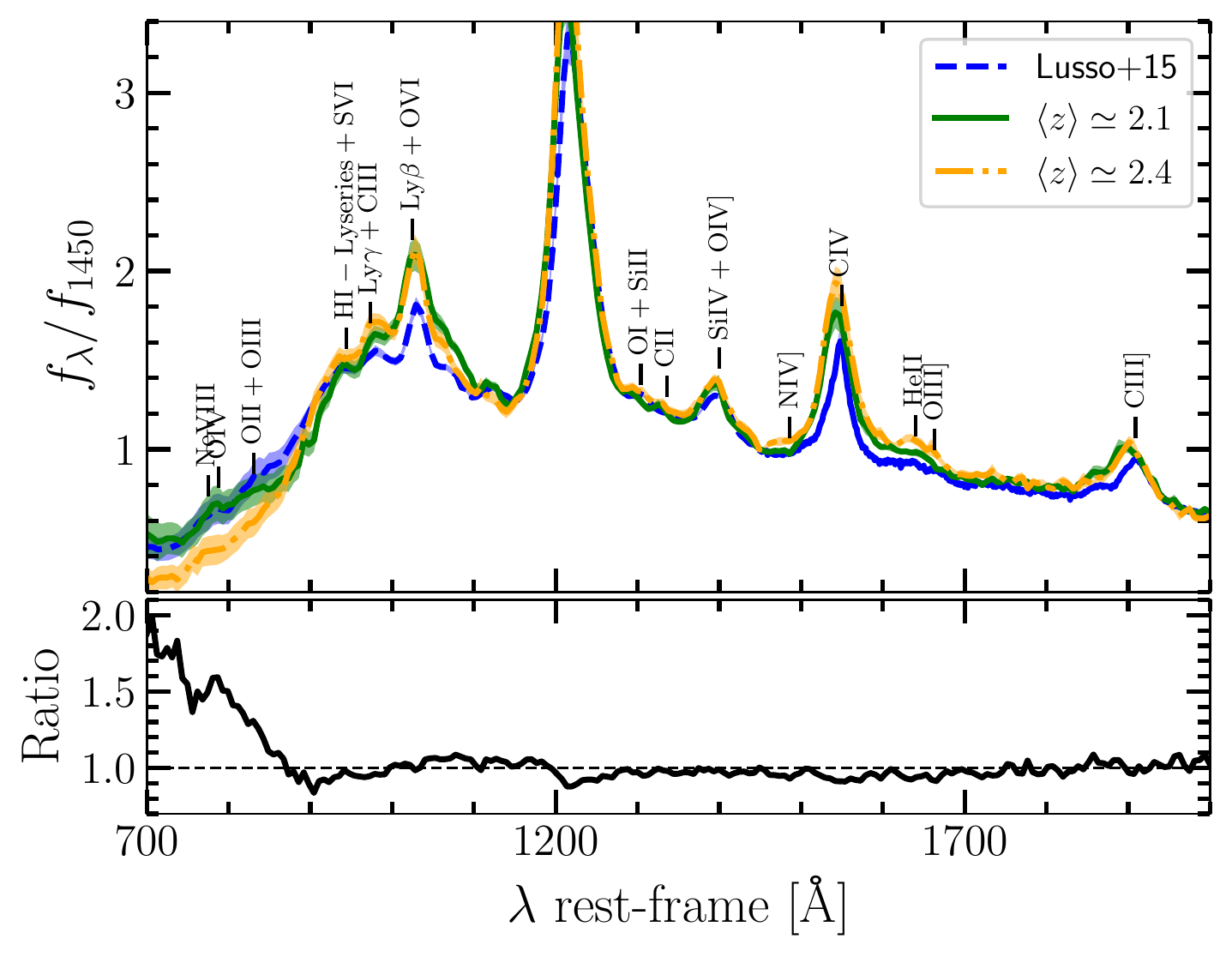}
\caption{Comparison between the mean observed quasar pair composite (from bootstrap) for the low redshift (48 quasars, $\langle z\rangle\simeq2.09$, green solid line) and the high redshift (46 quasars, $\langle z\rangle\simeq2.44$, orange dot-dashed line) quasar samples. The WFC3 and SDSS quasar composites by L15 are also shown as a reference. The ratio between the low-$z$ and the high-$z$ stack is shown in the bottom panel.}
 \label{obsstacks_lowz_highz}
\end{figure}
We further investigate how the spectral shape changes by splitting the new WFC3 quasar pair sample in two almost equally populated redshift bins: 48 quasars for the low redshift sample ($1.96\leq z \leq 2.24$, with a mean $\langle z\rangle\simeq2.09$), and 46 quasars for the high redshift sample ($2.26\leq z \leq 2.67$, $\langle z\rangle\simeq2.44$). The resulting composites are presented in Figure~\ref{obsstacks_lowz_highz}.  
Interestingly, the most notable difference between these two stacks at $\lambda<912$\AA, where the low redshift stack shows roughly a factor of two increase in flux compared to the high redshift one. 
Given that the higher redshift interval is similar in both source statistics and average redshift with respect to the L15 sample, one may expect to observe similar levels of absorption and shape of the composites. Yet, the high redshift quasar pair composite presents significant absorption (fluxes are dimmed by up to a factor of 2 at 700\AA) compared to the L15 one. On the other hand, the level of absorption of the pair composite in the low-$z$ stack at $z\sim2.09$ is overall similar to the L15 one, which is however at an average redshift of $\langle z\rangle\simeq2.44$.

\section{IGM absorption correction}
\label{IGM transmission correction}
Absorption from intergalactic \ion{H}{i} attenuates the quasar flux at wavelengths blueward of \ion{Ly}{$\alpha$}, both in the Lyman series
(creating the so-called Lyman forest) and in the Lyman continuum at rest $\lambda<912$\,\AA\ \citep[e.g.][]{1990A&A...228..299M}.
The significant abundance of neutral gas at $z>2$ is clear in our average quasar spectrum shown in the left panel of Fig.~\ref{obs_corr_comparison}. 

To recover the IGM corrected quasar emission, we considered the IGM transmission functions ($\Tl$) published by \citet[P14 and references therein]{2014MNRAS.438..476P} along with their uncertainties. These functions have been computed through a cubic Hermite spline model which describe the \ion{H}{i} absorber distribution function that, in turn, depends upon both redshift and column density ($\fnz=\partial^2n/\left(\partial N_\mathrm{HI}\partial z\right)$). P14 then performed a Monte Carlo Markov Chain (MCMC) analysis of existing constraints on $\fnz$ to derive the posterior probability distribution functions of seven spline points spaced at irregular logarithmic intervals in the range $\mnhi = 10^{12}$--$10^{22}$\,cm$^{-2}$.  

Here we consider 10,000 realizations of $\fnz$, and calculated $\Tl$ in the observed wavelength range with a semi-analytic technique (see Fig.~3 in L15). 
\rev{This modelling assumes that the \ion{H}{i} forest is composed of discrete ``lines'' with Doppler parameter $b = 24\,\mkms$ and that the normalization of $\fnz$ evolves as $\left(1+z\right)^{2.0}$ \citep{2009ApJ...705L.113P}. This redshift evolution is somewhat faster than what it has been found at lower redshifts. \citet{2016ApJ...817..111D} presented a COS survey of \ion{Ly}{$\alpha$} forest absorbers ($\log\nh>13$) at $z<0.47$ finding $\gamma=1.24\pm0.04$, whilst \citet{2017ApJ...849..106S} found $\gamma=1.14\pm0.89$ from a survey of LLSs and partial LLSs ($15.0\leq\log\nh\leq17.5$) at $0.24<z<0.48$. From an HST survey extended to higher redshifts, \citet{2011ApJ...736...42R} obtained $\gamma=1.19\pm0.56$ for LLSs with $\tau_{\rm LLS}\geq1$, also including partial LLSs, at $0.25\leq z\leq 2.59$ ($\gamma=1.33\pm0.61$ with $\tau_{\rm LLS}\geq2$).
Our assumed redshift evolution is somewhat consistent, within the uncertainties, with the $\gamma$ values obtained in high redshift surveys, and it is a reasonable value for the Ly$\alpha$ forest of quasars at $z\simeq2$, as it implies an increasing transmission for the lower-redshift Lyman series \citep{2009ApJ...705L.113P,2014MNRAS.438..476P}.}
Opacity due to metal line transitions was ignored since they contribute negligibly to the total absorption in the Lyman continuum.


The technique we followed is similar to the one described in L15. We briefly summarize the main steps below:

\begin{enumerate}
 \item We first generate a set of 20,000 mock quasar stacks, following the same 
 procedure as in Section~\ref{Composite construction}, by drawing randomly from the \rev{94 quasar spectra} to assess sample variance (allowing for duplications). 
 \item We then randomly draw one IGM transmission function from our suite of 10,000. We smoothed this to the WFC3 grism resolution (5 pixels), and we resampled the transmission function onto the rest-frame wavelength grid of our stacked quasar spectrum. This is repeated for each mock quasar stack.
 \item We divide the observed spectral flux ($f_{\lambda,\rm obs}$) by the IGM transmission curve, $f_{\lambda,{\rm corr}}=f_{\lambda,{\rm obs}}/T_{\lambda}$.
 \item The 20,000 mock stacks corrected from IGM absorption are then averaged to produce the stacked spectrum (normalized to unity at $\lambda=1450$\AA).
 \item The uncertainties on the corrected WFC3 stacked spectrum are estimated from the dispersion of these 20,000 mock stacks.
\end{enumerate}
The resulting stack for our WFC3 quasar pair sample is shown in right panel of Figure~\ref{obs_corr_comparison}, along with its 1$\sigma$ uncertainties and the WFC3 composite published by L15 as a comparison. 
The stacked spectra show somewhat similar shapes, with a softening at wavelengths $\lambda<912$\AA\ and several (mostly blended) emission lines.

The average IGM corrected WFC3 composites for the full WFC3 quasar sample (157 quasars) and for the pairs only \rev{(94 quasars)} are tabulated in Table~\ref{tab:spectra}.
\begin{table}
  \caption{WFC3 stacked spectrum corrected for IGM absorption.}
  \label{tab:spectra}
  \begin{center}
    \begin{tabular}{ccc|cc} \hline \hline  
    $\lambda^{\mathrm{a}}$   &   $f_{\lambda,\rm All}^{\mathrm{b}}$  & $\sigma(f_{\lambda,\rm All})^{\mathrm{c}}$   &   $f_{\lambda,\rm pairs}$  & $\sigma(f_{\lambda,\rm pairs})$ \\ 
    & All WFC3 &(157 quasars) & pairs only & (94 quasars) \\ \hline 
700.955  &  1.752  & 0.279  &  1.603  & 0.323    \\            
707.144  &  1.683  & 0.262  &  1.508  & 0.297    \\            
713.334  &  1.611  & 0.250  &  1.451  & 0.291    \\            
719.524  &  1.598  & 0.238  &  1.460  & 0.274    \\            
725.713  &  1.569  & 0.225  &  1.448  & 0.256    \\            
731.903  &  1.553  & 0.220  &  1.429  & 0.255    \\            
738.093  &  1.499  & 0.213  &  1.348  & 0.250    \\               
744.282  &  1.514  & 0.208  &  1.355  & 0.244    \\               
750.472  &  1.584  & 0.208  &  1.436  & 0.232    \\     
   \hline
    \end{tabular}
 \flushleft\begin{list}{}
 \item {\bf Notes.}
 \item${}^{\mathrm{a}}${ Rest-frame wavelength in Angstrom.}
 \item[]${}^{\mathrm{b}}${ Mean IGM corrected flux per \AA~normalized to the flux at 1450\AA.}
 \item[]${}^{\mathrm{c}}${ Flux uncertainties from our bootstrap analysis (see \S\ref{IGM transmission correction}).}
 \end{list}
 \vspace{0.2cm}
 (This table is available in its entirety in a machine-readable form in the online journal. A portion is shown here for guidance regarding its form and content.)
  \end{center}
\end{table}

\subsection{Spectral fit}
\label{Spectral fit}
We measured the properties of the most prominent emission lines and the spectral continuum by employing QSFit ({\it Quasar Spectral Fitting package}; \citealt{2017MNRAS.472.4051C}) that automatically performs the analysis of quasar spectra. This software provides, amongst other parameters, FWHM values, velocity offsets, and equivalent widths (EWs) of a number of emission lines. QSFit fits all the components simultaneously considering a single power law to describe the quasar continuum over the entire (rest-frame) wavelength coverage. We defer the interested reader to \citet{2017MNRAS.472.4051C} for details, here we briefly summarise the main features of this software which are relevant for our analysis.
We fitted the broad component of several emission lines such as \ion{Ly}{$\alpha$}, \ion{Si}{iv}, \ion{C}{iv}, and the semi-forbidden line of \ion{C}{iii}], as well as a combination of templates for the optical and UV iron emission \citep{2001ApJS..134....1V,2004A&A...417..515V}. We also considered a list of weaker lines that are not identified by QSFit (i.e. lines not associated with any known line, Section 2.7 in \citealt{2017MNRAS.472.4051C}). These additional components account for asymmetric profiles in known emission lines.

Lines and blends at $\lambda<1216$\AA\ from high-ionisation states such as \ion{O}{iv} 608, \ion{O}{v} 630, \ion{N}{iii} 685, \ion{O}{iii} 702, \ion{Ne}{Viii}+\ion{O}{iv} 772 and \ion{Ly}{$\gamma$}+\ion{C}{iii}] 873, may also be present, but it is impossible to reliably measure their strengths given the noise in our stacked spectrum at blue wavelengths. We thus fit our quasar composite only at $\lambda>1100$\AA. 
At the redshift and wavelength ranges probed by our WFC3 sample, the emission from the hosting galaxies and the Balmer continuum are negligible, we thus neglected both components in the fit.

\begin{figure*}
\includegraphics[width=9.cm,clip]{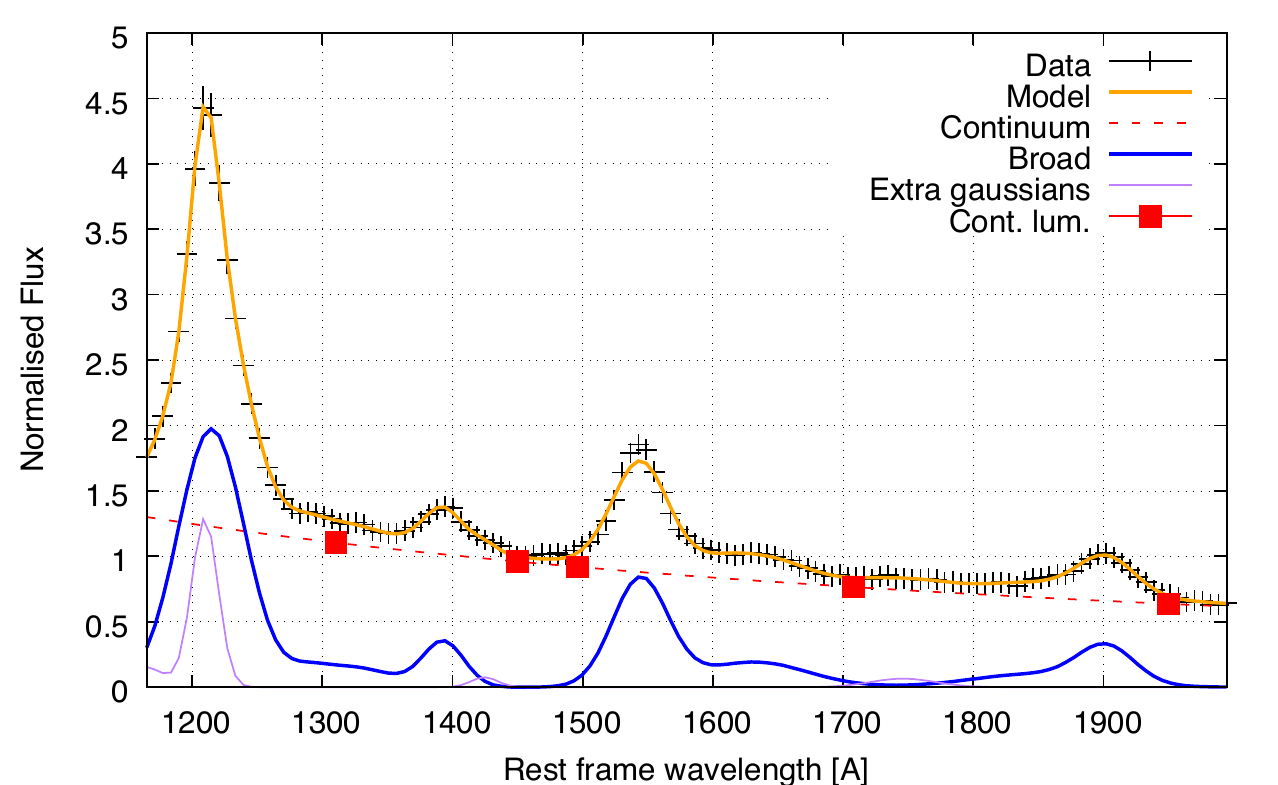}                                 
\includegraphics[width=9.cm,clip]{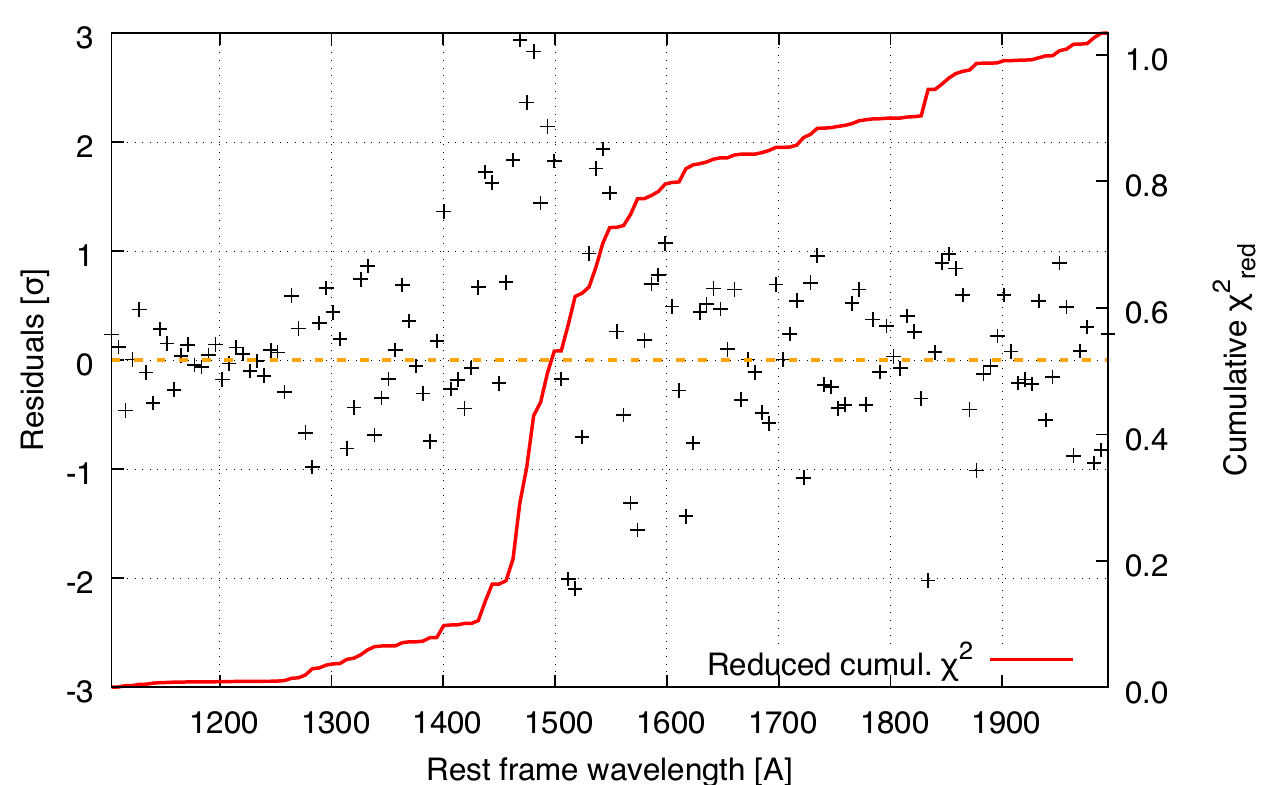}                                 
 \caption{{\it Left panel}: QSFit model (orange line) of the average spectrum of the \rev{94 quasars} normalized at 1450\AA. The individual components in the QSFit model are: the continuum component (dashed red line); the sum of all broad emission-line components is shown with a blue line. The sum of all ``Extra gaussian" lines (features not fitted with any known emission line) is shown with a purple solid line. The red squares are the continuum luminosity estimated by QSFit (see \S\ref{Spectral fit} for details). {\it Right panel}: Residuals (data -- model) in units of 1$\sigma$ uncertainties in the data (black cross symbol) and the cumulative reduced $\chi^2$ (red line, values on the right-hand axis).}
 \label{fit}
\end{figure*}

Figure~\ref{fit} (left panel) shows the rest-frame stacked spectrum for the \rev{94 quasars} extending from 1100 \AA~to 2000 \AA\ and the power-law fit to the continuum of the form $f_\lambda\propto\lambda^{\alpha_\lambda}$, where the best-fit power law index is $\alpha_\nu=-0.61\pm0.08$\footnote{In the following we will refer to $\alpha_\nu$ only. The relation between the fluxes in wavelength, $f_\lambda\propto\lambda^{\alpha_\lambda}$, and frequencies, $f_\nu\propto\nu^{\alpha_\nu}$, is $\alpha_\nu = - (2+\alpha_\lambda)$.} (dot-dashed line), in good agreement with previous works in the literature.
A summary of the spectral properties (i.e. full width at half-maximum, velocity offset, and equivalent widths) for the most prominent lines with no quality flag raised (i.e. ``good'', whose quality flag is 0) is provided in Table~\ref{tab:emlines}. 

To compare these findings with the ones of L15, we re-fitted their composite (WFC3 + SDSS) for single quasars using QSFIT with the same set-up. We find the same quasar continuum slope of $\alpha_\nu=-0.61\pm0.10$, whilst the FWHM, $v_{\rm obs}$, and EW are reported in  Table~\ref{tab:emlines}. 

\begin{figure}
\includegraphics[width=8.5cm,clip]{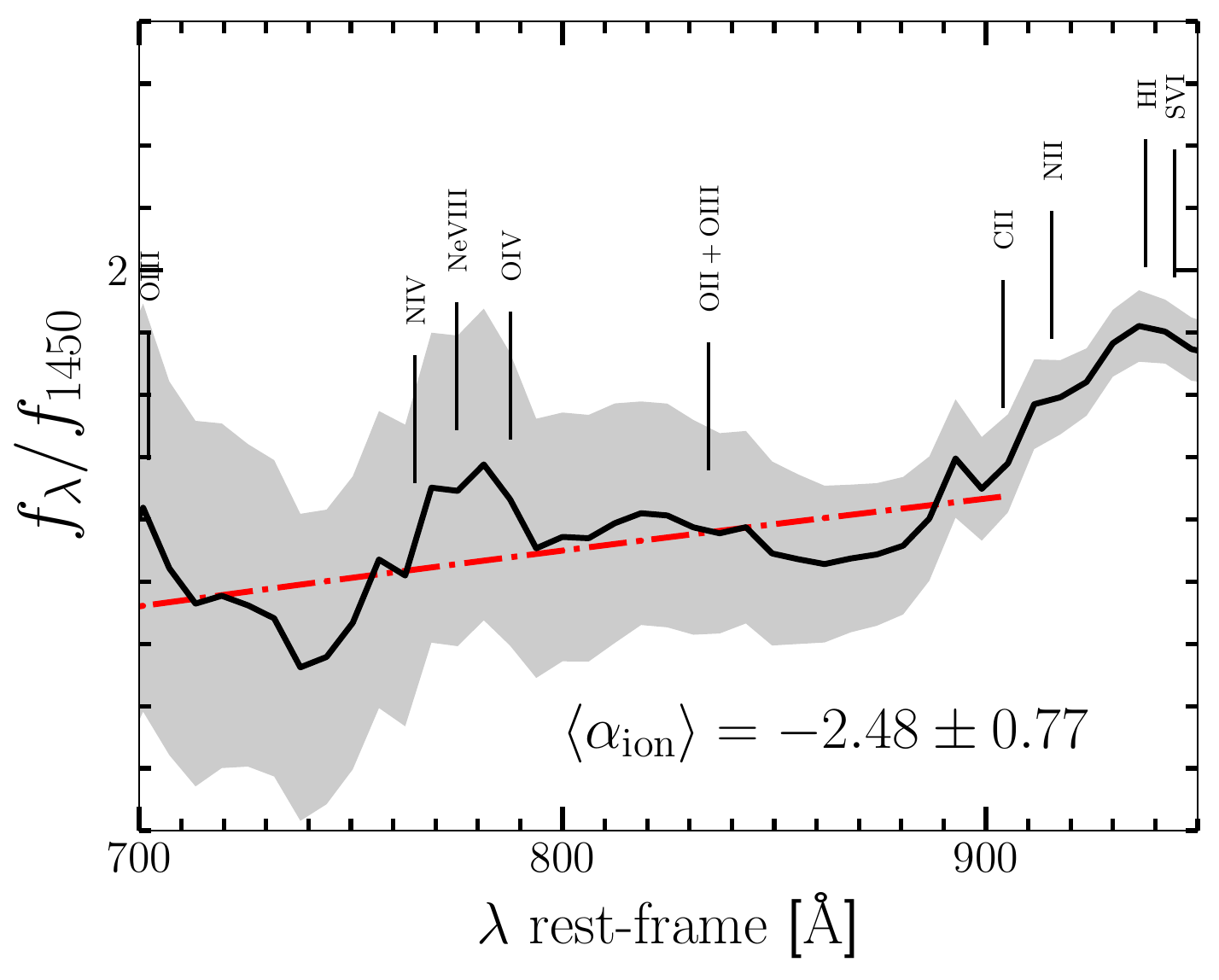}                                 
  \caption{The ionizing region of the average spectrum normalized at 1450\AA\ (700$-$912\AA) for the \rev{94 quasars}. The dashed line is the power-law continuum+lines obtained by fitting the rest-frame composite at $\lambda<912$\AA.}
 \label{fitfuv}
\end{figure}

The ionising slope is estimated by modelling continuum+lines with a simple single power-law in a similar fashion as done by L15. We computed the best-fit slope \rev{of the composite at $\lambda<912$\AA\ results} from a $\chi^2$ minimization in a each bootstrap realization as described in Section~\ref{IGM transmission correction}. The final value we quote for the  spectral slope (along with the 1$\sigma$ uncertainties) is estimated from the mean (and standard deviation) of all the bootstrap realizations. \rev{The resulting ionising slope is $\alpha_{\rm ion} = -2.48 \pm 0.77$. Figure~\ref{fitfuv} presents a zoom-in of the ionizing part of the composite with the resulting best-fit. We caution that, given the low WFC3 resolution and the high level of noise in the ionising region, we cannot identify weak lines that should be present at $\lambda<912$\AA, \rev{including \ion{Ne}{Viii} 775, \ion{O}{iv} 787.7, \ion{O}{ii}+\ion{O}{iii} 834.5, and \ion{Ly}{$\gamma$}+\ion{C}{iii} 873}. Blended lines from high-ionisation states such as \ion{O}{iv} 608, \ion{O}{v} 630, \ion{N}{iii} 685, and \ion{O}{iii} 702 (fundamental diagnostics for studying the physical conditions of broad emission line regions) may also be present, although it is impossible to reliably measure their strengths. For example, the ``dip" at $\lambda\simeq730-750$~\AA\ could also be a line-free region instead of a trough. However, the same dip is observed in the geometric mean (regarded as the better characterization of the AGN composite) by S14 (see the top panel of their Fig.~5) and the stack has a ionising slope of $-1.41\pm0.15$, whilst this dip disappears in their median (bottom panel of their Fig.~5, showing a slope of $-1.32\pm0.15$). This further highlight the challenge in estimating the ionising slope in quasar spectra.
Our ionising slope of the spectral fit shown in Figure~\ref{fitfuv}, although uncertain, should only be considered representative of the combined contribution of both continuum and emission lines of quasars at $z>2$ given the IGM transmission functions employed (possible caveats are discussed in \S~\ref{Caveat on the IGM trasmission function employed}).}

 \rev{We also computed the non-ionizing and ionizing spectral slopes by considering the IGM corrected spectral stack of the combined WFC3 sample from our previous survey (i.e. O11, O13, L15; 53 quasars) and the whole \hst quasar sample from our new WFC3 program (104 quasars): 157 quasar spectra in total.
We find spectral slopes of $\alpha_\nu=-0.52\pm0.04$ and $\alpha_{\rm ion}=-1.98\pm0.50$ for the non-ionizing and ionizing part of the spectrum, respectively, in statistical agreement with the values obtained considering only the WFC3 quasar pair sample.}

As discussed by L15, our analysis supports the results that a single power law does not seem to be a satisfactory description of the region below 912\AA, where the continuum exhibits a break with a flatter (softer) spectrum (see also \citealt{2002ApJ...565..773T}).
The shape of our new WFC3 stacks present a 20\% flux decrement around 912\AA\ and a very faint \rev{\ion{O}{ii}+\ion{O}{iii}$\lambda834.5$\AA\ blend}. Additionally, it is not trivial to interpret the feature at $\sim$ 730\AA\ as intrinsic continuum or absorption. Therefore, given the poor spectral resolution and the difficulties in fitting the ionizing spectral region, we refrain from employing more complicated models. 

%
\begin{table}
  \caption{Spectral properties of the WFC3 quasar pair composites}
  \label{tab:emlines}
  \begin{center}
    \leavevmode
    \begin{tabular}{lcccc} \hline \hline              
    Line  & $\lambda$ &  FWHM$^\mathrm{a}$  & $v_{\rm obs}^\mathrm{b}$ &  EW$^\mathrm{a}$ \\  
             & (\AA) &  (km/s)   & (km/s) &  (\AA) \\ \hline 
             & & \quad pairs &&\\ 
  \ion{Ly}{$\alpha$} &1215.7     & 10834.0$\pm$904.0  & 1518.2$\pm$175.2 &  68.3$\pm$4.6  \\
  \ion{Si}{iv} & 1399.8                  & 8220.7$\pm$953.1 & 1356.2$\pm$284.8 & 13.3$\pm$3.5    \\
  \ion{C}{iv} & 1549.48                & 8483.9$\pm$225.5 & 1067.7$\pm$105.2 & 50.0$\pm$2.0    \\
  \ion{C}{iii}] & 1908.734              & 8405.5$\pm$398.4 & 1356.3$\pm$182.2 & 22.0$\pm$6.7   \\
   \hline
   \hline
     \multicolumn{5}{c}{L15} \\
  \ion{Ly}{$\alpha$} & 1215.7     & 11848.0$\pm$804.7  & 148.1$\pm$125.2 &  57.3$\pm$3.2  \\
  \ion{Si}{iv} & 1399.8                  & 8428.0$\pm$782.2 & 568.0$\pm$258.1 & 12.3$\pm$0.7    \\
  \ion{C}{iv} & 1549.5                & 5857.2$\pm$354.7 & 301.8$\pm$72.7 & 19.2$\pm$4.6   \\
  \ion{C}{iii}] & 1908.7              & 7291.5$\pm$151.7 & -50.4$\pm$63.0 & 10.0$\pm$0.3    \\
  \multicolumn{5}{c}{Additional Lines} \\
  \ion{He}{ii} & 1640.4              & 14912.0$\pm$892.1 & 1486.8$\pm$301.2 & 9.5$\pm$0.7   \\
  \ion{O}{iii} & 1665.8              & 3534.4$\pm$913.4 & --860.1$\pm$288.4 & 0.65$\pm$0.3    \\
   \hline
    \end{tabular}
 \flushleft\begin{list}{}
 \item {\bf Notes.}
 \item${}^{\mathrm{a}}${Full width at half-maximum and equivalent widths of the emission lines in the WFC3 stack normalized at 1450\AA. Only the broad component of the emission line is reported.}
 \item${}^{\mathrm{b}}${Velocity offset with respect to the reference wavelength (only broad component).}
 \end{list}  \end{center}
\end{table}

\subsection{Caveat on the IGM trasmission function employed}
\label{Caveat on the IGM trasmission function employed}
\rev{The new WFC3 spectral stack (see Figures~\ref{obs_corr_comparison} and \ref{fitfuv}), corrected for the IGM absorption following the procedure outlined at the beginning of this Section, implicitly assumes that our employed $\Tl$ functions are, on average, representative of the IGM of quasar pairs. In other words, the environment of quasars pairs is not expected to be statistically different from the one of single quasars.
Nonetheless, the $\Tl$ function critically depends upon the parametrization of $\fnz$ \citep{1995ApJ...441...18M,2006MNRAS.365..807M,2014MNRAS.442.1805I} and its statistical nature is due to the stochasticity of Lyman limit systems \citep{1999ApJ...518..103B,2008MNRAS.387.1681I,2011ApJ...728...23W}. Our approach takes into account the stochasticity of Lyman limit absorption \citep{2011ApJ...728...23W}, therefore this is the best way to correct for Lyman series and Lyman continuum absorption of low-column density absorbers that cannot be identified and corrected by eye. In addition, LLSs could perhaps be masked by the low spectral resolution of WFC3, which prevents an unambiguous identification of weak partial Lyman limit systems in individual spectra without knowledge of the underlying quasar continuum.

The new WFC3 quasar pair spectral composite (with $\alpha_\nu\simeq-2.5\pm0.8$) is thus representative of the intrinsic shape (which we argue is in range $\alpha_\nu\simeq-1.4,-1.7$) plus any additional contribution of absorption associated with the quasar pair environment, which is not captured by our $\Tl$ functions. }
To investigate this further, in the following sections we will focus our attention on the ionizing region of the spectral composite to provide a better modelling of the IGM properties in proximity of these pairs.

\section{The Mean free path}
\label{Mean free path}
The most notable absorption features in quasar spectra are optically-thick absorption line systems, namely LLSs and damped \ion{Ly}{$\alpha$} absorbers (DLAs). These systems have a higher neutral hydrogen fraction than the IGM, and have column densities of $\nh>10^{17.2}$ cm$^{-2}$ (i.e. they are optically-thick to Lyman continuum photons). They play a major role in modulating the intensity of the extragalactic UV background, and in the determination of the mean free path to ionising photons in the IGM \citep[e.g.,][]{1997ApJ...489....7R,1998AJ....115.2206F,1999AJ....118.1450S,haardt12,faucher08,faucher09}. 

Our previous analysis has focused on the spectral properties of the quasar pairs once corrected for an average IGM absorption. We now assess more quantitatively the properties of the UV composite at $<910$\AA\ to investigate whether the quasar pair sample displays differences compared to single quasars, which could be ascribed to a different environment and/or a different ionization state of the IGM near these quasars. The first measurement we perform to this end is quantifying the mean free path of ionizing photons.

\subsection{Formalism}
\label{Formalism}
\begin{figure}
\includegraphics[width=8.5cm,clip]{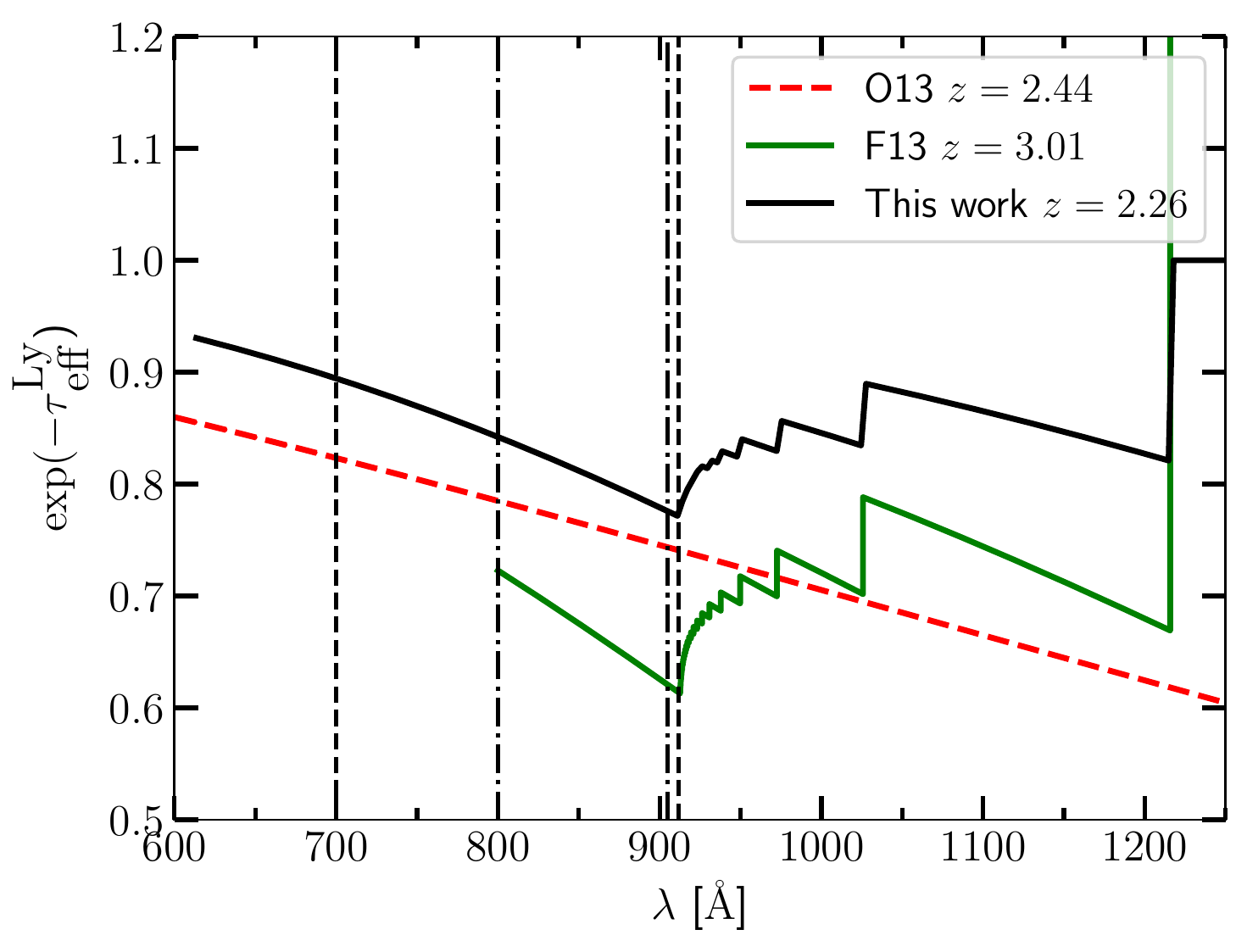}                                 
\caption{Our adopted Lyman series opacity as defined in equation~(\ref{taulyz}), where the employed $f(\nh,z)$ function is taken from \citet{2014MNRAS.438..476P}, and extrapolated at $z_{\rm qso}\simeq2.26$ for a fixed Doppler parameter $b=24$ km s$^{-1}$ (black solid line). The green line represents $\tauly$ for a redshift $z\simeq3$ (Fumagalli et al. 2013), while the red dashed line is the best-fit Lyman series opacity obtained by O13 given equation~(\ref{tauly}), with $\tauly(\lambda_{912})=0.3$ and $\gamma_\tau=1.64$. The vertical dashed and dot-dashed lines represent the wavelength ranges $700-912.76$~\AA\ and $800-905$~\AA, respectively. \label{taueff_ly}}
\end{figure}

To estimate the mean free path to ionising radiation $\lmfp$ we consider and review the model presented by
\citet[O13 hereafter, see also \citealt{2011ApJS..195...16O,2011ApJ...728...23W,2013ApJ...775...78F,worseck14,2014MNRAS.438..476P}]{2013ApJ...765..137O}.
The observed stacked quasar SED blueward of \ion{Ly}{$\alpha$}, i.e. $\lambda<1216$\AA, can be modelled as
\begin{equation}
\label{fobs_sed}
f_{\lambda,{\rm obs}} = a f_{\lambda,{\rm SED}} \lambda^{-\alpha} \exp(-\taue),
\end{equation}
where the term $a f_{\lambda,{\rm SED}} \lambda^{-\alpha}$ accounts for the intrinsic quasar ionising continuum ($f_{\lambda,{\rm intr}}$) and it will be discussed in detail in Section~\ref{quasar SED}. The $\taue$ parameter is the effective optical depth due to intervening absorbers and includes the contribution of absorption in the hydrogen Lyman series ($\tauly$) and the Lyman continuum ($\taull$) at $\lambda<912$\AA\ (i.e., $\taue=\tauly+\taull$).
The redshift evolution for $\tauly$ is usually defined as 
\begin{equation}
\label{tauly}
\tauly(\lambda) = \tauly(\lambda_{912}) \left(\frac{1+z_{912}}{1+z_{\rm qso}}\right)^{\gamma_\tau},
\end{equation}
where $z_{912}$ is the redshift at which a photon emitted at the redshift of the quasar ($z_{\rm qso}$) is absorbed at the Lyman limit
\begin{equation}
\label{z912}
z_{912} = \lambda(1+z_{\rm qso})/\lambda_{912} -1.
\end{equation}
As already discussed by \citet{2013ApJ...775...78F} and \citet{2011ApJ...728...23W}, one could consider both $\tauly(\lambda_{912})$ and $\gamma_\tau$ as free parameters, but the data between 700 and 912~\AA\ do not constrain them independently. To account for the Lyman series opacity we have thus estimated $\tauly$ numerically as
\begin{equation}
\label{taulyz}
\tauly(z) = \sum^\infty_{n=1} \int\int\int_z^{z_{\rm qso}} f(\nh,z^\prime,b)\exp(-\tau_\nu^n) d\nh dz^\prime db,
\end{equation}
where $\tau_\nu$ is the optical depth due to the incidence of the Ly series ($n=1,2,3,...$ corresponds to \ion{Ly}{$\alpha$}, \ion{Ly}{$\beta$}, \ion{Ly}{$\gamma$}, etc; \citealt{madau99}), and $f(\nh,z)$ is the column density distribution function from \citet{2014MNRAS.438..476P}. We have fixed the Doppler parameter $b$ for the Ly series to 24 km s$^{-1}$ and $z_{\rm qso}$ to the average redshift of the quasar sample ($z_{\rm qso}\simeq2.26$). The resulting Lyman series opacity is plotted in Figure~\ref{taueff_ly}. As a comparison, we also plotted the best-fit Lyman series opacity obtained by O13 given equation~(\ref{tauly}), with $\tauly(\lambda_{912})=0.3$ and $\gamma_\tau=1.64$ (see their eq.~(4) and Table~7), and the one employed by \citet{2013ApJ...775...78F} for $z\simeq3$ quasars.
The characteristic sawtoothed behaviour is due to the incidence of the  Ly series lines ($\tau_\nu$), while the shape at wavelengths bluer than 912\AA\ depends upon the adopted $f(\nh,z)$. Such a correction is anyways minor compared to the Lyman limit opacity, of the order of $\sim10-15$\% in the wavelength range of interest.

The last parameter we need to model is the Lyman limit optical depth, which is set by the opacity $\kappa^{\rm LL}$ seen at each redshift by the ionising photons emitted at $\lambda<912$\AA\ over a path-length $r$ from the quasar redshift to $z$,
\begin{equation}
\label{taullr}
\taull(\lambda) = \int_0^r \kappa^{\rm LL}(r^\prime,\lambda) dr^\prime.
\end{equation}
Following \citet{2013ApJ...775...78F} (see their Section~4), the opacity $\kappa^{\rm LL}(r,\lambda)$ can be re-written as a function of redshift
\begin{equation}
\kappa^{\rm LL}(r,\lambda) \simeq \kappa^{\rm LL}(z) = \kappa^{\rm LL}_{912}(z) \left(\frac{1+z}{1+z_{\rm qso}}\right)^{-2.75},
\end{equation}
which is defined by the product of the redshift-dependent opacity $\kappa^{\rm LL}_{912}(z_{\rm qso})$ and the \ion{H}{i} photoionization cross-section ($\sigma_{\rm ph}\propto \lambda^{2.75}$).
The dependence of $\kappa^{\rm LL}_{912}(z)$ with redshift can be parametrised as follows
\begin{equation}
\kappa^{\rm LL}_{912}(z) = \kappa^{\rm LL}_{912}(z_{\rm qso}) \left(\frac{1+z}{1+z_{\rm qso}}\right)^{\gamma_\kappa},
\end{equation}
but since $\kappa^{\rm LL}_{912}(z)$ is only weakly-dependent on redshift ($\gamma_\kappa\simeq0.4$ at $z\simeq2.4$, O13), we assume $\gamma_\kappa=0$ for our analysis.
For a given cosmology,
\begin{equation}
\label{cosmo}
\frac{dr}{dz} = \frac{c}{H_0(1+z)\sqrt{\Omega_{\rm M} (1+z)^3 + \Omega_\Lambda}} , 
\end{equation}
\rev{where we neglect the contribution of $\Omega_\Lambda$ given the redshift range probed by our quasar sample, resulting in error on the order of 3--5\% in our measurement}.
By combining the above equations, we can now define the final expression for the Lyman limit opacity as
\begin{equation}
\label{taull}
\taull = \frac{c}{H_0~\sqrt{\Omega_{\rm M}}} (1+z_{912})^{2.75} \kappa^{\rm LL}_{912}(z_{\rm qso}) \int_{z_{912}}^{z_{\rm qso}} (1+z^\prime)^{-5.25} dz^\prime.
\end{equation}

The final model for the normalized observed quasar SED is thus constructed by combining equations~(\ref{taulyz}) and (\ref{taull}) in equation~(\ref{fobs_sed}).

\subsection{The intrinsic quasar SED}
\label{quasar SED}
To quantify the ``extra" absorption observed in the quasar pair composite, we need to model the shape of the intrinsic quasar SED. This is a key assumption in our estimate of the mean free path and a necessary step in order to probe the foreground IGM. \rev{Previous works in the literature have found very similar slopes in the $1200-2000$\AA\ wavelength range, with spectral indexes roughly around $\alpha_\nu\simeq-0.61, -0.83$ (e.g. L15; Stevans et al. 2014), while the situation changes at much shorter wavelengths (e.g. the rest-frame range $500-1200$\AA) where the slope may vary significantly, from $\alpha_\nu\simeq-0.56, -0.72$ (Scott et al. 2004; Tilton et al. 2016), to $\alpha_\nu\simeq-1.41, -1.70$ (Shull et al. 2012; Stevans et al. 2014; L15).}

However, given the similarity of the redshift range between our quasar sample and the one presented by L15, we assume that the underlying intrinsic quasar SED is the L15 (corrected for IGM absorption) modulated with a power-law $\alpha=0.3$, \rev{i.e. $\alpha_\nu=-1.7$, we assumed that the intrinsic continuum slopes are the same for both pairs and singles quasars (the differences arise in the large-scale environment)},
\begin{equation}
\label{fintr_sed_l15}
f_{\lambda,{\rm intr}} = a f_{\lambda,{\rm L15}} \left(\frac{\lambda}{1450\textrm{\AA}}\right)^{-0.3},
\end{equation}
where $a$ allows for an offset between the assumed intrinsic SED and the WFC3 composite, which is a free parameter. Even though the quasar SED is normalised to 1450\AA, there may be some non-trivial difference in the flux measurement or emission line strength.
Given the different nature of our sample (i.e. $\sim90$\% are close  quasar pairs versus single field quasars), and the fact that the observed SED seems to show a mild level of absorption at $\sim800-850$\AA\ (see Figure~\ref{obs_corr_comparison}), we will discuss how our results change if we assume different intrinsic quasar SEDs in the following section.
\begin{figure}
\includegraphics[width=8.5cm,clip]{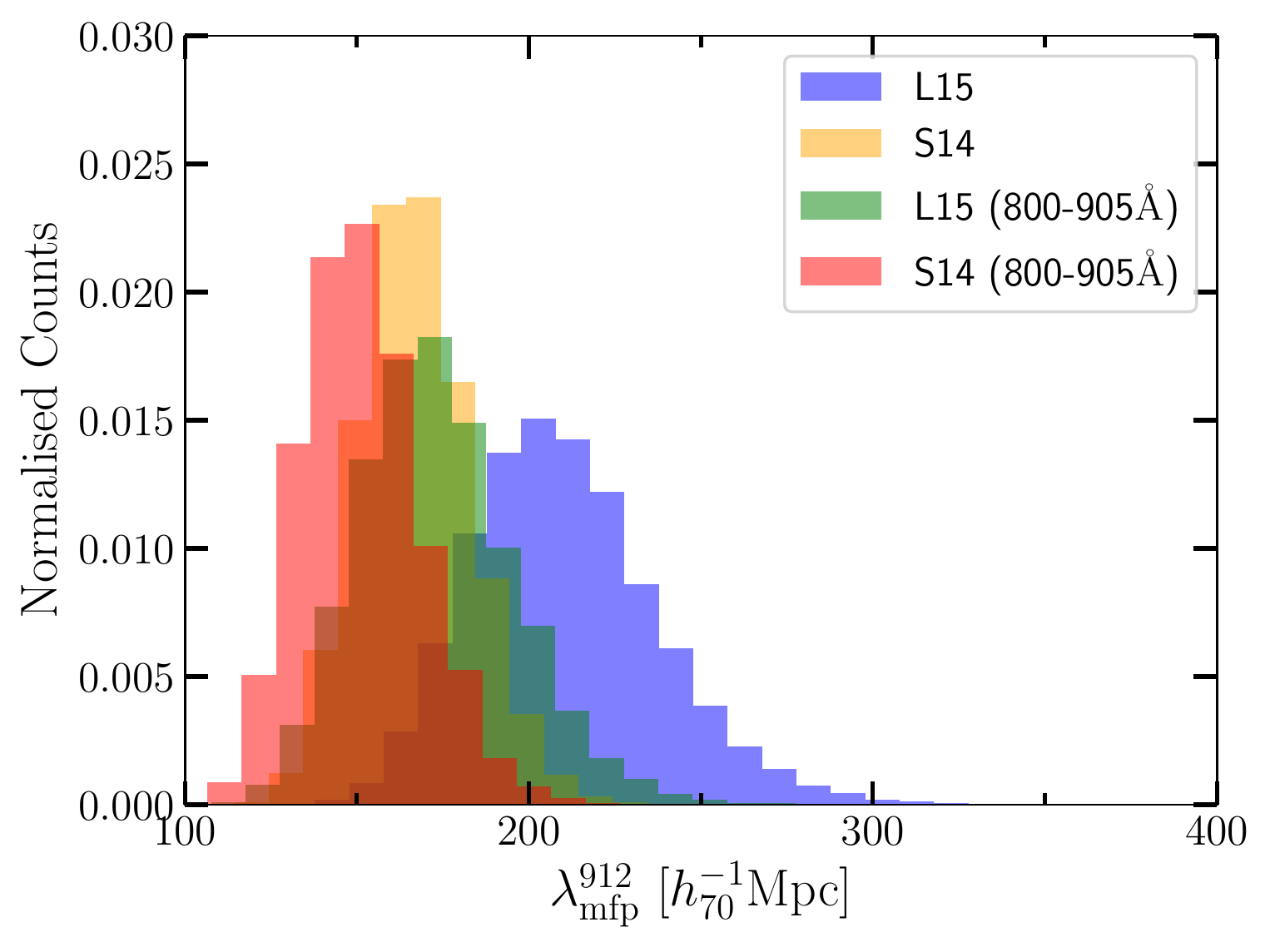}                                 
 \caption{Distributions of the best-fit $\lmfp$ values derived by modelling the 700--911.76~\AA\ wavelength range of 15,000 composite spectra via the bootstrap technique described in \S\ref{Mean free path}. We considered two different models for the intrinsic quasar SED: the IGM corrected L15 spectral stack (blue histogram) and the Stevans et al. (2014) composite (orange histogram). The same analysis has also been performed over a narrower wavelength range (i.e. 800--905~\AA).}
 \label{hist_lmfp}
\end{figure}
\begin{table}
  \caption{Mean $\lmfp$ values from the analysis of the quasar pair WFC3 sample.}
  \label{tab:lmfp}
  \begin{center}
    
    \begin{tabular}{lcc} \hline \hline  
SED model & $\langle\lmfp\rangle$ & $a$ \\ 
 & $(\h$ Mpc$)$ & \\ \hline 
L15 (700--911.76\AA) & 210.4$\pm$27.4 & 0.70$\pm$0.04 \\
S14 (700--911.76\AA) & 166.9$\pm$16.2 & 0.75$\pm$0.04 \\
L15 (800--905\AA) & 173.4$\pm$22.7 & 0.73$\pm$0.04 \\
S14 (800--905\AA) & 151.3$\pm$17.3 & 0.76$\pm$0.04 \\
   \hline
     \multicolumn{3}{c}{low-$z$ sample} \\
     L15 (700--911.76\AA) & 314.5$\pm$64.9 & 0.64$\pm$0.05 \\
   \hline
     \multicolumn{3}{c}{high-$z$ sample} \\
     L15 (700--911.76\AA) & 140.7$\pm$20.2 & 0.81$\pm$0.06 \\
   \hline
    \end{tabular}
  \end{center}
\end{table}
\begin{figure*}
 \plottwo{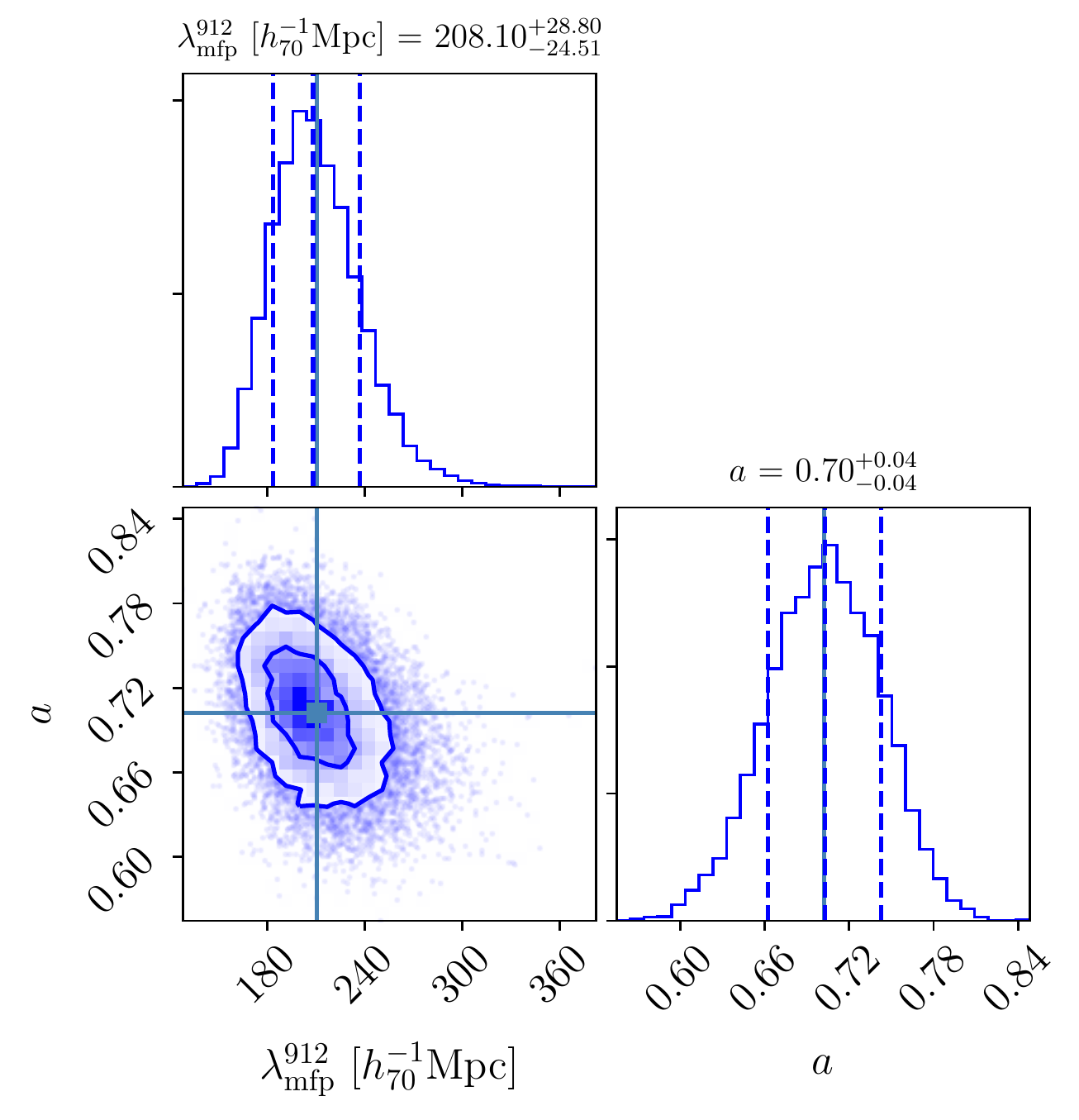}{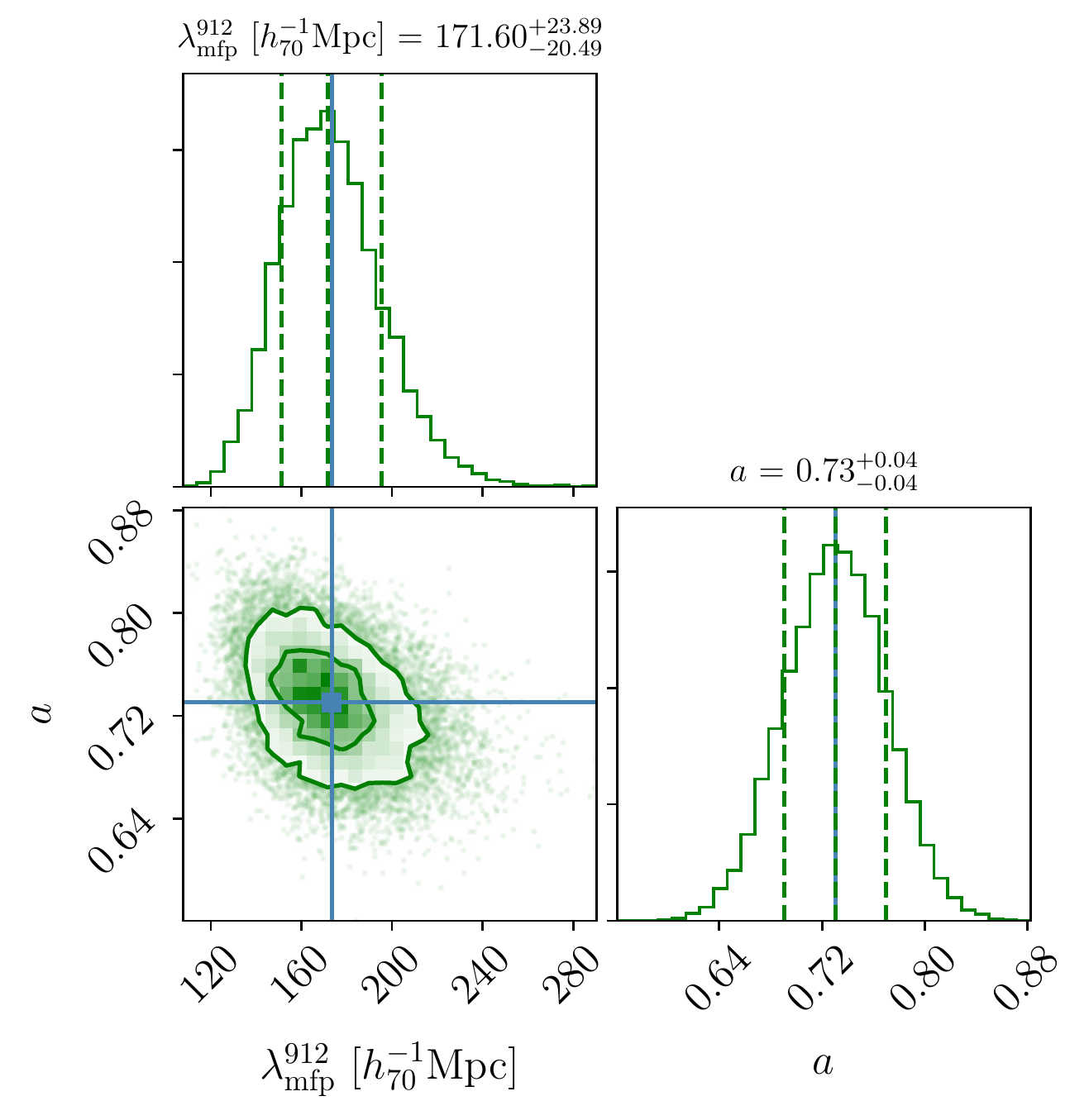}\\
 \plottwo{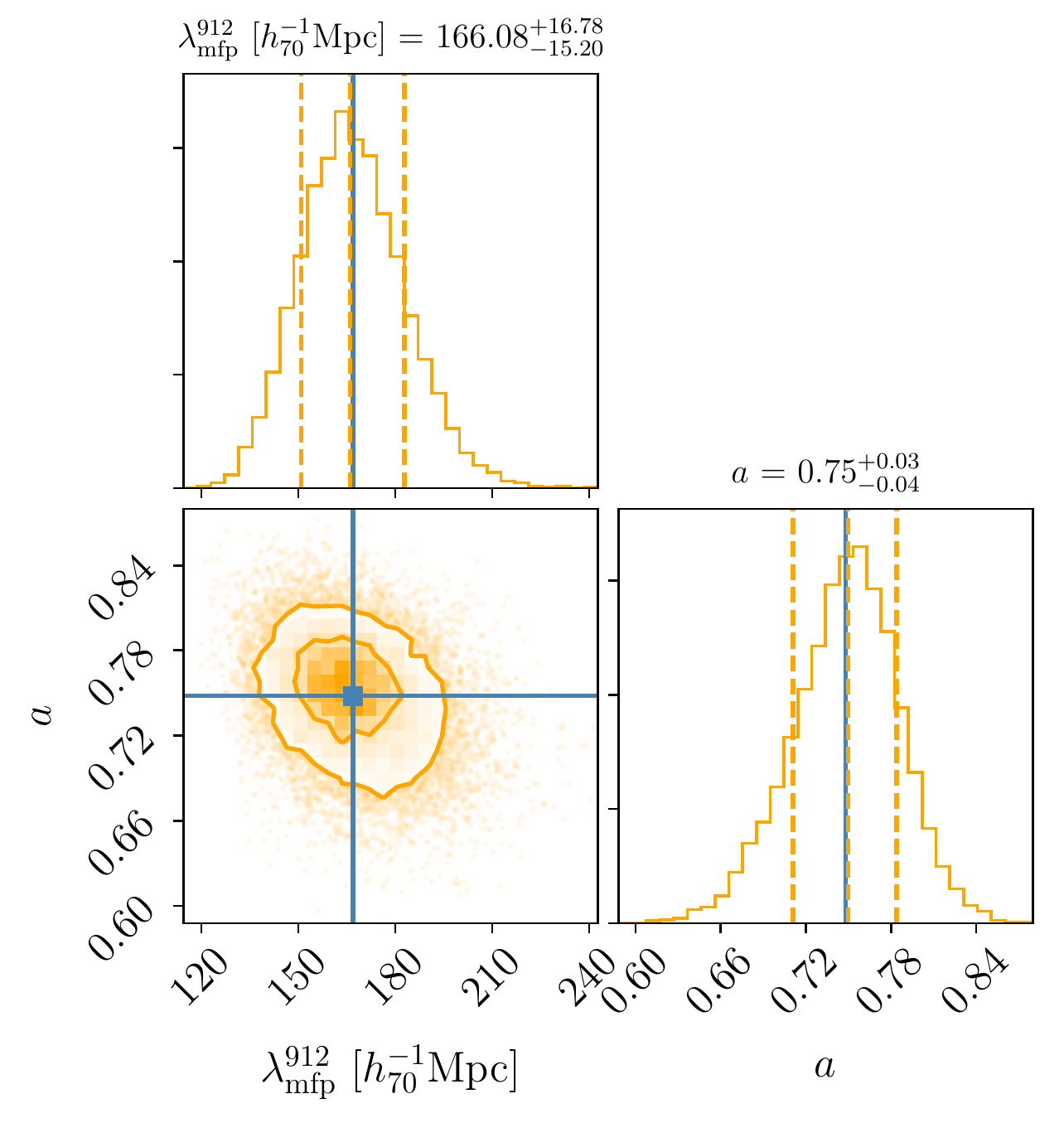}{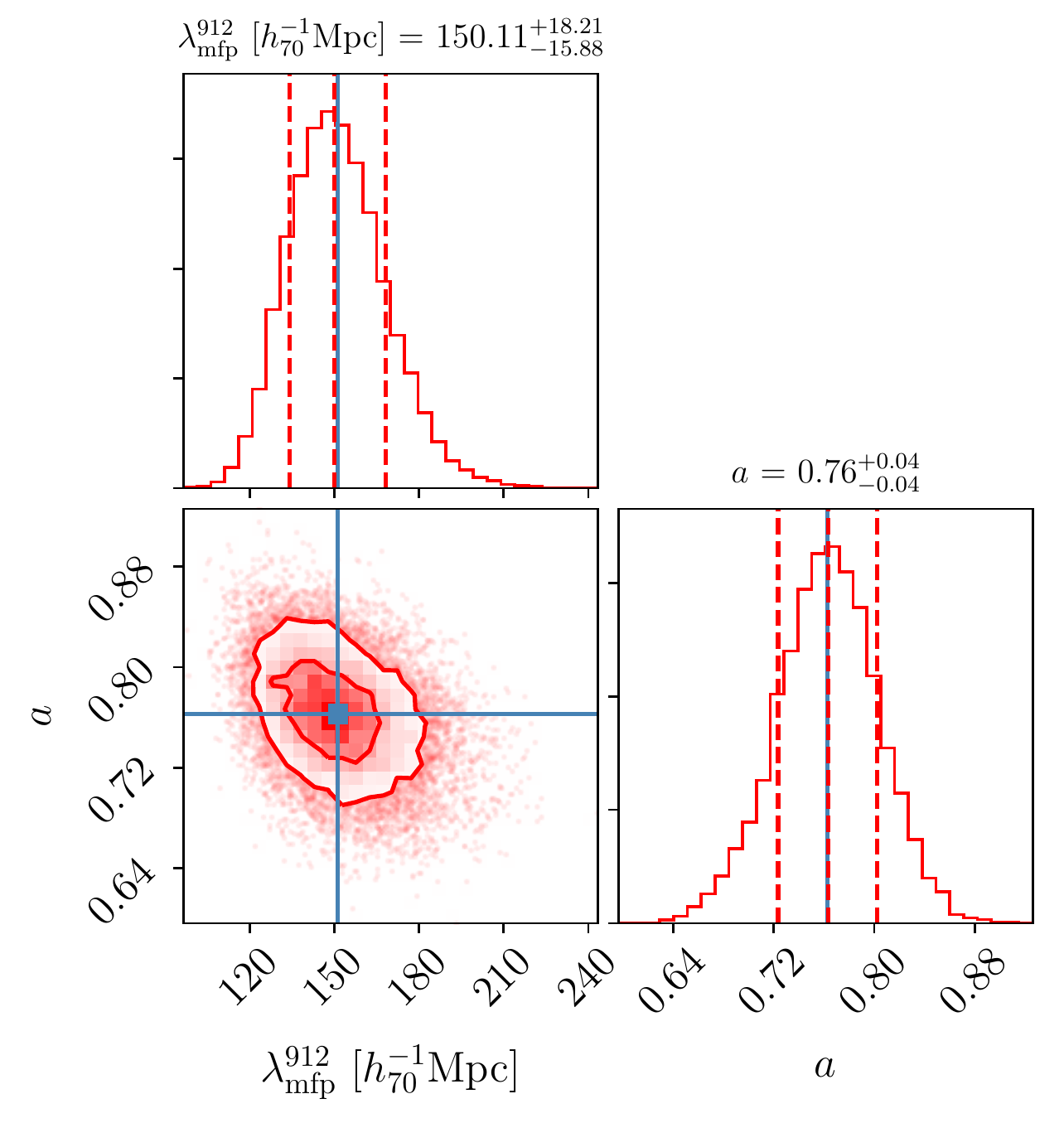}
 \caption{The 1 and 2$\sigma$ contours of the best $\lmfp$ and $a$ parameters estimated from 15,000 bootstraps where the analysis has been performed considering both the 700--911.76\AA\ (left panels) and 800-905\AA\ wavelength ranges (right panels). The median, 16$^{\rm th}$ and 84$^{\rm th}$ percentiles are shown on top of the histograms. The mean is shown as a solid line. {\it Top panels}: The assumed intrinsic quasar SED is the IGM corrected L15 spectral stack. {\it Bottom panels}: Results considering the tilted Stevans et al. (2014) COS quasar composite instead.}
 \label{model_lmfp_corner}
\end{figure*}
\begin{figure*}
 \plottwo{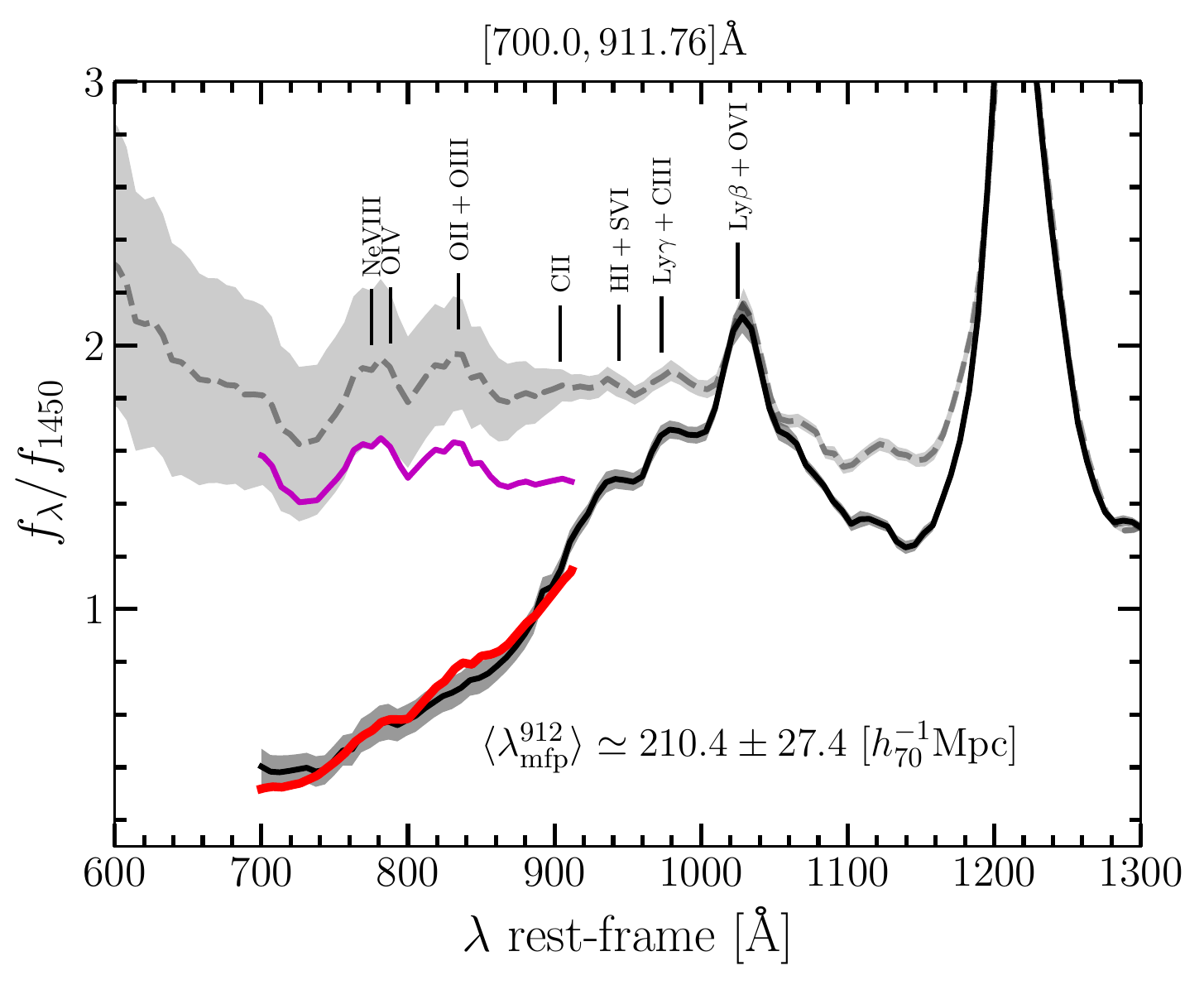}{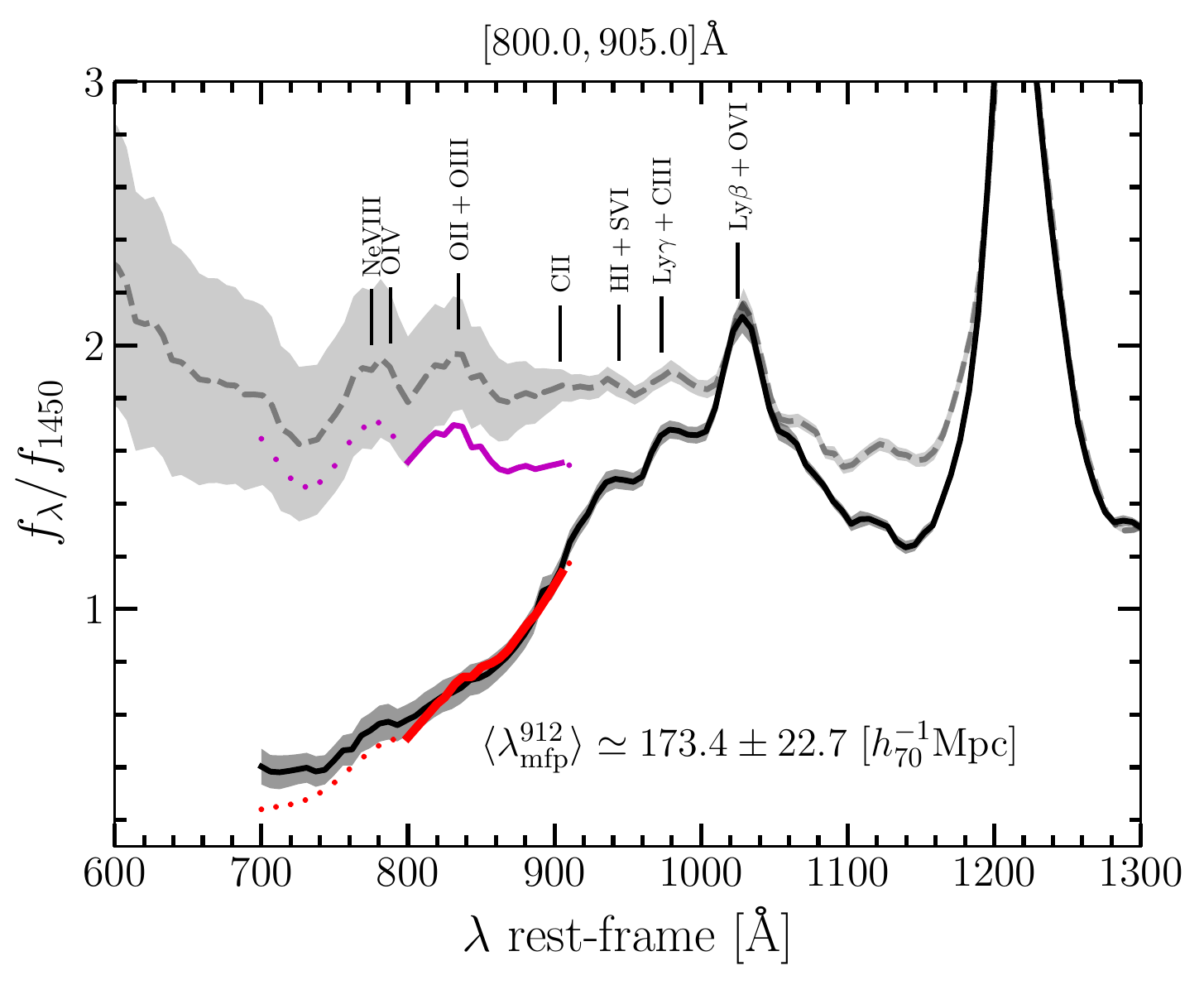}\\
 \plottwo{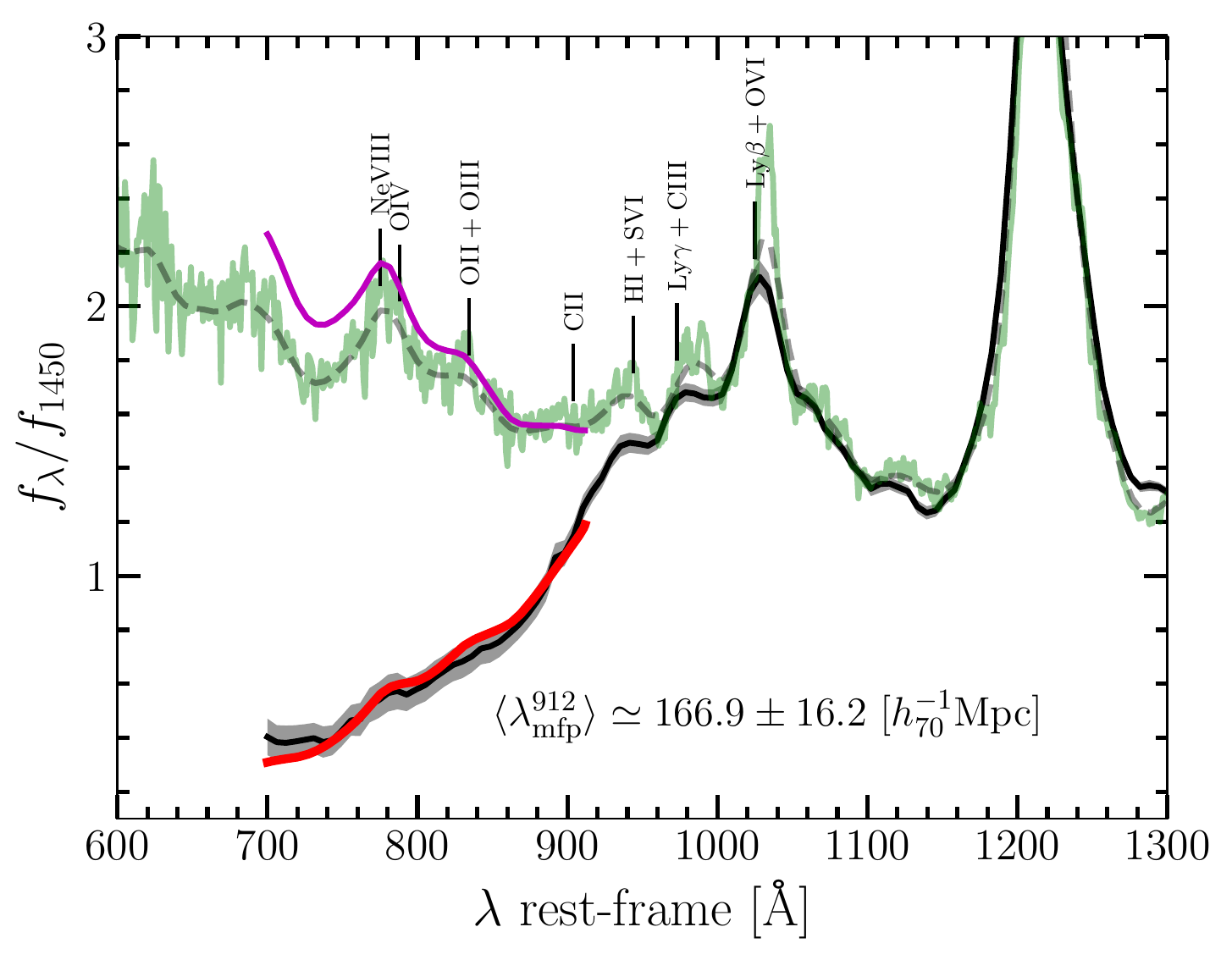}{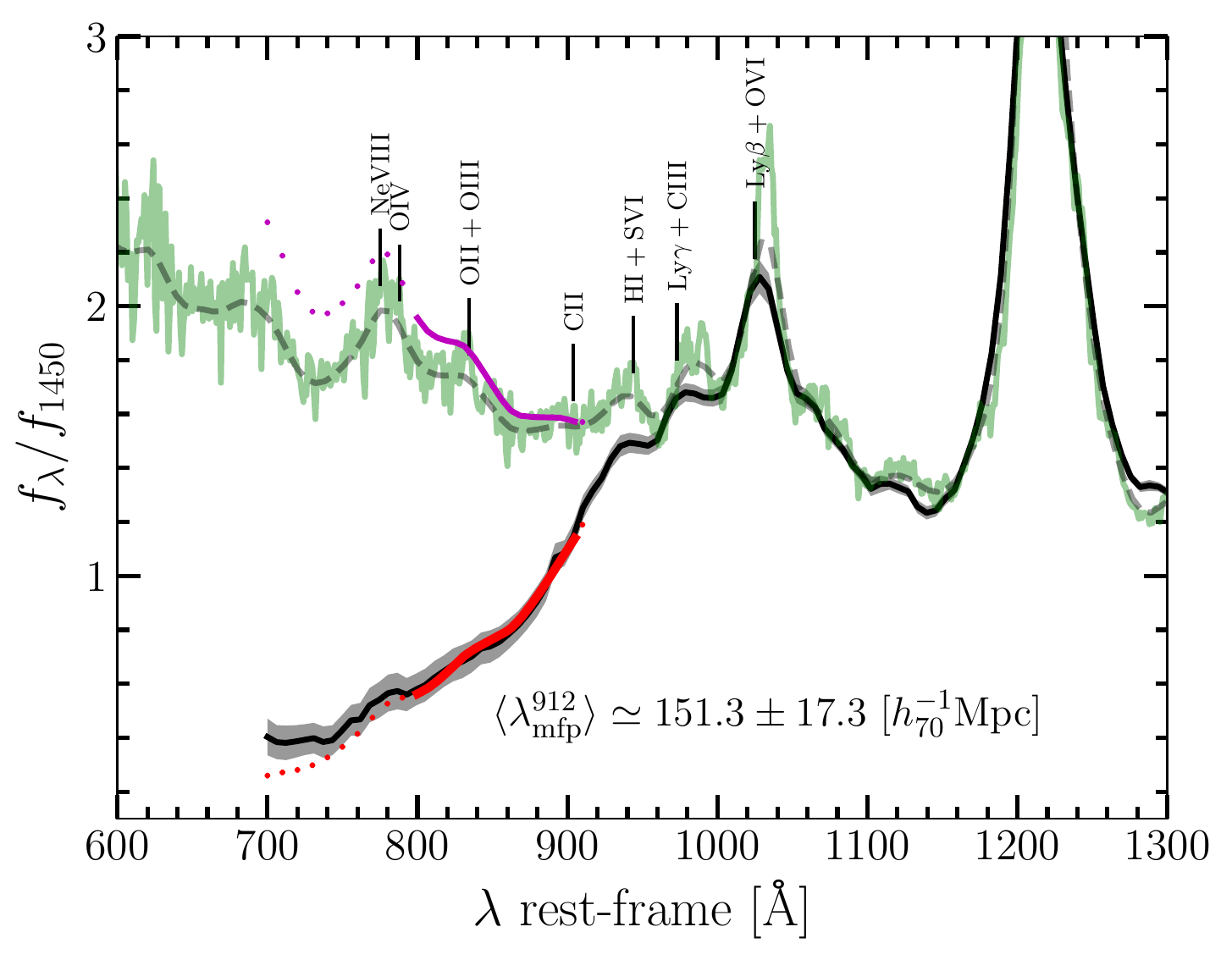}
 \caption{\rev{The left and right panels are obtained by fitting the observed quasar pair composite in the $[700,911.76]~$\AA\ and $[800,905]~$\AA, respectively (both ranges are shown on top of the figure).} {\it Top panels}: The black solid line is the observed WFC3 quasar pair composite, normalized to unity at $\lambda=1450$\AA. The dashed curve is the IGM corrected quasar stack by L15 assumed to be the underlying intrinsic quasar SED. The purple curve shows the best estimate of the intrinsic quasar continuum ($f_{\lambda,{\rm intr}}$), i.e., the scaled and tilted L15 spectrum as in equation~(\ref{fintr_sed_l15}), where the best-fit $a$ value is provided in Table~\ref{tab:lmfp}. The solid red curve is the complete model that includes both $\tauly$ and $\taull$. \rev{The dotted curve (only shown for completeness) represents the region of the spectra (purple: intrinsic assumed spectrum, red: model) that is not taken into account in our fitting procedure in the case of $[800,905]~$\AA}. {\it Bottom panels}: Same as above. The green curve represents the Stevans et al. (2014) quasar composite, where the dashed line is the COS stack rebinned to the dispersion solution of our WFC3 stacked spectrum and smoothed to the WFC3 spectral resolution.}
 \label{model_lmfp}
\end{figure*}
\subsection{Estimating the mean free path}
\label{Estimating the mean free path}
We apply the formalism described in \S\ref{Mean free path} to the observed WFC3 composite to obtain an estimate of $\lmfp$. We proceed by first building a set of stacked spectra with a standard bootstrap technique (allowing for repetition). For each quasar stack we then applied a maximum likelihood analysis in the wavelength interval 700--911.76~\AA\ where the free parameters are $a$ and $\kappa_{912}^{LL}(z_{\rm qso})$. The wavelength range for the model fit is chosen to be consistent with the one defined by O13 for comparison purposes, nonetheless we have also investigated how the slope changes if we consider a more conservative (narrower) wavelength range, 800--905~\AA. The higher wavelength is chosen to avoid the quasar proximity region at $>905$~\AA, while the lower bound is set to avoid the possible contribution of noisy data \citep{2013ApJ...775...78F,worseck14}.

\rev{Given the wide range of ionising spectral slopes published in the literature, we have also further examined the dependence of $\lmfp$ on the assumed intrinsic spectral shape}. As our quasar pair sample is at $z>2$, one possibility is to consider the composite SED published by Telfer et al. (2002). However, as already discussed by L15 and Scott et al. (2004), the IGM correction considered by Telfer et al. (2002) is basically negligible at $\lambda\leq1200$\AA, even if $z>2$ quasars are the main contributors at these wavelengths.
\rev{The more recent spectral composites published by Shull et al. (2012) and S14 are identical, with ionising spectral slopes (500--1000~\AA) of $\alpha_\nu = -1.41 \pm 0.21$ and $-1.41 \pm 0.15$, respectively.
These slope values are more precise than the ones we can compute from our WFC3 data, as they have been estimated by taking advantage of the higher spectral resolution of COS, which allows the authors to fit the local continua (correcting for identified LLSs and pLLSs), taking into account the contribution of quasar emission lines (see also \citealt{2017ApJ...849..106S}). The S14 spectral stack (covering the rest-frame range 475--1875~\AA) has been obtained from 159 AGNs at redshifts $0.001<z<1.476$ (with an average redshift of $\langle z\rangle=0.34$), and probes both lower redshifts and optical magnitudes (see Fig.~\ref{magi_z}) with respect to the objects analysed here. Nonetheless, being their composite at much higher resolution than our WFC3 one and directly corrected for both LLSs and pLLSs, which we do statistically, we decided to consider also the intrinsic quasar SED published by S14.
We scaled and tilted the COS composite considering their observed spectral slope} at $\lambda<912$\AA\ as
\begin{equation}
\label{fintr_sed_s14}
f_{\lambda,{\rm intr}} = a f_{\lambda,{\rm S14}} \left(\frac{\lambda}{1450\textrm{\AA}}\right)^{-0.6}.
\end{equation}
The COS stack is also rebinned to the dispersion solution of our stacked spectrum and smoothed to the WFC3 spectral resolution. The best estimate of the mean free path ($\lmfp\propto 1/\kappa_{912}^{LL}(z_{\rm qso})$) is derived by the mean (median) of 15,000 different realizations along with its uncertainties.

\begin{figure*}
 \plottwo{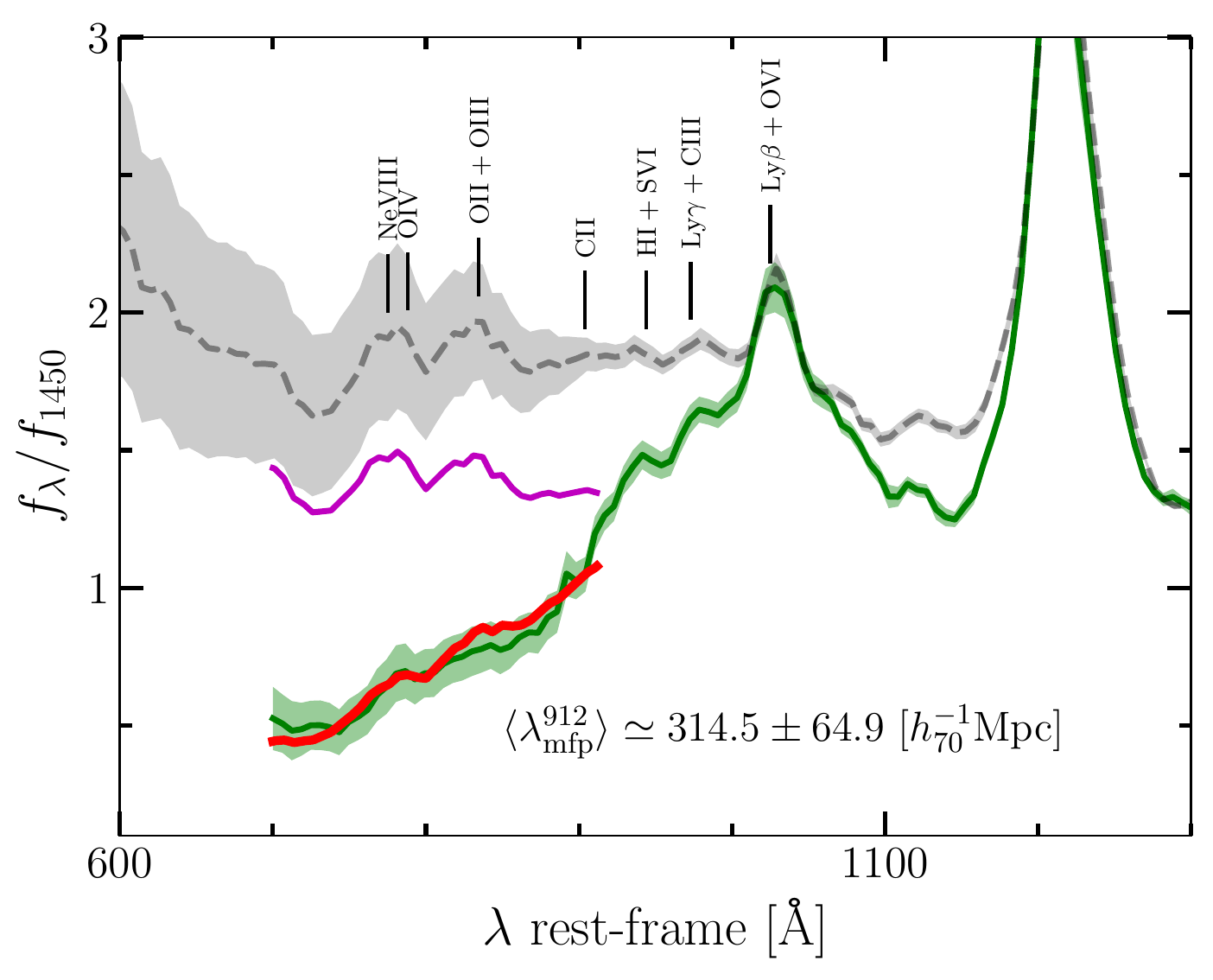}{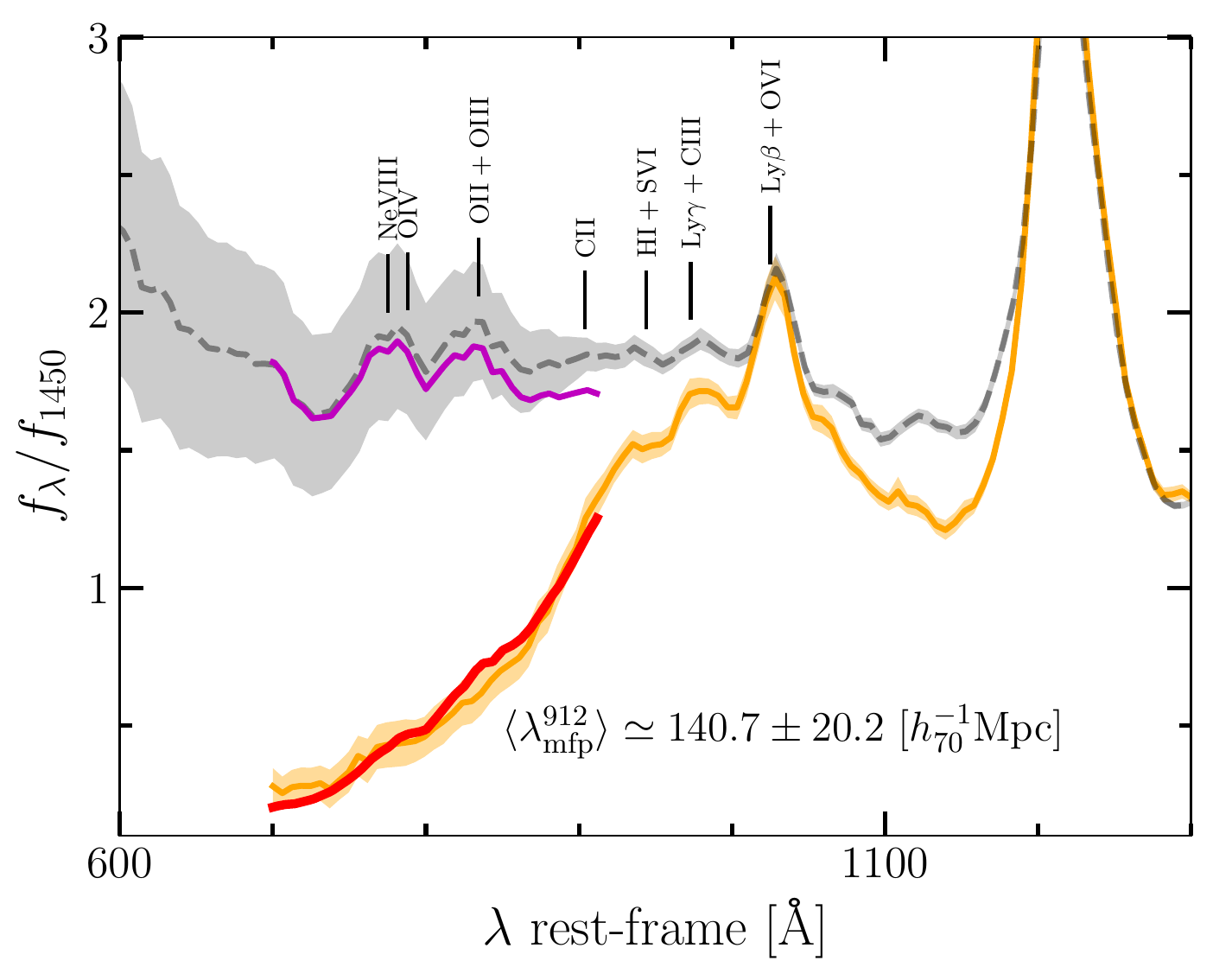}
 \caption{The green and orange solid lines are the observed WFC3 quasar pair composites for the low- ($\langle z\rangle=2.09$) and high-$z$ ($\langle z\rangle=2.44$) subsamples, respectively. Keys as in Fig.~\ref{model_lmfp}.}
 \label{model_lmfp_hl}
\end{figure*}

The normalised distribution of the best fit $\lmfp$ values for the 15,000 different realizations of our quasar pair sample computed by assuming the L15 and S14 stacks as the underlying continua and the 700--911.76~\AA\ and 800--905~\AA\ wavelength intervals are shown in Figure~\ref{hist_lmfp}. \rev{The $1$ and $2\sigma$ contours of the best $\lmfp$ and $a$ values resulting from our maximum likelihood analysis are presented in Figure~\ref{model_lmfp_corner}}. The $\lmfp$ distributions show a large dispersion, with the one obtained by assuming the S14 SED and the narrow wavelength interval being the one probing the smaller $\lmfp$ values. Nonetheless, they are all consistent within the $1.5\sigma$ level. A summary of the the best-fit $a$ and $\lmfp$ values is provided in Table~\ref{tab:lmfp}.

The best-fit models are shown in Figure~\ref{model_lmfp}. The left and right panels are obtained by fitting the observed quasar pair composite in the $[700,911.76]$\AA\ and $[800,905]$\AA, respectively, whilst the top and bottom panels present our findings by assuming the IGM corrected L15 and the S14 composites as the underlying intrinsic quasar SED. The purple curve shows the best estimate of the intrinsic quasar continuum ($f_{\lambda,{\rm intr}}$), i.e., the scaled and tilted L15 and S14 spectra as defined in equation~(\ref{fintr_sed_l15}). The solid red curve represents the complete model, which includes both $\tauly$ and $\taull$ in the rest-frame wavelength range considered. 

The quasar SED defined in equation~(\ref{fintr_sed_s14}) seems to be a better representation of the observed WFC3 stack, which is probably mainly due to the lower contribution of the \rev{\ion{O}{ii}+\ion{O}{iii}$\lambda834.5$\AA\ blend} (see Figure~\ref{obs_corr_comparison}).
Yet, in the 800--905\AA\ interval, the extrapolation of the model to bluer wavelengths clearly shows that it significantly under predicts the observed quasar flux. 
We thus consider the SED that includes the broader wavelength range as the most representative.

Overall, we find that the $\lmfp$ measurements are all consistent within the 1$\sigma$ level, with the ones obtained by making use of the tilted Stevans et al. composite \rev{predicting slightly steeper intrinsic quasar SEDs at $\lambda<912$\AA\ ($\alpha_\nu\simeq-1.41$). We caution that the $\lmfp$ values we found are sensitive to the adopted underlying continuum as well as the contribution of prominent emission lines. Given the results of our analysis, we cannot favour a scenario for quasar pairs having a different ionising continuum with respect to single quasars in a similar redshift range.} We thus argue that the most representative $\lmfp$ estimate ranges between 167--210 $\h$Mpc.

As a comparison, the value of the mean free path obtained by O13 for the WFC3 sample of single quasars is $\lmfp\simeq242\pm42~\h$Mpc (\citealt{2014MNRAS.438..476P}), which is in good agreement with our findings within the uncertanties. We stress here that the formalism considered by O13 for the $\lmfp$ measurement is rather different from ours (see their Section~5). Our adopted Lyman series opacity is a factor of $\sim12\%$ lower than O13 (see Fig.~\ref{taueff_ly}), and O13 considered the Telfer et al. (2002) as the intrisic quasar template spectrum. The O13's model has six free parameters: two for the quasar SED (i.e. the tilt and normalization), two for the Lyman series opacity ($\gamma_\tau$ and $\tauly(\lambda_{912})$), and two to model the Lyman limit opacity ($\gamma_\kappa$ and $\kappa^{\rm LL}_{912}(z_{\rm qso})$); whilst ours has only two (i.e. the slope of the intrinsic SED, $a$, and $\kappa^{\rm LL}_{912}(z_{\rm qso})$).

We have thus re-fitted the O13 WFC3 quasar sample using our formalism to establish possible systematics amongst different assumptions. By assuming the intrinsic L15 quasar SED with no tilt, as this stack was constructed with the same data, and fit the spectra in the same wavelength interval as the one adopted by O13 (i.e. 700--911.76~\AA), we find $\lmfp\simeq213.8\pm28.0~\h$Mpc and $a=0.95\pm0.04$. 

From this comparison we could conclude that quasar pairs and single quasars seem to trace similar IGM distributions. Nonetheless, our sample covers a broad redshift range (see Fig.~\ref{magi_z}), thus the $\tauly$ function employed (see \S\ref{Formalism}) may have a tendency to over(under) estimate the correction for quasars at low (high) redshifts with respect to the mean redshift of the sample (i.e. $z\simeq2.26$). We have thus performed the whole analysis by assuming the 700--911.76\AA\ wavelength range, the tilted L15 SED, and the $\tauly$ function relative for the low- and the high-$z$ samples at the average redshift of $z\simeq2.09$ and $z\simeq2.44$, respectively. The resulting mean $\lmfp$ values are $\lmfp\simeq 315\pm65~\h$Mpc and $\lmfp\simeq 140\pm20~\h$Mpc for the low and high-$z$ samples respectively, and the best fit models are presented in Figure~\ref{model_lmfp_hl}. 

The difference between the $\lmfp$ value of O13 and the one we measured for the high-$z$ quasar pair subsample is at the 2$\sigma$ level. Despite the large uncertainties, our analysis of the $\lmfp$ suggests a possible difference in the environment of pairs with respect to single quasars at similar redshifts, in line with the comparisons of the observed UV stacks we have discussed in previous sections.

\begin{figure}
\includegraphics[width=8.5cm,clip]{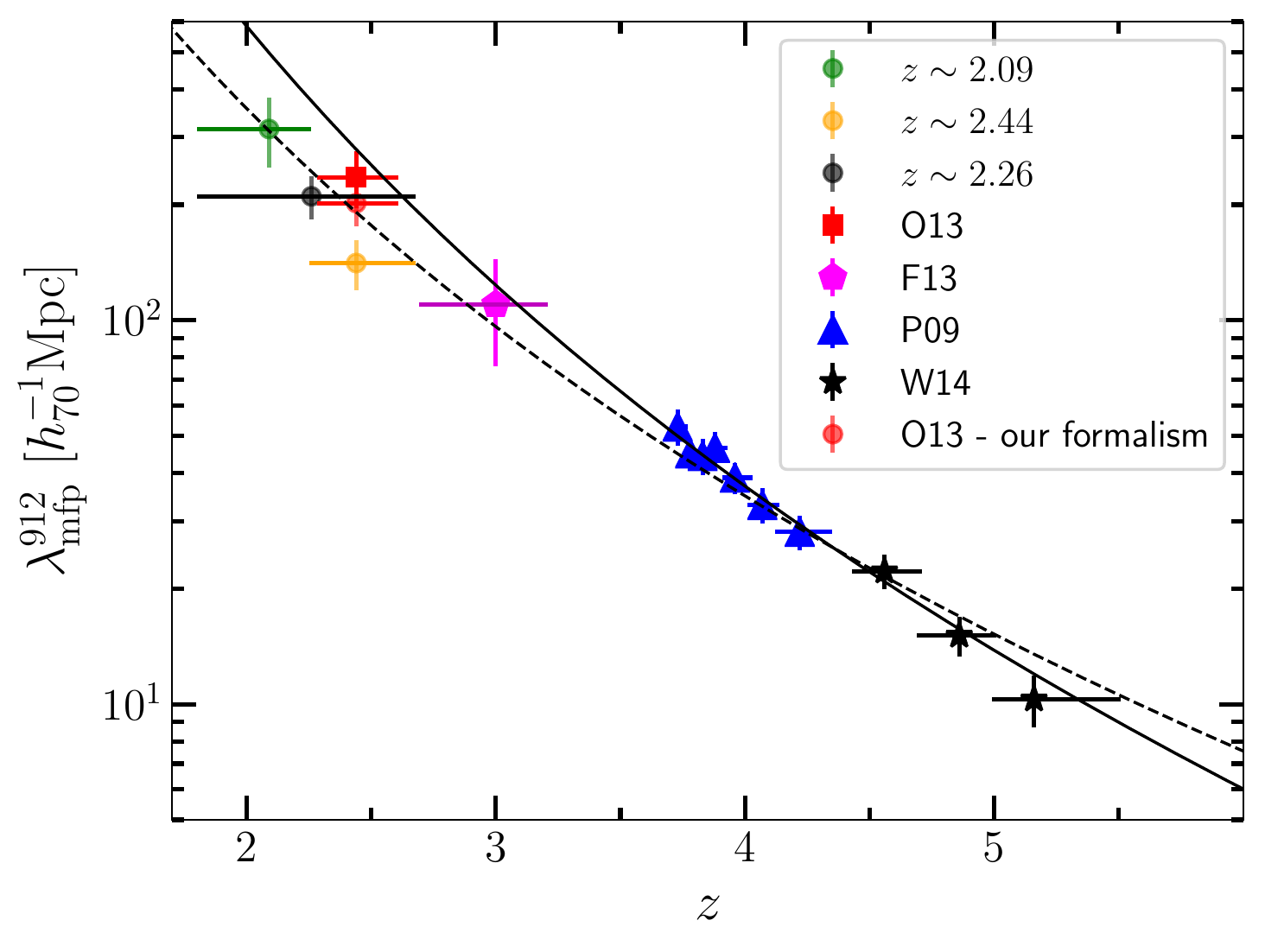}                                 
\caption{The mean free path to ionising photons in the IGM as a function of redshift. The data points show the direct measurements of $\lmfp$ through the quasar spectral stacking analysis by \citet[black stars]{worseck14}, \citet[blue triangles]{2009ApJ...705L.113P}, \citet[magenta pentagon]{2013ApJ...775...78F}, O13 (red square). 
The black filled circle represents the $\lmfp$ value estimated using the WFC3 quasar pair sample, while the red circle represents the best fit $\lmfp$ value obtained by applying our formalism to the O13 WFC3 sample. The green and orange circles represent the mean $\lmfp$ obtained by splitting the WFC3 sample into low and high redshift bins. We also overplotted the best-fit function published by \citet{worseck14} to model the observed decrease of $\lmfp$ with increasing redshift: $(1+z)^\eta$ with $\eta\simeq-5.4$. The dashed-line is the best-fit obtained by also including the WFC3 low- and high-$z$ $\lmfp$ values obtained with our analysis ($\eta\simeq-4.5\pm0.2$).}
 \label{lmfp_z}
\end{figure}

\subsection{The redshift evolution of the mean free path}
\label{The redshift evolution of the mean free path}
The evolution of the mean free path as a function of redshift provides insights on the cosmological distribution of gas around galaxies that dominates the hydrogen Lyman limit opacity. 
The most comprehensive collection of direct $\lmfp$ measurements is provided by \citet[W14 hereafter, see their Table~4; see also \citealt{2013ApJ...769..146R}]{worseck14}. They found that $\lmfp$ increases by an order of magnitude from $z=5$ to 2.5, where most of the measurements are at $z>3$ and only two $\lmfp$ values are currently being estimated at $z=2.0-2.5$.
Our survey adds two additional data points on the $\lmfp-z$ relation at $z<3$, and allows us to directly compare the distribution of \ion{H}{i} LL absorbers of quasars pairs with the ones of single quasars.

Figure~\ref{lmfp_z} shows the mean free path to ionising photons as a function of redshift with our additional measurements at $z=2.0-2.5$. 
For a comparison, we overplot the complete set of $\lmfp$ estimates and their uncertainties published by W14.
We overplot the best-fit function obtained by W14 to model the observed decrease of $\lmfp$ with increasing redshift: $A(1+z)^\eta$ with $\eta=-5.4\pm0.4$ and $A=37\pm2\h$Mpc. The red circle represents the best fit value of $\lmfp$ obtained by applying our formalism to the O13 WFC3 sample, i.e. $\lmfp\simeq213.8\pm28.0~\h$Mpc, to be compared to $235.8\pm40.3~\h$Mpc obtained by W14 using the same data. 

Overall, we find that our $\lmfp$ estimates are systematically lower with respect to the best fit relation by W14 and that the $\lmfp$ value of the high redshift quasar pair sample is statistically different at the 2$\sigma$ level with respect to the one for single quasar obtained by O13 when using the same fitting technique. By considering a two-parameter model in a similar fashion to the one adopted by W14, $\lmfp(z)=A\left[(1+z)/5\right]^\eta$, we find $\eta=-4.5\pm0.2$ and $A=35.0\pm1.2\h$Mpc (dashed line in Fig.~\ref{lmfp_z}), which is different at the 2$\sigma$ level with respect to the slope found by W14. 

Estimates of the $\lmfp$ are very sensitive to the spectral shape at $\lambda< 910\AA$. Given that our WFC3 spectral composite displays a somewhat steeper slope at $\lambda<912$\AA\ compared to the L15 one, our low $\lmfp$ values can be due to either a different evolution of the optically-thick gas on cosmological scales, i.e. changes in the gas accretion rate on to galactic haloes at $z<3$, or a different CGM/IGM distribution for pairs with respect to single quasars (i.e. a more dense or neutral environment for quasar pairs).
We regard the first scenario unlikely, as a change in the relative contribution of optically-thick absorbers to the mean free path would show as a break in the $\lmfp-z$ relation at $z<3$ \citep{2013ApJ...775...78F}. Yet, our data do not require a break in the power law presented in Figure~\ref{lmfp_z}.
To explore the latter scenario, in the next section we analyse the WFC3 spectra together with the high-resolution ones (when available) to study the incidence of optically-thick absorbers along the line of sight.

\begin{figure*}
\plotone{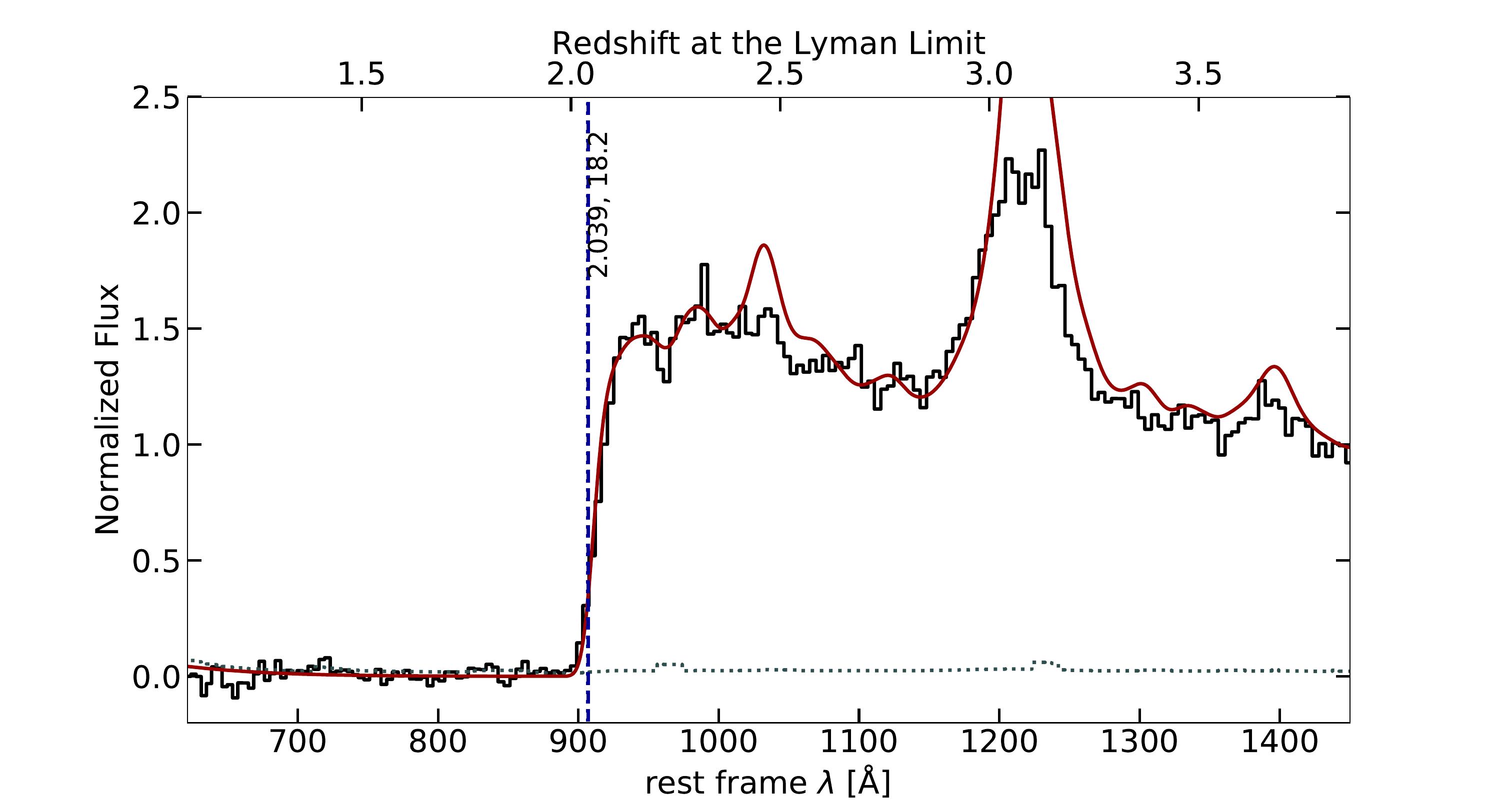}
\caption{WFC3 quasar spectrum (black histogram) and the adopted model (red solid line) to identify the Lyman limit absorbers along the line of sight \rev{for SDSSJ002424.21$-$012825.7}. The dashed line reports the redshift of the quasar at the Lyman limit, whilst the dotted lines mark the redshift and the $\log\nh$ for each LLS.\\(The complete set of figures (94 images) are available in the online journal.)}
 \label{lls}
\end{figure*}

\section{Fitting for LLS absorbers}
\label{Fitting for LLS absorbers}
To further investigate whether the quasar pairs in our survey tend to live in a denser and/or more neutral environment than single quasars, we fit for absorbers by modelling each quasar spectrum using a similar approach as the one adopted by O13. We summarise the main steps.
First, we need to estimate the level of continuum and the slope at $\lambda>1200$\AA. For comparison purposes, we considered the same template as the one employed by O13, i.e. a Telfer et al. (2002) quasar template spectrum, modulated by a scaled normalization and a tilt. The values of the normalization and the tilt have been determined utilising a custom GUI that allows one to visually compare the quasar template for each WFC3 spectrum\footnote{https://github.com/pyigm/pyigm}. We then add one or more systems to model any drop in the observed flux at $\lambda<912$ \AA\ yielding significant opacity. 
To robustly identify absorption systems using low resolution spectroscopy, spectral data with high S/N and quasars without complex ionising continua are usually preferred. As our dataset has relatively modest S/N, we also considered the additional SDSS data and the high resolution spectroscopy from our on-going follow-up campaign (at the time of writing, $\sim50\%$ of the sample has additional spectroscopy from ground-based facilities, see Section~\ref{Redshift estimates}) to identify and/or confirm strong ($\log \nh\geq17.2$ or $\taulls\geq1$) absorbers through the presence of saturated \ion{H}{I} absorption lines, and narrow absorption line doublets such as \ion{C}{iv}$\lambda\lambda$1548,1551. 
The analysis for each quasar spectrum was performed independently by E. L. and M. F., then the models were visually inspected and compared. 
This exercise is meant to provide an indicative (but quantitative) estimate of the incidence of absorbers in our sample that we can then compare with the results from single quasars using similar data. A more in depth analysis of the absorbers in our sample is not the purpose of this paper and will be presented in a forthcoming publication.

An example of this analysis is shown in Figure~\ref{lls}, where we present the final adopted model superimposed to the WFC3 spectrum (the redshift of the identified absorbers are also displayed). 
As already discussed by O13, $\chi^2-$minimization algorithms usually assign unrealistically small statistical errors ($<2\%$), we thus considered an uncertainty which is based upon the comparison between the values for the redshift of the absorbers, $\zabs$, and the column density obtained from the different authors. \rev{For systems with $\log\nh>17.2$, the uncertainty on the $\zabs$ is in the range $\sim0.02-0.05$ ($\simeq1,745-4,400$ km/s)}. 
Our survey is not complete for multiple absorbers with $\log \nh<17.2$ in the range 5,000-10,000km/s from the Lyman limit. Our main aim is however to identify strong LLSs, and only use weak absorbers ($\log\nh<17.2$) to model the continuum. 

When two absorbers are located within $|\delta v|<$10,000 km/s ($\delta z\simeq0.1$) and the redshifts are based on WFC3 spectra only ($\sim40\%$ of the quasars in our sample), we tend to consider such complexes as a single LLS with a summed optical depth, similar to what is done in the O13 analysis. This choice over-estimates the incidence of systems with $\taulls\simeq1$ by roughly 10\% (see discussion in O13 and \citealt{prochaska10}), and it affects both quasar samples.
Yet, whilst in our study we also based our identification on additional spectra with higher resolution data (e.g. ESI, Mage, BOSS, LIRIS) when available, the O13 analysis was based on WFC3 spectra only. Therefore, we may find more systems especially at $\log \nh\simeq17.2$ or lower with respect to O13. 
As this may introduce biases in our comparison, we believe that considering only those absorbers with $\log \nh\geq17.2$ at $|\delta v|<5,000$ km/s is a conservative choice that minimises possible systematics. 

\begin{figure*}
\plottwo{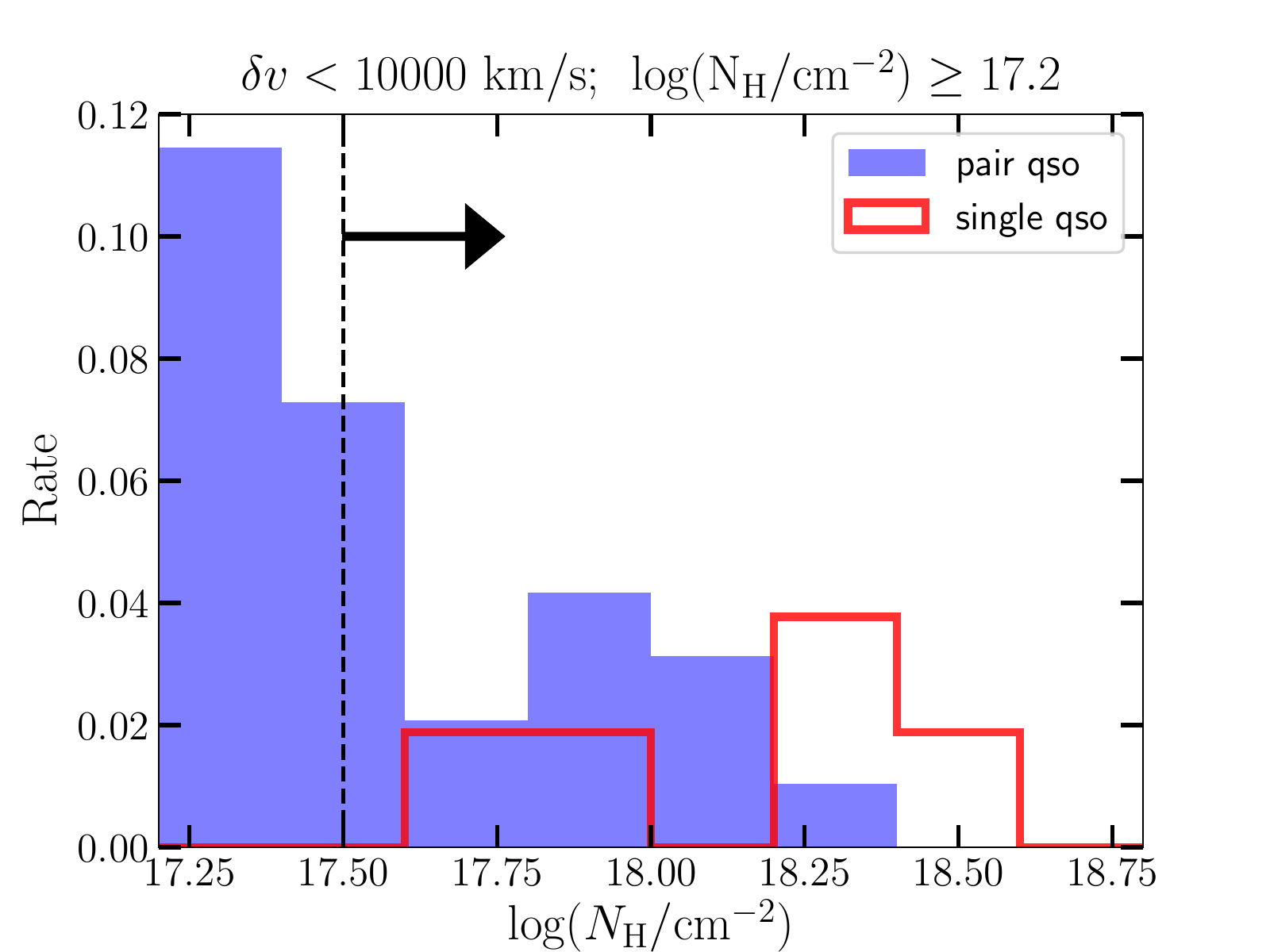}{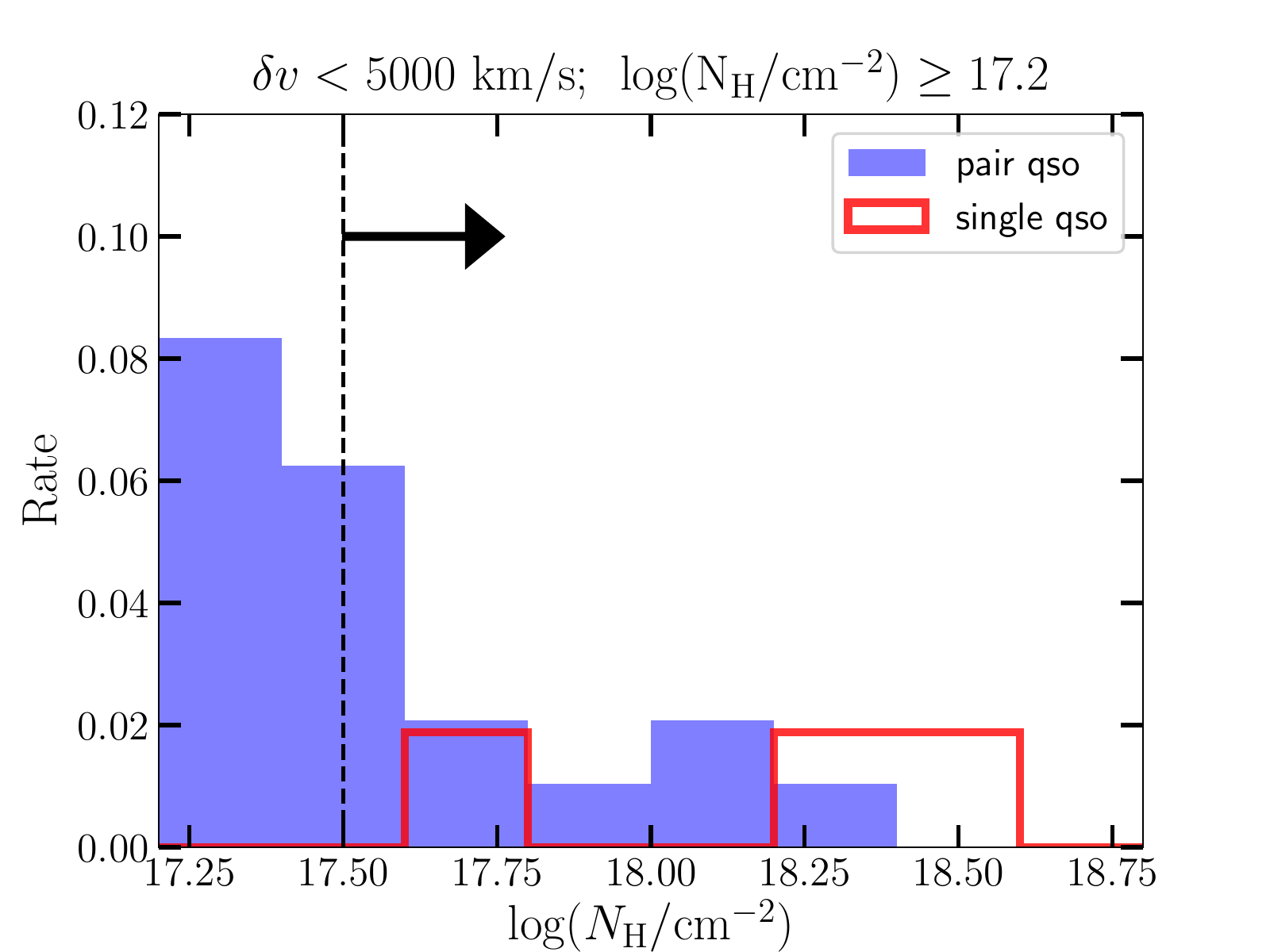}
\caption{Absorber column density distribution for our quasar pair sample (blue filled histogram) and the one by O13 of single quasars (red open histogram) for a velocity cut of $|\delta v|=$10,000 km/s and $|\delta v|=$5,000 km/s, for the left and right panel respectively. Each bin has been normalized to the total number of quasars in the two samples, i.e. $N_{\rm TOT}=$94 and 53 for our sample and the O13 one, respectively. All $\nh$ values higher than $10^{17.5}$ cm$^{-2}$ should be considered lower limits.}
 \label{nh}
\end{figure*}

During our search, we identified 28 (20) systems having $\log \nh\geq17.2$ with $|\delta v|=$10,000 km/s (5,000 km/s), 14 (9) of which with $\log \nh\geq17.5$ and $|\delta v|=$10,000 km/s (5,000 km/s).
Figure~\ref{nh} presents the $\nh$ distributions of the absorbers for $|\delta v|<$5,000 km/s and 10,000 km/s for our WFC3 quasar pair sample and for the O13 sample of single quasars, where they identified 109 absorbers overall (see their Table~2). The bins have been normalised to the total numbers of quasars in each sample (i.e. 53 sources for O13 and 94 in our analysis).
Six percent of the O13 total quasar sample have $\log \nh\geq17.2$ within $|\delta v|=$5,000 km/s (3 absorbers), which is roughly a factor of 3 lower compared to our 20\% (20 absorbers). If we assume an uncertainty on the identification of $\pm2$ in both samples (as we cannot resolve multiple strong absorbers within 10,000 km/s, see also Section~4.1 in O13), these two rates are different at the 5$\sigma$ level. 
This result further suggests that the quasar pair at 10--150 kpc (projected) separation observed in our survey tend to live in environments with denser/more neutral gas than single quasars, in line with the results of our mean free path analysis.

\section{Discussion and Conclusion}
\label{Discussion and Conclusion}
In this study we characterize the spectral shape and environmental properties of close ($R_\perp\leq 150$ kpc) quasar pairs at $z=2.0-2.5$, and compare the results to studies of single quasars at similar redshifts. Below we summarise our main findings.
\begin{enumerate}
\item \rev{Our quasar pair sample leads to an ionizing power-law index of $\alpha_{\rm ion}\simeq-2.5\pm0.8$ (see \S\ref{Spectral fit}), which, taken at face value, is softer (i.e. steeper) than the one obtained by L15 for single, much brighter, quasars ($\alpha_{\rm ion}=-1.7\pm0.6$), although both characterized by high uncertainties.} This result assumes that the IGM distribution for the pair quasar sample is the same as that of single quasars at a similar redshift range. As our WFC3 quasar pair sample spans a broad redshift range, the assumption of an average IGM absorption function at the mean redshift of the sample may over(under) estimate the correction for quasars at low (high) redshifts with respect to the mean redshift of the sample (i.e. $z\simeq2.26$). To check this point, we analysed the quasar pair sample in two redshifts bins.

\item When we create two composite samples binned by redshift, we find that the high redshift quasar pair composite ($\langle z\rangle\simeq2.44$) shows a factor of $\sim$2 lower fluxes at rest-wavelengths below $\sim$900\AA\ compared to the L15 one, which consider quasars at the same average redshift. On the other hand, the level of absorption of the pair composite in the low redshift interval ($\langle z\rangle\simeq2.09$) is overall similar to the L15 stack (see Figure~\ref{obsstacks_lowz_highz}). We thus conclude that our assumption of a similar IGM distribution for the two samples at similar average redshift is likely incorrect and uncover differences in the gas distribution between pairs and single quasars.

\item We find lower values of $\lmfp$ for pairs ($\lmfp\simeq140.7\pm20.2~\h$Mpc) compared to single quasars ($\lmfp\simeq213.8\pm28~\h$Mpc) at similar redshifts. Yet, uncertainties are large given current data (the difference is at the 2$\sigma$ level). As a result of our $\lmfp$ analysis and its evolution with time, we cannot rule out a difference in the relative contribution of absorbers (DLAs, LLSs) along the line of sight. 

\item When we search the spectra for strong absorbers, we find that 6\% of the O13 total quasar sample have absorbers with $\log \nh\geq17.2$ within $|\delta v|=$5,000 km/s (3 absorbers), which is roughly a factor of 3 lower compared to our 20\% (20 absorbers). These two rates are significantly different at the 5$\sigma$ level.

\end{enumerate}



\subsection{Implication for the quasar environment}
\label{Implication for the quasar environment}
Studies of quasar environments at various scales (from a few kpc to Mpc) and at different cosmic epochs are fundamental to understand the role of the large-scale environment in driving matter into the centre of galaxies, possibly triggering the quasar activity. One way to investigate quasars' environment is to measure the incidence of strong \ion{H}{i} absorbers in quasar sightlines, which trace the distribution of cool, dense gas that forms structures.

In the case of a single quasar, the expectation for the incidence of Lyman limit absorption systems per unit path length \rev{(i.e. the number density of LLS per unit redshift), $N(z)$, is $\simeq0.96^{+0.23}_{-0.29}$ at $z\simeq2.2$ \citep[see their Fig.~5 and Table~5]{2013ApJ...765..137O}. In other words, a quasar will hit on average roughly one absorber within a path $\Delta z=1$ along the line of sight (see also \citealt{2017ApJ...849..106S} for similar results at $z<0.5$)}. 
As absorbers along the line of sight are often found to be intervening, 
optically-thick absorbers tend to be located at relatively large distances from luminous quasars, and not in close proximity to the quasars themselves.
This effect can be attributed to the elevated radiation field near quasars. Indeed, \citet{2006ApJ...651...61H} argued that for a quasar at $z=2.5$ with an r-band magnitude of $r=19$, the continuum ionizing flux is 130 times higher than that of the extragalactic UV background at an angular separation of 60\arcsec, which corresponds to a proper distance of 340$h^{-1}$kpc (see also \citealt{2007ApJ...655..735H}). This enhanced ionization of clouds near the quasar gives rise to the {\it proximity effect} \citep{1988ApJ...327..570B}.

The effects of ionization, however, are expected to be non-isotropic due to quasar beaming. Indeed, \citet{2006ApJ...651...61H} presented a technique for studying the clustering of absorbers near luminous quasars
in the transverse direction, perpendicular to the direction in which the quasar is radiating. To this end, they made use of alignments of background quasar sight lines to search for optically-thick absorption in the vicinity of foreground quasars at $1.9 < z_{\rm fg} < 4.0$, finding a high-incidence of optically-thick gas in the circumgalactic medium of the quasar host, much in excess with observations along the line of sight.
This result provides significant evidence that these absorbers are, indeed, strongly clustered around quasars (see also \citealt{hennawi2006,2007ApJ...655..735H,hennawi2010}), but likely photoionized along the line of sight. 
By using a background quasar sightline to study the foreground quasar's environment in absorption (where the pairs are thus at different redshift), \citet{2013ApJ...776..136P} also found an excess of \ion{H}{i} \ion{Ly}{$\alpha$} absorption in the $30{\rm kpc}<R_\perp\leq1$Mpc environment transverse of 650 projected quasar pairs at $z\sim2$. In agreement with previous studies, their analysis is consistent with quasars being hosted by massive dark matter halos $M_{\rm halo}\simeq10^{12.5} M_\odot$ at $z\sim2.5$, where the transverse direction is much less likely to be illuminated by ionizing radiation than the line-of-sight (see also \citealt{2012MNRAS.424..933W}). 


The synergy between these two effects (quasar photoionization vs excess of \ion{H}{i} absorption surrounding the quasars) is complex and depends upon many factors, such as the mass of the host galaxy (i.e. the mass of the halo), and the luminosity and opening angle of the quasar (e.g. Faucher-Giguere et al. 2008). 
The case of closely separated quasar pairs ($R_\perp\leq 150$ kpc) at similar redshifts ($\Delta z \leq 10$Mpc, Figure~\ref{deltavrp}) is thus of particular interest as, differently from projected pairs, these systems are excellent tracers of the environment of quasars that share the same large-scale structure, or in some cases even the same dark matter halos.

Our WFC3 pair sample shows an enhancement of LLSs with $\log \nh\simeq17.2$ at $|\delta v|<$5,000 km/s compared to single quasars at the 5$\sigma$ level along the line of sight.
This higher incidence of optically-thick gas along the pairs' sightlines explains the \rev{softer shape} of the composite quasar pair stack when compared to the one of single quasars at similar redshift (see left panel of Figure~\ref{obs_corr_comparison} and Figure~\ref{obsstacks_lowz_highz}). Amongst the 20 strong absorbers we found, six of them lie in correlated pairs (all in the low redshift bin), while the rest is equally shared between the low and high redshift bin (see Figure~\ref{deltavrp}). 
Moreover, we find that the location of these 20 absorbers is equally shared between foreground and background quasars, with 9 absorbers identified in background quasars and 11 absorbers in foreground sources. 
We still retrieve an equal fraction of strong absorbers in the foreground and background quasar if we consider absorbers at $|\delta v|<$3,000 km/s (14 absorbers).
There is no obvious trend with magnitudes, as these 20 absorbers are located in quasars having average $r^\ast$ magnitudes typical of the overall sample ($r^\ast\simeq20$).

Despite the small statistics, this implies that both foreground and background quasars are embedded in the same, and equally dense and (partially) neutral, environment at $<15$Mpc (Figure~\ref{deltavrp}).
This is different from what seen in single quasars, and also from the results of the projected pairs, where the background sightlines 
show a clear excess of absorbers at the redshift of the foreground 
quasars. Given that the mean luminosity is about a factor of 3 fainter than the typical quasars considered by \citet{2006ApJ...651...61H}, we argue that the higher fraction of optically-thick absorbers in quasar pairs is not primarily driven by a lower radiation field.
Instead, based on this analysis, we argue that the gas absorbing Lyman limit photons in our WFC3 sample of closely-projected quasar pairs is likely to arise mostly within structures located in denser regions within the CGM or IGM where both quasars reside.


\acknowledgments
We are grateful to the referee, Prof. Michael Shull, for his thorough reading and for useful comments and suggestions which have significantly improved the clarity of the paper.
E.L. thanks G. Calderone for having modified the code QSFit to handle low-resolution WFC3 spectra and for feedback and suggestions on the fitting procedure.
E.L. is supported by a European Union COFUND/Durham Junior Research Fellowship (under EU grant agreement no. 609412). M.F. acknowledges support by the Science and Technology Facilities Council [grant number ST/P000541/1]. This project has received funding from the European Research Council (ERC) under the European Union's Horizon 2020 research and innovation programme (grant agreement No 757535).
Support for HST Program GO-14127 was provided by NASA through grants from the Space Telescope Science Institute, which is operated by the Association of Universities for Research in Astronomy, Inc., under NASA contract NAS526555. 
M.R. acknowledges support by HST-GO-14127.011 and a NASA Keck PI Data Award, administered by the NASA Exoplanet Science Institute. Some of the data presented herein were obtained at the W. M. Keck Observatory from telescope time allocated to the National Aeronautics and Space Administration through the agency's scientific partnership with the California Institute of Technology and the University of California. The Observatory was made possible by the generous financial support of the W. M. Keck Foundation. The authors wish to recognize and acknowledge the very significant cultural role and reverence that the summit of Maunakea has always had within the indigenous Hawaiian community.  We are most fortunate to have the opportunity to conduct observations from this mountain.

%

\vspace{5mm}
\facilities{HST (WFC3);~ Keck:II (ESI)}


\software{astropy \citep{2013A&A...558A..33A}, corner \citep{corner}, matplotlib \citep{2007CSE.....9...90H}, emcee \citep{2013PASP..125..306F}, QSFit \citep{2017MNRAS.472.4051C}.}

\bibliographystyle{aasjournal}
\bibliography{bibl.bib} 

\begin{thebibliography}{}
\expandafter\ifx\csname natexlab\endcsname\relax\def\natexlab#1{#1}\fi
\providecommand{\url}[1]{\href{#1}{#1}}

\bibitem[{{Alexander} \& {Hickox}(2012)}]{2012NewAR..56...93A}
{Alexander}, D.~M., \& {Hickox}, R.~C. 2012, \nar, 56, 93

\bibitem[{{Anderson} \& {Bedin}(2010)}]{2010PASP..122.1035A}
{Anderson}, J., \& {Bedin}, L.~R. 2010, \pasp, 122, 1035

\bibitem[{{Anguita} {et~al.}(2008){Anguita}, {Faure}, {Yonehara}, {Wambsganss},
  {Kneib}, {Covone}, \& {Alloin}}]{2008A&A...481..615A}
{Anguita}, T., {Faure}, C., {Yonehara}, A., {et~al.} 2008, \aap, 481, 615

\bibitem[{{Astropy Collaboration} {et~al.}(2013){Astropy Collaboration},
  {Robitaille}, {Tollerud}, {Greenfield}, {Droettboom}, {Bray}, {Aldcroft},
  {Davis}, {Ginsburg}, {Price-Whelan}, {Kerzendorf}, {Conley}, {Crighton},
  {Barbary}, {Muna}, {Ferguson}, {Grollier}, {Parikh}, {Nair}, {Unther},
  {Deil}, {Woillez}, {Conseil}, {Kramer}, {Turner}, {Singer}, {Fox}, {Weaver},
  {Zabalza}, {Edwards}, {Azalee Bostroem}, {Burke}, {Casey}, {Crawford},
  {Dencheva}, {Ely}, {Jenness}, {Labrie}, {Lim}, {Pierfederici}, {Pontzen},
  {Ptak}, {Refsdal}, {Servillat}, \& {Streicher}}]{2013A&A...558A..33A}
{Astropy Collaboration}, {Robitaille}, T.~P., {Tollerud}, E.~J., {et~al.} 2013,
  \aap, 558, A33

\bibitem[{{Bajtlik} {et~al.}(1988){Bajtlik}, {Duncan}, \&
  {Ostriker}}]{1988ApJ...327..570B}
{Bajtlik}, S., {Duncan}, R.~C., \& {Ostriker}, J.~P. 1988, \apj, 327, 570

\bibitem[{{Baldwin}(1977)}]{baldwin1977}
{Baldwin}, J.~A. 1977, \apj, 214, 679

\bibitem[{{Bershady} {et~al.}(1999){Bershady}, {Charlton}, \&
  {Geoffroy}}]{1999ApJ...518..103B}
{Bershady}, M.~A., {Charlton}, J.~C., \& {Geoffroy}, J.~M. 1999, \apj, 518, 103

\bibitem[{{Brotherton} {et~al.}(2001){Brotherton}, {Tran}, {Becker}, {Gregg},
  {Laurent-Muehleisen}, \& {White}}]{2001ApJ...546..775B}
{Brotherton}, M.~S., {Tran}, H.~D., {Becker}, R.~H., {et~al.} 2001, \apj, 546,
  775

\bibitem[{{Calderone} {et~al.}(2017){Calderone}, {Nicastro}, {Ghisellini},
  {Dotti}, {Sbarrato}, {Shankar}, \& {Colpi}}]{2017MNRAS.472.4051C}
{Calderone}, G., {Nicastro}, L., {Ghisellini}, G., {et~al.} 2017, \mnras, 472,
  4051

\bibitem[{{Cisternas} {et~al.}(2011)}]{2011ApJ...726...57C}
{Cisternas}, M., {et~al.} 2011, \apj, 726, 57

\bibitem[{{Claeskens} {et~al.}(1996){Claeskens}, {Surdej}, \&
  {Remy}}]{1996A&A...305L...9C}
{Claeskens}, J.-F., {Surdej}, J., \& {Remy}, M. 1996, \aap, 305, L9

\bibitem[{{Compostella} {et~al.}(2013){Compostella}, {Cantalupo}, \&
  {Porciani}}]{compostella13}
{Compostella}, M., {Cantalupo}, S., \& {Porciani}, C. 2013, \mnras, 435, 3169

\bibitem[{{Corbin}(1990)}]{1990ApJ...357..346C}
{Corbin}, M.~R. 1990, \apj, 357, 346

\bibitem[{{Corbin}(2000)}]{2000ApJ...536L..73C}
---. 2000, \apjl, 536, L73

\bibitem[{{Czerny} \& {Elvis}(1987)}]{1987ApJ...321..305C}
{Czerny}, B., \& {Elvis}, M. 1987, \apj, 321, 305

\bibitem[{{Danforth} {et~al.}(2016){Danforth}, {Keeney}, {Tilton}, {Shull},
  {Stocke}, {Stevans}, {Pieri}, {Savage}, {France}, {Syphers}, {Smith},
  {Green}, {Froning}, {Penton}, \& {Osterman}}]{2016ApJ...817..111D}
{Danforth}, C.~W., {Keeney}, B.~A., {Tilton}, E.~M., {et~al.} 2016, \apj, 817,
  111

\bibitem[{{Deane} {et~al.}(2014){Deane}, {Paragi}, {Jarvis}, {Coriat},
  {Bernardi}, {Fender}, {Frey}, {Heywood}, {Kl{\"o}ckner}, {Grainge}, \&
  {Rumsey}}]{2014Natur.511...57D}
{Deane}, R.~P., {Paragi}, Z., {Jarvis}, M.~J., {et~al.} 2014, \nat, 511, 57

\bibitem[{{Di Matteo} {et~al.}(2005){Di Matteo}, {Springel}, \&
  {Hernquist}}]{2005Natur.433..604D}
{Di Matteo}, T., {Springel}, V., \& {Hernquist}, L. 2005, \nat, 433, 604

\bibitem[{{Dietrich} {et~al.}(2002){Dietrich}, {Hamann}, {Shields},
  {Constantin}, {Vestergaard}, {Chaffee}, {Foltz}, \&
  {Junkkarinen}}]{2002ApJ...581..912D}
{Dietrich}, M., {Hamann}, F., {Shields}, J.~C., {et~al.} 2002, \apj, 581, 912

\bibitem[{{Djorgovski} {et~al.}(2007){Djorgovski}, {Courbin}, {Meylan},
  {Sluse}, {Thompson}, {Mahabal}, \& {Glikman}}]{2007ApJ...662L...1D}
{Djorgovski}, S.~G., {Courbin}, F., {Meylan}, G., {et~al.} 2007, \apjl, 662, L1

\bibitem[{{Eftekharzadeh} {et~al.}(2017){Eftekharzadeh}, {Myers}, {Hennawi},
  {Djorgovski}, {Richards}, {Mahabal}, \& {Graham}}]{2017MNRAS.468...77E}
{Eftekharzadeh}, S., {Myers}, A.~D., {Hennawi}, J.~F., {et~al.} 2017, \mnras,
  468, 77

\bibitem[{{Espey} {et~al.}(1989){Espey}, {Carswell}, {Bailey}, {Smith}, \&
  {Ward}}]{1989ApJ...342..666E}
{Espey}, B.~R., {Carswell}, R.~F., {Bailey}, J.~A., {Smith}, M.~G., \& {Ward},
  M.~J. 1989, \apj, 342, 666

\bibitem[{{Fardal} {et~al.}(1998){Fardal}, {Giroux}, \&
  {Shull}}]{1998AJ....115.2206F}
{Fardal}, M.~A., {Giroux}, M.~L., \& {Shull}, J.~M. 1998, \aj, 115, 2206

\bibitem[{{Farina} {et~al.}(2013){Farina}, {Montuori}, {Decarli}, \&
  {Fumagalli}}]{2013MNRAS.431.1019F}
{Farina}, E.~P., {Montuori}, C., {Decarli}, R., \& {Fumagalli}, M. 2013,
  \mnras, 431, 1019

\bibitem[{{Faucher-Gigu{\`e}re} {et~al.}(2008){Faucher-Gigu{\`e}re}, {Lidz},
  {Hernquist}, \& {Zaldarriaga}}]{faucher08}
{Faucher-Gigu{\`e}re}, C.-A., {Lidz}, A., {Hernquist}, L., \& {Zaldarriaga}, M.
  2008, \apj, 688, 85

\bibitem[{{Faucher-Gigu{\`e}re} {et~al.}(2009){Faucher-Gigu{\`e}re}, {Lidz},
  {Zaldarriaga}, \& {Hernquist}}]{faucher09}
{Faucher-Gigu{\`e}re}, C.-A., {Lidz}, A., {Zaldarriaga}, M., \& {Hernquist}, L.
  2009, \apj, 703, 1416

\bibitem[{{Findlay} {et~al.}(2018){Findlay}, {Prochaska}, {Hennawi},
  {Fumagalli}, {Myers}, {Bartle}, {Chehade}, {DiPompeo}, {Shanks}, {Lau}, \&
  {Rubin}}]{2018arXiv180408624F}
{Findlay}, J.~R., {Prochaska}, J.~X., {Hennawi}, J.~F., {et~al.} 2018, ArXiv
  e-prints, arXiv:1804.08624

\bibitem[{{Fitzpatrick}(1999)}]{1999PASP..111...63F}
{Fitzpatrick}, E.~L. 1999, \pasp, 111, 63

\bibitem[{{Fontanot} {et~al.}(2014){Fontanot}, {Cristiani}, {Pfrommer},
  {Cupani}, \& {Vanzella}}]{fontanot14}
{Fontanot}, F., {Cristiani}, S., {Pfrommer}, C., {Cupani}, G., \& {Vanzella},
  E. 2014, \mnras, 438, 2097

\bibitem[{{Fontanot} {et~al.}(2012){Fontanot}, {Cristiani}, \&
  {Vanzella}}]{fontanot12}
{Fontanot}, F., {Cristiani}, S., \& {Vanzella}, E. 2012, \mnras, 425, 1413

\bibitem[{{Foreman} {et~al.}(2009){Foreman}, {Volonteri}, \&
  {Dotti}}]{2009ApJ...693.1554F}
{Foreman}, G., {Volonteri}, M., \& {Dotti}, M. 2009, \apj, 693, 1554

\bibitem[{{Foreman-Mackey}(2016)}]{corner}
{Foreman-Mackey}, D. 2016, The Journal of Open Source Software, 1, 24

\bibitem[{{Foreman-Mackey} {et~al.}(2013){Foreman-Mackey}, {Hogg}, {Lang}, \&
  {Goodman}}]{2013PASP..125..306F}
{Foreman-Mackey}, D., {Hogg}, D.~W., {Lang}, D., \& {Goodman}, J. 2013, \pasp,
  125, 306

\bibitem[{{Francis} {et~al.}(1991){Francis}, {Hewett}, {Foltz}, {Chaffee},
  {Weymann}, \& {Morris}}]{1991ApJ...373..465F}
{Francis}, P.~J., {Hewett}, P.~C., {Foltz}, C.~B., {et~al.} 1991, \apj, 373,
  465

\bibitem[{{Fumagalli} {et~al.}(2014){Fumagalli}, {Hennawi}, {Prochaska},
  {Kasen}, {Dekel}, {Ceverino}, \& {Primack}}]{2014ApJ...780...74F}
{Fumagalli}, M., {Hennawi}, J.~F., {Prochaska}, J.~X., {et~al.} 2014, \apj,
  780, 74

\bibitem[{{Fumagalli} {et~al.}(2013){Fumagalli}, {O'Meara}, {Prochaska}, \&
  {Worseck}}]{2013ApJ...775...78F}
{Fumagalli}, M., {O'Meara}, J.~M., {Prochaska}, J.~X., \& {Worseck}, G. 2013,
  \apj, 775, 78

\bibitem[{{Furlanetto}(2009)}]{furlanetto09}
{Furlanetto}, S.~R. 2009, \apj, 703, 702

\bibitem[{{Gaskell}(1982)}]{1982ApJ...263...79G}
{Gaskell}, C.~M. 1982, \apj, 263, 79

\bibitem[{{Green}(1996)}]{1996ApJ...467...61G}
{Green}, P.~J. 1996, \apj, 467, 61

\bibitem[{{Haardt} \& {Madau}(1996)}]{haardt96}
{Haardt}, F., \& {Madau}, P. 1996, \apj, 461, 20

\bibitem[{{Haardt} \& {Madau}(2012)}]{haardt12}
---. 2012, \apj, 746, 125

\bibitem[{{Hennawi} \& {Prochaska}(2007)}]{2007ApJ...655..735H}
{Hennawi}, J.~F., \& {Prochaska}, J.~X. 2007, \apj, 655, 735

\bibitem[{{Hennawi} {et~al.}(2015){Hennawi}, {Prochaska}, {Cantalupo}, \&
  {Arrigoni-Battaia}}]{2015Sci...348..779H}
{Hennawi}, J.~F., {Prochaska}, J.~X., {Cantalupo}, S., \& {Arrigoni-Battaia},
  F. 2015, Science, 348, 779

\bibitem[{{Hennawi} {et~al.}(2006{\natexlab{a}}){Hennawi}, {Strauss}, {Oguri},
  {Inada}, {Richards}, {Pindor}, {Schneider}, {Becker}, {Gregg}, {Hall},
  {Johnston}, {Fan}, {Burles}, {Schlegel}, {Gunn}, {Lupton}, {Bahcall},
  {Brunner}, \& {Brinkmann}}]{hennawi2006}
{Hennawi}, J.~F., {Strauss}, M.~A., {Oguri}, M., {et~al.} 2006{\natexlab{a}},
  \aj, 131, 1

\bibitem[{{Hennawi} {et~al.}(2006{\natexlab{b}}){Hennawi}, {Prochaska},
  {Burles}, {Strauss}, {Richards}, {Schlegel}, {Fan}, {Schneider}, {Zakamska},
  {Oguri}, {Gunn}, {Lupton}, \& {Brinkmann}}]{2006ApJ...651...61H}
{Hennawi}, J.~F., {Prochaska}, J.~X., {Burles}, S., {et~al.}
  2006{\natexlab{b}}, \apj, 651, 61

\bibitem[{{Hennawi} {et~al.}(2010){Hennawi}, {Myers}, {Shen}, {Strauss},
  {Djorgovski}, {Fan}, {Glikman}, {Mahabal}, {Martin}, {Richards}, {Schneider},
  \& {Shankar}}]{hennawi2010}
{Hennawi}, J.~F., {Myers}, A.~D., {Shen}, Y., {et~al.} 2010, \apj, 719, 1672

\bibitem[{{Hopkins} {et~al.}(2005){Hopkins}, {Hernquist}, {Cox}, {Di Matteo},
  {Martini}, {Robertson}, \& {Springel}}]{2005ApJ...630..705H}
{Hopkins}, P.~F., {Hernquist}, L., {Cox}, T.~J., {et~al.} 2005, \apj, 630, 705

\bibitem[{{Hopkins} {et~al.}(2006){Hopkins}, {Hernquist}, {Cox}, {Di Matteo},
  {Robertson}, \& {Springel}}]{2006ApJS..163....1H}
---. 2006, \apjs, 163, 1

\bibitem[{{Hopkins} {et~al.}(2008){Hopkins}, {Hernquist}, {Cox}, \&
  {Kere\v{s}}}]{2008ApJS..175..356H}
{Hopkins}, P.~F., {Hernquist}, L., {Cox}, T.~J., \& {Kere\v{s}}, D. 2008,
  \apjs, 175, 356

\bibitem[{{Hunter}(2007)}]{2007CSE.....9...90H}
{Hunter}, J.~D. 2007, Computing in Science and Engineering, 9, 90

\bibitem[{{Inoue} \& {Iwata}(2008)}]{2008MNRAS.387.1681I}
{Inoue}, A.~K., \& {Iwata}, I. 2008, \mnras, 387, 1681

\bibitem[{{Inoue} {et~al.}(2014){Inoue}, {Shimizu}, {Iwata}, \&
  {Tanaka}}]{2014MNRAS.442.1805I}
{Inoue}, A.~K., {Shimizu}, I., {Iwata}, I., \& {Tanaka}, M. 2014, \mnras, 442,
  1805

\bibitem[{{Jiang} {et~al.}(2008){Jiang}, {Fan}, {Annis}, {Becker}, {White},
  {Chiu}, {Lin}, {Lupton}, {Richards}, {Strauss}, {Jester}, \&
  {Schneider}}]{jiang08}
{Jiang}, L., {Fan}, X., {Annis}, J., {et~al.} 2008, \aj, 135, 1057

\bibitem[{{Laor} {et~al.}(1997){Laor}, {Fiore}, {Elvis}, {Wilkes}, \&
  {McDowell}}]{1997ApJ...477...93L}
{Laor}, A., {Fiore}, F., {Elvis}, M., {Wilkes}, B.~J., \& {McDowell}, J.~C.
  1997, \apj, 477, 93

\bibitem[{{Liu} {et~al.}(2011){Liu}, {Shen}, \&
  {Strauss}}]{2011ApJ...736L...7L}
{Liu}, X., {Shen}, Y., \& {Strauss}, M.~A. 2011, \apjl, 736, L7

\bibitem[{{Lusso} {et~al.}(2015){Lusso}, {Worseck}, {Hennawi}, {Prochaska},
  {Vignali}, {Stern}, \& {O'Meara}}]{2015MNRAS.449.4204L}
{Lusso}, E., {Worseck}, G., {Hennawi}, J.~F., {et~al.} 2015, \mnras, 449, 4204

\bibitem[{{MacLeod} {et~al.}(2012){MacLeod}, {Ivezi{\'c}}, {Sesar}, {de Vries},
  {Kochanek}, {Kelly}, {Becker}, {Lupton}, {Hall}, {Richards}, {Anderson}, \&
  {Schneider}}]{2012ApJ...753..106M}
{MacLeod}, C.~L., {Ivezi{\'c}}, {\v Z}., {Sesar}, B., {et~al.} 2012, \apj, 753,
  106

\bibitem[{{Madau}(1995)}]{1995ApJ...441...18M}
{Madau}, P. 1995, \apj, 441, 18

\bibitem[{{Madau} \& {Haardt}(2015)}]{2015ApJ...813L...8M}
{Madau}, P., \& {Haardt}, F. 2015, \apjl, 813, L8

\bibitem[{{Madau} {et~al.}(1999){Madau}, {Haardt}, \& {Rees}}]{madau99}
{Madau}, P., {Haardt}, F., \& {Rees}, M.~J. 1999, \apj, 514, 648

\bibitem[{{Malkan} \& {Sargent}(1982)}]{1982ApJ...254...22M}
{Malkan}, M.~A., \& {Sargent}, W.~L.~W. 1982, \apj, 254, 22

\bibitem[{{Massey} {et~al.}(2010){Massey}, {Stoughton}, {Leauthaud}, {Rhodes},
  {Koekemoer}, {Ellis}, \& {Shaghoulian}}]{2010MNRAS.401..371M}
{Massey}, R., {Stoughton}, C., {Leauthaud}, A., {et~al.} 2010, \mnras, 401, 371

\bibitem[{{McQuinn} {et~al.}(2009){McQuinn}, {Lidz}, {Zaldarriaga},
  {Hernquist}, {Hopkins}, {Dutta}, \& {Faucher-Gigu{\`e}re}}]{mcquinn09}
{McQuinn}, M., {Lidz}, A., {Zaldarriaga}, M., {et~al.} 2009, \apj, 694, 842

\bibitem[{{Meiksin}(2005)}]{meiksin05}
{Meiksin}, A. 2005, \mnras, 356, 596

\bibitem[{{Meiksin}(2006)}]{2006MNRAS.365..807M}
---. 2006, \mnras, 365, 807

\bibitem[{{Meiksin} \& {White}(2003)}]{meiksin03}
{Meiksin}, A., \& {White}, M. 2003, \mnras, 342, 1205

\bibitem[{{Miralda-Escud{\'e}} {et~al.}(2000){Miralda-Escud{\'e}}, {Haehnelt},
  \& {Rees}}]{miralda00}
{Miralda-Escud{\'e}}, J., {Haehnelt}, M., \& {Rees}, M.~J. 2000, \apj, 530, 1

\bibitem[{{Moller} \& {Jakobsen}(1990)}]{1990A&A...228..299M}
{Moller}, P., \& {Jakobsen}, P. 1990, \aap, 228, 299

\bibitem[{{Myers} {et~al.}(2008){Myers}, {Richards}, {Brunner}, {Schneider},
  {Strand}, {Hall}, {Blomquist}, \& {York}}]{2008ApJ...678..635M}
{Myers}, A.~D., {Richards}, G.~T., {Brunner}, R.~J., {et~al.} 2008, \apj, 678,
  635

\bibitem[{{Netzer} {et~al.}(1992){Netzer}, {Laor}, \&
  {Gondhalekar}}]{1992MNRAS.254...15N}
{Netzer}, H., {Laor}, A., \& {Gondhalekar}, P.~M. 1992, \mnras, 254, 15

\bibitem[{{O'Meara} {et~al.}(2011){O'Meara}, {Prochaska}, {Chen}, \&
  {Madau}}]{2011ApJS..195...16O}
{O'Meara}, J.~M., {Prochaska}, J.~X., {Chen}, H.-W., \& {Madau}, P. 2011,
  \apjs, 195, 16

\bibitem[{{O'Meara} {et~al.}(2013){O'Meara}, {Prochaska}, {Worseck}, {Chen}, \&
  {Madau}}]{2013ApJ...765..137O}
{O'Meara}, J.~M., {Prochaska}, J.~X., {Worseck}, G., {Chen}, H.-W., \& {Madau},
  P. 2013, \apj, 765, 137

\bibitem[{{Prochaska} {et~al.}(2014){Prochaska}, {Madau}, {O'Meara}, \&
  {Fumagalli}}]{2014MNRAS.438..476P}
{Prochaska}, J.~X., {Madau}, P., {O'Meara}, J.~M., \& {Fumagalli}, M. 2014,
  \mnras, 438, 476

\bibitem[{{Prochaska} {et~al.}(2010){Prochaska}, {O'Meara}, \&
  {Worseck}}]{prochaska10}
{Prochaska}, J.~X., {O'Meara}, J.~M., \& {Worseck}, G. 2010, \apj, 718, 392

\bibitem[{{Prochaska} {et~al.}(2009){Prochaska}, {Worseck}, \&
  {O'Meara}}]{2009ApJ...705L.113P}
{Prochaska}, J.~X., {Worseck}, G., \& {O'Meara}, J.~M. 2009, \apjl, 705, L113

\bibitem[{{Prochaska} {et~al.}(2013){Prochaska}, {Hennawi}, {Lee}, {Cantalupo},
  {Bovy}, {Djorgovski}, {Ellison}, {Lau}, {Martin}, {Myers}, {Rubin}, \&
  {Simcoe}}]{2013ApJ...776..136P}
{Prochaska}, J.~X., {Hennawi}, J.~F., {Lee}, K.-G., {et~al.} 2013, \apj, 776,
  136

\bibitem[{{Rafelski} {et~al.}(2015){Rafelski}, {Teplitz}, {Gardner}, {Coe},
  {Bond}, {Koekemoer}, {Grogin}, {Kurczynski}, {McGrath}, {Bourque}, {Atek},
  {Brown}, {Colbert}, {Codoreanu}, {Ferguson}, {Finkelstein}, {Gawiser},
  {Giavalisco}, {Gronwall}, {Hanish}, {Lee}, {Mehta}, {de Mello},
  {Ravindranath}, {Ryan}, {Scarlata}, {Siana}, {Soto}, \&
  {Voyer}}]{2015AJ....150...31R}
{Rafelski}, M., {Teplitz}, H.~I., {Gardner}, J.~P., {et~al.} 2015, \aj, 150, 31

\bibitem[{{Rauch} {et~al.}(1997){Rauch}, {Miralda-Escud{\'e}}, {Sargent},
  {Barlow}, {Weinberg}, {Hernquist}, {Katz}, {Cen}, \&
  {Ostriker}}]{1997ApJ...489....7R}
{Rauch}, M., {Miralda-Escud{\'e}}, J., {Sargent}, W.~L.~W., {et~al.} 1997,
  \apj, 489, 7

\bibitem[{{Ribaudo} {et~al.}(2011){Ribaudo}, {Lehner}, \&
  {Howk}}]{2011ApJ...736...42R}
{Ribaudo}, J., {Lehner}, N., \& {Howk}, J.~C. 2011, \apj, 736, 42

\bibitem[{{Richards} {et~al.}(2002){Richards}, {Vanden Berk}, {Reichard},
  {Hall}, {Schneider}, {SubbaRao}, {Thakar}, \& {York}}]{2002AJ....124....1R}
{Richards}, G.~T., {Vanden Berk}, D.~E., {Reichard}, T.~A., {et~al.} 2002, \aj,
  124, 1

\bibitem[{{Richards} {et~al.}(2006)}]{2006AJ....131.2766R}
{Richards}, G.~T., {et~al.} 2006, \aj, 131, 2766

\bibitem[{{Rudie} {et~al.}(2013){Rudie}, {Steidel}, {Shapley}, \&
  {Pettini}}]{2013ApJ...769..146R}
{Rudie}, G.~C., {Steidel}, C.~C., {Shapley}, A.~E., \& {Pettini}, M. 2013,
  \apj, 769, 146

\bibitem[{{Sandrinelli} {et~al.}(2014){Sandrinelli}, {Falomo}, {Treves},
  {Farina}, \& {Uslenghi}}]{2014MNRAS.444.1835S}
{Sandrinelli}, A., {Falomo}, R., {Treves}, A., {Farina}, E.~P., \& {Uslenghi},
  M. 2014, \mnras, 444, 1835

\bibitem[{{Sandrinelli} {et~al.}(2018){Sandrinelli}, {Falomo}, {Treves},
  {Scarpa}, \& {Uslenghi}}]{2018MNRAS.474.4925S}
{Sandrinelli}, A., {Falomo}, R., {Treves}, A., {Scarpa}, R., \& {Uslenghi}, M.
  2018, \mnras, 474, 4925

\bibitem[{{Satyapal} {et~al.}(2014){Satyapal}, {Ellison}, {McAlpine}, {Hickox},
  {Patton}, \& {Mendel}}]{2014MNRAS.441.1297S}
{Satyapal}, S., {Ellison}, S.~L., {McAlpine}, W., {et~al.} 2014, \mnras, 441,
  1297

\bibitem[{{Satyapal} {et~al.}(2017){Satyapal}, {Secrest}, {Ricci}, {Ellison},
  {Rothberg}, {Blecha}, {Constantin}, {Gliozzi}, {McNulty}, \&
  {Ferguson}}]{2017ApJ...848..126S}
{Satyapal}, S., {Secrest}, N.~J., {Ricci}, C., {et~al.} 2017, \apj, 848, 126

\bibitem[{{Schlafly} \& {Finkbeiner}(2011)}]{2011ApJ...737..103S}
{Schlafly}, E.~F., \& {Finkbeiner}, D.~P. 2011, \apj, 737, 103

\bibitem[{{Scott} {et~al.}(2004){Scott}, {Kriss}, {Brotherton}, {Green},
  {Hutchings}, {Shull}, \& {Zheng}}]{2004ApJ...615..135S}
{Scott}, J.~E., {Kriss}, G.~A., {Brotherton}, M., {et~al.} 2004, \apj, 615, 135

\bibitem[{{Shankar} \& {Mathur}(2007)}]{shankar07}
{Shankar}, F., \& {Mathur}, S. 2007, \apj, 660, 1051

\bibitem[{{Shen} {et~al.}(2011){Shen}, {Richards}, {Strauss}, {Hall},
  {Schneider}, {Snedden}, {Bizyaev}, {Brewington}, {Malanushenko},
  {Malanushenko}, {Oravetz}, {Pan}, \& {Simmons}}]{2011ApJS..194...45S}
{Shen}, Y., {Richards}, G.~T., {Strauss}, M.~A., {et~al.} 2011, \apjs, 194, 45

\bibitem[{{Shull} {et~al.}(2017){Shull}, {Danforth}, {Tilton}, {Moloney}, \&
  {Stevans}}]{2017ApJ...849..106S}
{Shull}, J.~M., {Danforth}, C.~W., {Tilton}, E.~M., {Moloney}, J., \&
  {Stevans}, M.~L. 2017, \apj, 849, 106

\bibitem[{{Shull} {et~al.}(1999){Shull}, {Roberts}, {Giroux}, {Penton}, \&
  {Fardal}}]{1999AJ....118.1450S}
{Shull}, J.~M., {Roberts}, D., {Giroux}, M.~L., {Penton}, S.~V., \& {Fardal},
  M.~A. 1999, \aj, 118, 1450

\bibitem[{{Shull} {et~al.}(2012){Shull}, {Stevans}, \&
  {Danforth}}]{2012ApJ...752..162S}
{Shull}, J.~M., {Stevans}, M., \& {Danforth}, C.~W. 2012, \apj, 752, 162

\bibitem[{{Springel} {et~al.}(2005){Springel}, {Di Matteo}, \&
  {Hernquist}}]{2005MNRAS.361..776S}
{Springel}, V., {Di Matteo}, T., \& {Hernquist}, L. 2005, \mnras, 361, 776

\bibitem[{{Stevans} {et~al.}(2014){Stevans}, {Shull}, {Danforth}, \&
  {Tilton}}]{2014ApJ...794...75S}
{Stevans}, M.~L., {Shull}, J.~M., {Danforth}, C.~W., \& {Tilton}, E.~M. 2014,
  \apj, 794, 75

\bibitem[{{Surdej} {et~al.}(1997){Surdej}, {Claeskens}, {Remy}, {Refsdal},
  {Pirenne}, {Prieto}, \& {Vanderriest}}]{1997A&A...327L...1S}
{Surdej}, J., {Claeskens}, J.-F., {Remy}, M., {et~al.} 1997, \aap, 327, L1

\bibitem[{{Telfer} {et~al.}(2002){Telfer}, {Zheng}, {Kriss}, \&
  {Davidsen}}]{2002ApJ...565..773T}
{Telfer}, R.~C., {Zheng}, W., {Kriss}, G.~A., \& {Davidsen}, A.~F. 2002, \apj,
  565, 773

\bibitem[{{Tilton} {et~al.}(2016){Tilton}, {Stevans}, {Shull}, \&
  {Danforth}}]{2016ApJ...817...56T}
{Tilton}, E.~M., {Stevans}, M.~L., {Shull}, J.~M., \& {Danforth}, C.~W. 2016,
  \apj, 817, 56

\bibitem[{{Vanden Berk} {et~al.}(2001)}]{vandenberk2001}
{Vanden Berk}, D.~E., {et~al.} 2001, \aj, 122, 549

\bibitem[{{Vasei} {et~al.}(2016){Vasei}, {Siana}, {Shapley}, {Quider}, {Alavi},
  {Rafelski}, {Steidel}, {Pettini}, \& {Lewis}}]{2016ApJ...831...38V}
{Vasei}, K., {Siana}, B., {Shapley}, A.~E., {et~al.} 2016, \apj, 831, 38

\bibitem[{{V{\'e}ron-Cetty} {et~al.}(2004){V{\'e}ron-Cetty}, {Joly}, \&
  {V{\'e}ron}}]{2004A&A...417..515V}
{V{\'e}ron-Cetty}, M.-P., {Joly}, M., \& {V{\'e}ron}, P. 2004, \aap, 417, 515

\bibitem[{{Vestergaard} \& {Wilkes}(2001)}]{2001ApJS..134....1V}
{Vestergaard}, M., \& {Wilkes}, B.~J. 2001, \apjs, 134, 1

\bibitem[{{Weston} {et~al.}(2017){Weston}, {McIntosh}, {Brodwin}, {Mann},
  {Cooper}, {McConnell}, \& {Nielsen}}]{2017MNRAS.464.3882W}
{Weston}, M.~E., {McIntosh}, D.~H., {Brodwin}, M., {et~al.} 2017, \mnras, 464,
  3882

\bibitem[{{White} {et~al.}(2012){White}, {Myers}, {Ross}, {Schlegel},
  {Hennawi}, {Shen}, {McGreer}, {Strauss}, {Bolton}, {Bovy}, {Fan},
  {Miralda-Escude}, {Palanque-Delabrouille}, {Paris}, {Petitjean}, {Schneider},
  {Viel}, {Weinberg}, {Yeche}, {Zehavi}, {Pan}, {Snedden}, {Bizyaev},
  {Brewington}, {Brinkmann}, {Malanushenko}, {Malanushenko}, {Oravetz},
  {Simmons}, {Sheldon}, \& {Weaver}}]{2012MNRAS.424..933W}
{White}, M., {Myers}, A.~D., {Ross}, N.~P., {et~al.} 2012, \mnras, 424, 933

\bibitem[{{Willott} {et~al.}(2010){Willott}, {Delorme}, {Reyl{\'e}}, {Albert},
  {Bergeron}, {Crampton}, {Delfosse}, {Forveille}, {Hutchings}, {McLure},
  {Omont}, \& {Schade}}]{willott10}
{Willott}, C.~J., {Delorme}, P., {Reyl{\'e}}, C., {et~al.} 2010, \aj, 139, 906

\bibitem[{{Worseck} \& {Prochaska}(2011)}]{2011ApJ...728...23W}
{Worseck}, G., \& {Prochaska}, J.~X. 2011, \apj, 728, 23

\bibitem[{{Worseck} {et~al.}(2014){Worseck}, {Prochaska}, {O'Meara}, {Becker},
  {Ellison}, {Lopez}, {Meiksin}, {M{\'e}nard}, {Murphy}, \&
  {Fumagalli}}]{worseck14}
{Worseck}, G., {Prochaska}, J.~X., {O'Meara}, J.~M., {et~al.} 2014, \mnras,
  445, 1745

\bibitem[{{Zheng} {et~al.}(1997){Zheng}, {Kriss}, {Telfer}, {Grimes}, \&
  {Davidsen}}]{zheng97}
{Zheng}, W., {Kriss}, G.~A., {Telfer}, R.~C., {Grimes}, J.~P., \& {Davidsen},
  A.~F. 1997, \apj, 475, 469

\bibitem[{{Zheng} \& {Malkan}(1993)}]{1993ApJ...415..517Z}
{Zheng}, W., \& {Malkan}, M.~A. 1993, \apj, 415, 517

\end{thebibliography}

\appendix

\section{Comparison of foreground and background spectral stacks}
\label{appendix A}
\begin{figure*}
\plottwo{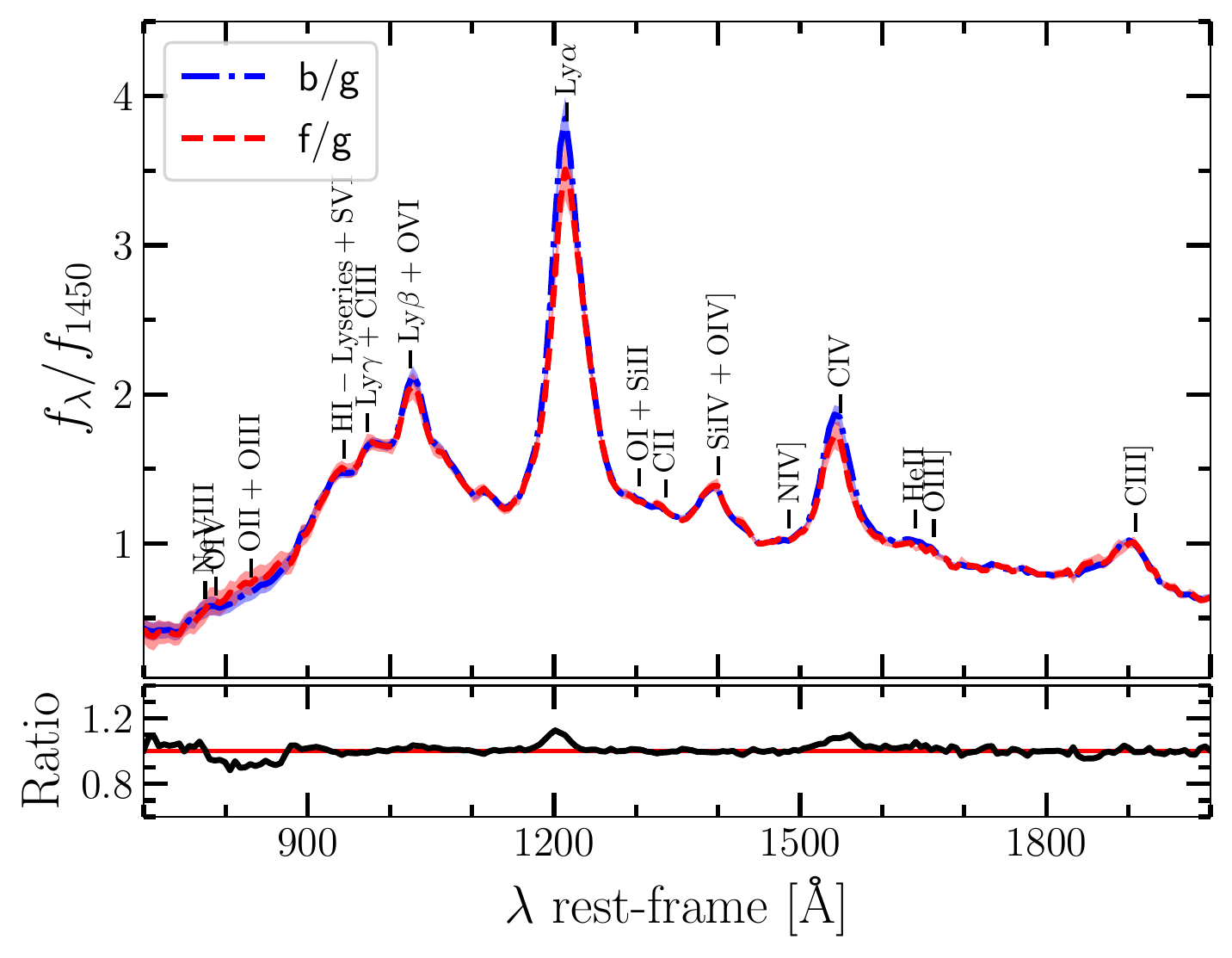}{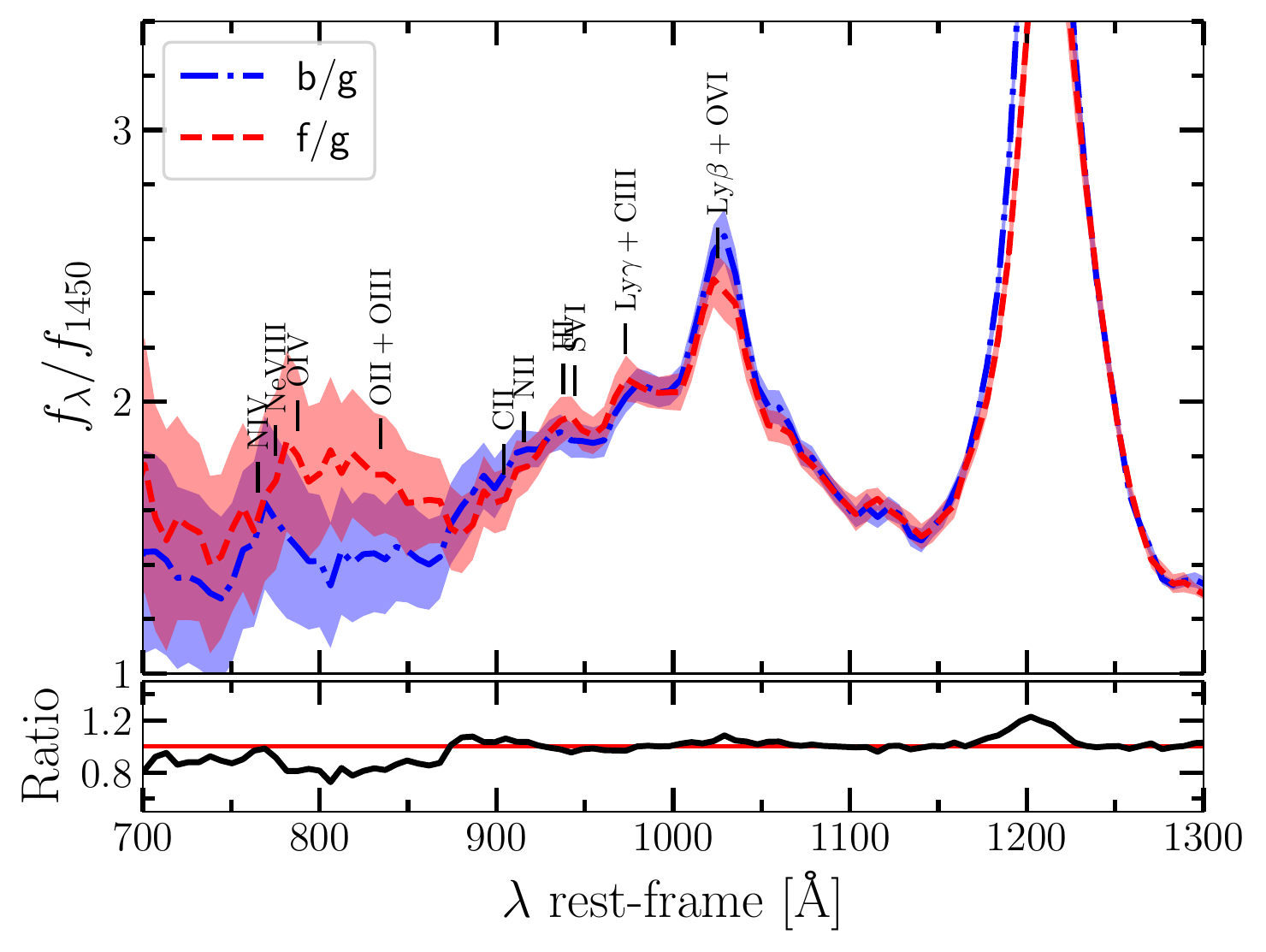}
\caption{{\it Left panel:} Mean observed quasar spectrum (from bootstrap) for the background quasars (blue dot-dashed line) compared to the one obtained for the foreground quasars (red dashed line). The background to foreground spectral ratio is shown in the bottom panel.{\it Right panel}: Mean IGM corrected quasar spectrum with uncertainties from bootstrap (shaded area) for the background quasars (blue dot-dashed line) compared to the one obtained for the foreground quasars (red dashed line)}
 \label{fg_bg}
\end{figure*}
In Figure~\ref{fg_bg} we present the mean observed and corrected for IGM absorption composite normalized to unity at 1450\AA\ for the background (mean/median redshift is 2.275/2.249) and foreground (mean/median redshift is 2.255/2.236) quasar samples separately. It is clear from this comparison that both samples have very similar ionizing continua, whilst differences arise in the most prominent broad emission lines (\ion{C}{iv} and \ion{Ly}{$\alpha$}), in the form of a $\sim20-25$\% emission line flux increase for the background quasar stack compared to the foreground one. Such a feature is likely due to a mild Baldwin effect as the foreground quasars are, on average, somewhat brighter (mean/median $g^\ast=20.10/20.24$) than the background (mean/median $g^\ast=20.21/20.35$) sources.

Moreover, a $\sim20$\% flux decrement at $\lambda\simeq760-880$\AA\ in the background composite with respect to the foreground one is also present, \rev{falling in the midst of prominent quasar emission lines (i.e. NeVIII$\lambda\lambda775$, OIV$\lambda$788, OII+OIII$\lambda\lambda834.5$). We note that this is not highly significant as uncertainties on the average spectra are of the order of $\sim10-12$\% at $\lambda\simeq800$\AA.} We should also point out that our estimates of the quasar redshifts can be fairly uncertain ($\sigma_z\sim500-1000$ km/s), as they have been evaluated from \ion{C}{iv} for the majority of the sources in the sample (\rev{thus subject to offsets from the systemic; e.g. 500--1000~km/s; \citealt{1990ApJ...357..346C}}), while only a few cases \ion{Mg}{ii} ($\sigma_z\sim200-500$~km/s) was available\footnote{\rev{\citealt{1990ApJ...357..346C} found that the velocity difference between the Mg II and C IV lines could exceed 4000 km/s, whilst the one from of CIV from CIII] exceeds 2000 km/s.}}.
Given that the pair separations are very small, it is then possible to confuse a background with a foreground pair, and vice versa.

\end{document}